\documentclass[preprint]{aastex}
\usepackage{natbib,subfig}

\newcommand   {\av}    {\mbox{${\rm A_v}$}}

\renewcommand {\deg}   {\mbox{$^\circ$}}

\newcommand   {\kms}   {\mbox{km\,s$^{-1}$}}
\renewcommand {\ga}    
{\mbox{\rlap{\hbox{\lower5pt\hbox{$\sim$}}}\hbox{$>$}}}
\renewcommand {\la}    
{\mbox{\rlap{\hbox{\lower5pt\hbox{$\sim$}}}\hbox{$<$}}}

\newcommand{\mstar}{M_{\star}}

\newcommand{\lstar}{L_{\star}}
\newcommand{\mdote}{\dot{M}_{\rm env}}

\newcommand{\mdisk}{M_{\rm disk}}

\begin{document}



\def\kms {\hbox{km{\hskip0.1em}s$^{-1}$}} 

\def\msol{\hbox{$\hbox{M}_\odot$}}
\def\lsol{\hbox{$\hbox{L}_\odot$}}
\def\kms{km s$^{-1}$}
\def\Blos{B$_{\rm los}$}
\def\etal   {{\it et al.}}                     
\def\psec           {$.\negthinspace^{s}$}
\def\pasec          {$.\negthinspace^{\prime\prime}$}
\def\pdeg           {$.\kern-.25em ^{^\circ}$}
\def\degree{\ifmmode{^\circ} \else{$^\circ$}\fi}
\def\ut #1 #2 { \, \textrm{#1}^{#2}} 
\def\u #1 { \, \textrm{#1}}          
\def\nH {n_\mathrm{H}}

\def\ddeg   {\hbox{$.\!\!^\circ$}}              
\def\deg    {$^{\circ}$}                        
\def\le     {$\leq$}                            
\def\sec    {$^{\rm s}$}                        
\def\msol   {\hbox{$M_\odot$}}                  
\def\i      {\hbox{\it I}}                      
\def\v      {\hbox{\it V}}                      
\def\dasec  {\hbox{$.\!\!^{\prime\prime}$}}     
\def\asec   {$^{\prime\prime}$}                 
\def\dasec  {\hbox{$.\!\!^{\prime\prime}$}}     
\def\dsec   {\hbox{$.\!\!^{\rm s}$}}            
\def\min    {$^{\rm m}$}                        
\def\hour   {$^{\rm h}$}                        
\def\amin   {$^{\prime}$}                       
\def\lsol{\, \hbox{$\hbox{L}_\odot$}}
\def\sec    {$^{\rm s}$}                        
\def\etal   {{\it et al.}}                     

\def\xbar   {\hbox{$\overline{\rm x}$}}         

\slugcomment{resubmitted to ApJ}
\shorttitle{}
\shortauthors{}


\title{Star Formation in the Central 400 pc of the Milky Way:\\ 
Evidence for a Population of Massive YSOs}
\author{F. Yusef-Zadeh\footnote{Department of Physics and Astronomy,
Northwestern University, Evanston, Il. 60208
(zadeh@northwestern.edu)},
J. W. Hewitt$^1$, 
R. G. Arendt
\footnote{Science Systems and Applications, Inc. and University of Mayland - 
Baltimore County, GSFC, Code 665, Greenbelt, MD 20771},  
B. Whitney
\footnote{Space Science Institute, 4750 Walnut Street, Suite 205, Boulder, CO 80301}, 
G. Rieke
\footnote{Steward Observatory, University of Arizona, 933 North Cherry Avenue, 
Tucson, AZ 85721}, 
M. Wardle
\footnote{Department of Physics and Engineering, Macquarie University, Sydney,  
NSW 2109, Australia }, 
J. L. Hinz$^4$, 
S. Stolovy
\footnote{Spitzer Science Center, California Institute of Technology, Mail Code 220-6, 1200 
East California Boulevard, Pasadena, CA 91125}, 
C. C. Lang
\footnote{Department of Physics \& Astronomy, University of Iowa, Iowa City, IA 52245, USA}, 
M. G. Burton
\footnote{School of Physics, University of New South Wales, Sydney, NSW 2052, Australia}, 
S. Ramirez$^6$}


\begin{abstract} 

The central kpc of the Milky Way might be expected to differ significantly from the 
rest of 
the Galaxy with regard to gas dynamics and the formation of young stellar objects (YSOs). 
We probe this possibility with mid-infrared observations obtained with IRAC and MIPS on 
Spitzer and with MSX. We use color-color diagrams and spectral energy distribution (SED) 
fits to explore the nature of YSO candidates (including objects with 4.5$\mu$m excesses 
possibly due to molecular emission).  There is an asymmetry in the distribution of the 
candidate YSOs, which tend to  be found at negative Galactic longitudes; this behavior 
contrasts with that of the molecular gas,  approximately 2/3 of which is at positive longitudes. 
The small scale height of these objects suggests that they are within the Galactic center 
region and are dynamically young. They lie between two layers of infrared dark clouds 
(IRDCs) and may have originated from these clouds. We identify new sites for this recent 
star formation by comparing the mid-IR, radio, submillimeter, and methanol maser data. The 
methanol masers appear to be associated with young, embedded YSOs characterized by 
4.5$\mu$m excesses. We use the SEDs of these sources to estimate their physical 
characteristics; their masses appear to range from $\sim$ 10 to $\sim$ 20 \msol. Within 
the central $400 \times 50$ pc ($|l|<1.3\degr$ and $|b|<10'$) the star formation rate 
based on the identification of Stage I evolutionary phase of YSO candidates is about 
0.14 
\msol\, yr$^{-1}$. Given that the majority of the sources in the population of YSOs are 
classified as Stage I objects, we suggest that a recent burst of star formation took place 
within the last 10$^5$ years. This suggestion is also  consistent with estimates of star 
formation rates within the last 
 $\sim10^7$ years showing  a peak around 10$^5$ years ago. 
Lastly, we find
that the Schmidt-Kennicutt Law applies well in the central 400 pc of the Galaxy. 
This implies that 
star formation 
does not appear to be dramatically  affected by the  extreme physical conditions
in the Galactic center region.  
\end{abstract}

\keywords{Galaxy: center - clouds  - ISM: general - ISM - 
radio continuum - masers - cosmic rays - stars: formation - galaxies: starburst}

\section{Introduction}
\label{introduction} 

Understanding the processes occurring in the nuclear disk of our own Galaxy is 
interesting not only for insight into our own Milky Way Galaxy, but also because its 
study can potentially provide a template to the closest galactic nuclei.  This 
important region of the Galaxy hosts several sources of energetic activity, and is the 
site of massive molecular clouds with pockets of past and present massive star 
formation. On a larger scale, the distribution of molecular clouds is asymmetric in the 
so-called ``central molecular zone'' showing 2/3 of the molecular gas on the positive 
longitude side of the Galactic center region 
\citep{bania77,bally88,oka05,tsuboi99,martin04,MorrisSerabyn96}.
 The nature of this asymmetry in the context of gas dynamics 
is not understood. This unique region is also distinguished from the rest of the Galaxy 
by the presence of a bar that is responsible for feeding gas into the central few hundred 
parsecs of the Galaxy. In particular, the non-circular motion of the gas in this region 
is thought to be induced by the bar's potential combined with dynamical friction 
leading to migration of molecular clouds to the Galactic center 
\citep{binney91,stark91}.
 The collision of these migrating molecular clouds with each 
other in a small volume of the Galactic center is thought to be the driving mechanism 
for star formation on a wide range of scales \citep{wardlefyz08}. Thus, it is 
possible that the star formation process depends on the Galactic center 
environment 
including the effects of the bar on the gas dynamics.

Although there is a high concentration  of dense molecular clouds distributed in
the Galactic center region, the star formation rate has shown extreme values. On
the one hand, the massive star forming region Sgr B2 points to the closest example
of starburst activity in our Galaxy. This region contains some of the most
spectacular young and dense stellar clusters 
having  an age of a few million
years \citep{figer99,paumard06,lu06}.
In the context of recently formed stars in the
nuclear disk, detailed study using Infrared Space  Observatory (ISO) at 7 and 
15$\mu$m indicates  
the presence of a population of young stellar object (YSO) candidates in 
the
region  $-0.424^\circ < l < -0.115^\circ$ and  $-0.194^\circ < b <
0.089^\circ $ \citep{omont03,schuller06}. On the other hand, 
quiescent giant molecular clouds such as  G0.25-0.01 \citep{lis94,lis98}
appear to contain highly inefficient star formation. 

One intriguing question is whether the star formation in this region proceeds 
analogously to that elsewhere in the Galaxy, or whether the central region 
provides a unique environment and might provide a template for alternative 
modes of star formation in galactic nuclei generally. For example, the 
molecular clouds in the Galactic center region have elevated temperatures. 
Molecular line 
studies yield temperature estimates of 75 to 200K (H\"uttemeister et al. 1993), 
while far-infrared data indicate dust temperatures $\leq$ 30K (Odenwald \& Fazio 
1984; Cox \& Laureijs 1989; Pierce-Price et al. 2000). This temperature 
discrepancy suggests direct heating of the molecular gas either by cosmic rays 
or shocks, which may reduce the efficiency of star formation in the region (Lis 
et al. 2001b; Yusef-Zadeh et al. 2007b).

In order to  census  past and present star formation activity and to gain
insight
into the nature of star formation in the unique environment of the Galactic
center, we carried out a 24$\mu$m survey of the central molecular zone and beyond
between $|l|<4\degr$ and $|b|<1\degr$.  
This is the most recent in a series of papers that have presented sensitive surveys 
of the 
Galactic center region at near-IR and mid-IR wavelengths. 
Our MIPS survey
provides an important data point when  combined with the 
$ Spitzer\ Infrared\ Array\ Camera$  survey of the 
Galactic
center region \citep{ramirez08,stolovy06,arendt08}. This   allows us to
distinguish   YSOs and AGBs by  fitting the 
spectral energy
distribution (SED)  of infrared point sources as well as by constructing color
magnitude diagrams (CMDs). Furthermore,  24$\mu$m observations reveal dense,
cold dust clouds through their absorption against background emission and
identify warm dust clouds through their emission without being contaminated by
PAH emission.  The presence of large-scale 
IRDCs
in the Galactic 
center at this wavelength traces  potential sites of the 
earliest phase of active and quiescent  
star formation.   
In particular,  the correlation of 24$\mu$m point sources distributed toward 
infrared dark clouds with 4.5$\mu$m 
excess sources, 
thought to trace shock excited molecular emission, 
can be used to identify the earliest  phase of massive star formation 
(Chambers et al. 2009). 
Lastly,  24$\mu$m emission points to HII regions with embedded
thermal sources of heating as well as to sites where synchrotron emission from
relativistic particles is produced. The ratio of
24$\mu$m to radio emission 
 can separate thermal from nonthermal sources 
\citep{furst87}.

The structure of this paper is as follows. Section 2 describes details of 
observations using MIPS, IRAC, and radio data, while section 3 presents the 
large scale distribution of extended clouds and compact stellar sources. Then 
in section 4 we focus on the stellar distribution of 24$\mu$m sources and the 
color magnitude diagram of the sources covered by the MIPS and IRAC surveys. 
Section 5 discusses the SEDs of candidate YSOs followed by a brief description of 
individual candidate sites of on-going star formation activity. The new sites 
of young, massive star formation are found by the detection 
of sources with excess 4.5$\mu$m emission. 
There is  a rich collection of prominent infrared sources toward the Galactic 
center, some of which such as  Sgr A, the radio Arc and Sgr C, are saturated.  
We discuss elsewhere the distribution of 24$\mu$m emission from Sgr A 
(G0.0-0.0) and the Galactic center radio Arc (G0.2-0.0), the Sgr B HII complex 
(G0.7-0.0) and G0.6-0.0, Sgr C (G359.4-0.1), Sgr E (G358.7-0.0) and the Tornado 
nebula. Here we assume the distance to the Galactic center is 8.5 kpc (Ghez et al. 
2008).

\section{Observations and Data Reductions}

Apart from the 24$\mu$m survey, several  other surveys of this region have recently 
been  carried  out in other wavelength bands, as described briefly below. 
We have correlated these data  with our MIPS survey data. 

\subsection{MIPS Survey at 24$\mu$m}

The Galactic Center region is very bright at mid-IR wavelengths. Therefore, 
to cover the full 8 $\times$ 2 \degr survey area most efficiently, we used 
MIPS in fast scan mode with large cross-scan steps of 302$''$.
Because of the permitted 
scan orientation during the Galactic center visibility window, the map was a 
parallelogram. 
This fast scan mode achieved enough 
individual source sightings for high reliability and reached the specified 1 mJy 
detection limit. Eight rectangular scans were performed in sequence, each taking 
approximately 2.3 hours, giving a total observing time of $\sim$ 18 hours.  
 The diffraction limit of the 24$\mu$m image is 5.8$''^2$.  These data 
 represent the highest spatial resolution and sensitivity large-scale map 
 made of the inner few degrees of the Galactic center at 24$\mu$m. The source 
 photometry was based on the  24$\mu$m catalog compiled  by 
Hinz et al. (2009).

Our data are complemented by the {\it Spitzer} Legacy program MIPS Inner Galactic 
Plane Survey (MIPSGAL) which observed 220 deg$^2$ of the inner Galactic plane, 
$65\degr > l > 10\degr$ and $-10\degr > l > -65\degr$ for $|b| < 1\degr$, at 24 
and 70\,$\micron$  \citep{carey09} which  avoided 
duplicating coverage of the Galactic center 
region. MIPSGAL-II post-BCD data were used to fill gaps around the high latitude 
edges. This was done by reprojecting the MIPSGAL II data over the whole field. 
The MIPSGAL data are only used where there is zero coverage at the edges of 
the mosaic.

There are  several HII  
regions where  the 24$\micron$ emission is saturated. The 
most extended saturated regions lie toward Sgr A, the radio Arc and Sgr C, near $l 
\sim 0\arcdeg, 0.2\arcdeg, 0.6\arcdeg$, respectively. The point source saturation 
limit, depending on the background level, ranges between 4 and 4.5 Jy which 
correspond to 0.60 and 0.51 magnitudes at 24 $\micron$. In order to remedy the 
saturation artifacts, we replaced the saturated regions by MSX Band E data (21.34 
$\micron$) using the following steps. (1) The MSX data are reprojected to match 
the pixel scale and alignment of the MIPS image. A relative distortion between the 
positions of sources in MSX and MIPS images was  corrected by a 4th-order 2-D 
polynomial warping of the MSX images. This correction reduced positional 
differences to $\sim$ 0.5 pixels ($2.49''$ pixels). (2) A linear fit to the 
correlation between pixel intensities is performed to derive a photometric 
scale 
factor and offset to be applied to match MSX Band E intensities with MIPS 
24$\micron$ intensities. The correlation is performed with bright (but not 
saturated) data by restricting pixel intensities to $> 500$ MJy sr$^{-1}$ for the 
MIPS data and in the range of $[1-4.5]\times10^{-5}$ W m$^{-2}$ sr$^{-1}$. The 
derived scale factor was $\sim$40\% larger than the quoted factor for 
converting  MSX data from W m$^{-2}$ sr$^{-1}$ to MJy sr$^{-1}$. The reason for 
such a discrepancy is not clear, but part of the difference may be caused by the 
slightly different nominal wavelengths of the instruments. (3) A mask is defined 
to identify the saturated regions. The mask is set to 1.0 where the rescaled MSX 
data are brighter than 1000 MJy sr$^{-1}$ and then ramped down to 0.0 for any 
pixels $>3$ pixels away from these regions. (4) The mask, and 1.0 minus the mask, 
are then applied as weights in the linear combination of the MSX and MIPS images. 
Thus, in the resulting image, 99.6\% of the pixels contain unmodified MIPS 
intensities, 0.3\% of the pixels contain scaled MSX intensities, and in between 
0.1\% of the pixels contain a weighted average of MIPS and scaled MSX intensities. 
(5) Lastly, we interpolate over $\sim$10 small MIPS coverage gaps that occupy 
0.03\% of the image.

Despite the weighted averages, differences in the PSF widths ($\sim18.3''$ 
for 
MSX, $\sim6''$ for MIPS) create artifacts where MSX data replace saturated MIPS 
point sources. Therefore, any quantitative analysis on MIPS  saturated sources 
should be performed on the original MSX images, rather than the combined MIPS+MSX 
images. The combined images are useful for qualitative evaluation of the structure 
and color of bright extended emission, and of course, for any unsaturated 
emission.

\subsection{MIPS Survey at 70$\mu$m}

In order to obtain more complete SEDs of some of the high latitude  24$\mu$m 
sources 
that show
excess emission at 4.5$\mu$m,  we selected a  
total of nine
such
sources distributed at high Galactic latitudes 
within
our survey. These sources were  
observed separately with MIPS at 70$\mu$m.  The low-latitude sources were not selected 
due  their possible saturation at low latitudes in the plane.  
These targets  were observed in MIPS photometry mode 
with the exposure time for 
each object of  3 seconds and default pixel scale of 10\,$\arcsec$.
All observations were obtained between 2007 September 21 and 23.  These
data were reduced using the MIPS Instrument Team's Data Analysis Tool v.
3.06 \citep{gordon07}.  Aperture photometry was performed
utilizing a set of custom IDL scripts.  The data were flux calibrated using
the conversion factor of 702$\pm35$ MJy sr$^{-1}$ per raw MIPS units (5\%
uncertainty) following Gordon et al. (2007). 
An aperture of $35\arcsec$ was used, 
with a sky radius of 39$\arcsec-65\arcsec$.  The aperture
correction for this configuration is 1.185.

\subsection{Radio Surveys}

Radio continuum observations at $\lambda$20cm surveyed the  
$-2^\circ < l < 5^\circ$ and $|b| < 40'$ region  with a
spatial resolution of $\approx30''$ and 10$''$ using two different array
configurations of  the Very 
Large Array (VLA) of the
National Radio Astronomy Observatory\footnote{The National Radio
Astronomy Observatory is a facility of the National Science Foundation,
operated under a cooperative agreement by Associated Universities, Inc.} \citep{fyz04}.
Another recent survey at 20cm has extended 
this region to higher positive latitudes of up to 1\degr \citep{law08}. 
We obtained 32 overlapping images at 20cm from these two 
surveys and  convolved individual images  with a Gaussian having  a 
FWHM=12.8$''\times12.8''$
before the images  were  mosaiced.

This region has also been observed at 20 and 6cm with the VLA 
\citep{becker94gps,zoonematkermani90}.
 However, these surveys are  
short snapshot observations of a region dominated by bright sources 
with a wide range of angular scales and poor sensitivity
and so are not useful
for comparison with infrared sources  due to 
the non-uniform {\it uv} coverage. 
 The detection of  point 
sources at 90 and 20cm \citep{Nord04,fyz04,law08} does 
suffer from the artifacts introduced by the lack of {\it uv} coverage. More 
recently, Lang et al. (2009) have 
completed a 6cm survey of the Galactic center region with a uniform {\it uv} 
coverage. We have used this data set  to measure the fluxes of several point 
sources 
in the G359.43+0.02 cluster, as described below. 

\subsection{Other Surveys}

A survey of Spitzer/IRAC observations of the central 2\degr$\times$ 1.4\degr\
($\sim290\times210$ pc) was  recently conducted 
in the four 
wavelength
bands at 3.6, 4.5, 5.8 and 8$\mu$m \citep{stolovy06,ramirez08,arendt08}. 
The catalog presented in Ramirez et al. (2008) includes point source
 photometry from the IRAC survey and correlates those sources with
previously published  photometry from 2MASS at J, H, and K$_s$
bands \citep{Skrutskie97}.

A Submillimeter Common-User Bolometer Array (SCUBA) survey of the Galactic 
center covered a region of 2\degr$\times$0\ddeg5  at 450 and 850$\mu$m with a
spatial resolution of 8\asec and 15\asec, respectively \citep{pierceprice00}.
 Given
the low resolution of the submillimeter data, the detected submillimeter 
flux  is used  as   upper limits
at 450$\mu$m and 850$\mu$m \citep{difrancesco08} in the spectral energy distribution of 
individual point sources. 
However, large-scale submillimeter maps of extended sources  at 450 and 850$\mu$m have 
been used for comparison with radio and mid-IR images.   

\section{Large-Scale Distribution of  24$\mu$m Emission} 

\subsection{Extended Sources} 

Figure \ref{fig:1}a shows the large scale view of the surveyed 
region at 24$\mu$m. In order
to bring out the weak sources, the  sources distributed  between  
--1.8$\degr<
l < 0.8\degr$  are burned out in this figure. The mean  brightness 
of this 
region is roughly 4 to 5 times higher than the region beyond the inner few degrees
of the Galactic center. Several extended HII regions   are identified to have
radio continuum and submillimeter  counterparts \citep{fyz04,pierceprice00}. An
enlarged view of the  burned-out  3\degr$\times$2\degr region is shown in  
Figure \ref{fig:1}b.
Prominent Galactic center  HII complexes along the Galactic plane are associated
with Sgr A -- E, the Arches and the Pistol, all of which  are 
labeled, as  shown in Figure
1c. We note two extended 24$\mu$m emitting features G0.23-0.05 and G0.31-0.07 
lying between the 
radio Arc and Sgr B1. 
The radio Arc,  which  includes the Arches cluster and  the Pistol nebula,    
 consists of a network of nonthermal radio 
 filaments running perpendicular to the Galactic plane \citep{fyz84,lang99}. 
We note a number of parallel filamentary features at 24$\mu$m associated with G0.31-0.07 
which will be 
discussed elsewhere. 
At positive latitudes, two extended clouds known as the western and eastern
Galactic center lobes are shown prominently in Figure \ref{fig:1}b near  
$l\sim -0.5\degr$ and
$\sim 0.2\degr$, respectively. The striking structure  of the  western lobe with its
strong 24$\mu$m emission coincides  with  AFGL 5376 \citep{uchida90}. In the 
eastern lobe, two columns of 24$\mu$m 
features with an extension of about 30$'$ run  parallel to each other away from
the Galactic plane.  These features lie in the region of the Galactic center lobe
that is known to emit strong synchrotron emission \citep{law08}. The
``double-helix'' nebula \citep{morris06} lies along the northern extension of  the
eastern linear feature in the Galactic center lobe. The southern ends  of these
linear  features appear to be  pointed toward the large-scale  ionized features
associated with the Arches cluster G0.12+0.02 and a cluster of HII regions 
(H1--H5) \citep{fyz87morris,zhao93}. In addition,  there are several 
counterparts
to foreground H${\alpha}$ emission line nebulae,  RCW 137, 141,  and
142, which are prominent at $|b| >$ 0\ddeg2.

At  positive longitudes, a string of IRDCs is 
concentrated between l$\sim0.2\degr$ 
and Sgr B2 near l=0.7$\degr$.
Sub-millimeter emission from these clouds is prominent  at 450 and 850$\mu$m
\citep{lis94,pierceprice00}. Figure \ref{fig:2}a  displays an enlarged view of 24$\mu$m
emission from dust  clouds and stellar sources distributed at  positive
longitudes.  With careful adjustment of the overall background level, the ratio 
of IRAC images, such as  
 $I(4.5) / [I(3.6)^{1.4}*I(5.8)]^{0.5}$, 
 are  used to highlight the location and 
structure of the IR dark clouds. This ratio is high for the IRDCs  (Figure 
\ref{fig:2}b).
The bright feature in this figure 
corresponds to a dust lane that  consists of  a  string of dust clouds. These
clouds appear narrow  near l=0.25\degr  and become increasingly  wider extending
toward negative latitudes near Sgr B2. The kinematics of this dust lane have  
been
studied in numerous molecular line surveys  of this region \citep{oka98,martin04}.

At negative latitudes, the distribution of dust emission at 24$\mu$m and the
ratio map  are shown in Figure \ref{fig:3}a,b, 
respectively. 
IRDCs are detected toward
the well-known molecular clouds M-0.13-0.08 (20 \kms) and
M-0.02-0.07 (50 \kms) \citep{HerrnsteinHo05,ArmstrongBarrett85}.
Both clouds are located near the Galactic center. One of the
  largest  structures in the ratio map is the pair of two
 dust lanes that 
run parallel to each other along the Galactic plane between
l=0\degr  and l=-40$'$ on the positive and negative latitudes.  The previous CS 
(1-0)
line study of this region has shown  the counterpart to 
the positive latitude dust lane to have   a
coherent kinematic structure.  We also note an elongated dust cloud (called ``Vertical 
IRDC'' in Fig. 1c) running toward
positive latitudes at l=-40$'$, b=4$'$  with a latitude  extent of 12$'$. 
The prominence of this feature with respect to the dust lanes in the ratio map suggests 
that it is a foreground cloud.

\subsection{Color Distribution of Extended Emission}


In order to analyze the color
distribution at different bands, a close-up view of the central region of the
24$\mu$m survey is cropped to the exact size of the 8$\mu$m survey
\citep{stolovy06}. Figure \ref{fig:13}a shows the combined 24, 8 and 4.5$\mu$m images with
red, green and blue colors, respectively. Figure \ref{fig:13}b shows only the 
3.6, 4.5, and 8$\mu$m images in blue, green and red colors, respectively.  Overall,
there is a great deal of non-uniformity in the color distribution detected between
IRAC and MIPS images, as shown in Figure \ref{fig:13}a.  In contrast, uniform color
distribution is noted in IRAC images, as shown in Figure \ref{fig:13}b
\citep{arendt08}. One of the main reasons for the non-uniformity of colors 
is
that 24$\mu$m emission is an excellent diagnostic of  local sources of
heating of large dust grains. This is in part due to the fact that thermal emission 
from 
large dust grains at 24$\mu$m becomes
significantly enhanced  when the radiation field is only 100 times that of the 
local interstellar radiation field (ISRF) in the solar vicinity
\citep{arendt08}. In contrast, 8$\mu$m emission mainly arises  from 
aromatics (hereafter PAHs) that are excited by the interstellar radiation 
field and nearby OB stars.  Thermal emission from small
dust grains becomes strong enough to compete with the PAH emission in the 8$\mu$m
band only when the local  radiation field  is about 10$^4$ times that of
the solar neighborhood \citep{LiDraine01}. Thus, there is much less variation in 
the 8$\mu$m  to 5.8$\mu$m color than in the 24$\mu$m  to 8$\mu$m color.
These color distribution images based on IRAC and MIPS data  place a lower 
limit of $\sim$100 times the ISRF throughout the Galactic center region with 
the exception of the regions known to have a very strong 
radiation field such as the Arches and  Sgr A (see Arendt et al. 2008).


Several extended sources show a distinct pattern of excess 24$\mu$m emission 
(red in Fig. 4a) surrounded by 8$\mu$m emission (green and yellow in Fig. 4a). 
The 24$\mu$m excess (or 8$\mu$m deficit) can be accounted for 
 by the
destruction of PAHs by local sources of UV radiation or shocks 
\citep{watson08,churchwell07,povich07}. In contrast
to small grains and PAHs that can be destroyed, the large dust grains re-radiate
the absorbed UV radiation at long wavelengths, 24$\mu$m.  The most striking
example that shows these effects is the optical H${\alpha}$ line emission nebulae
RCW 137,  141 and 142 which are foreground sources and are internally heated.
PAHs are destroyed close to the source of heating, but are excited at the edge of
the cloud by the external ISM radiation field. The 24$\mu$m diffuse emission along
the Galactic plane traces local sources of heating by HII regions.


There is another type of 24$\mu$m source that shows neither 8$\mu$m 
counterpart nor any evidence for embedded sources of heating.  AFGL 5376 
\citep{uchida90} is one example that is located at high latitudes. 
G0.85-0.44 is another 
source that shows   similar characteristics.  Alternative mechanisms 
must 
be  responsible for heating these clouds. One 
candidate is the high flux of cosmic rays that could be 
responsible 
for heating large dust grains but not PAHs.


\subsection{Infrared Dark Clouds}

We note several IRDCs in Figure \ref{fig:13}a with different levels of darkness. 
These cold, dark clouds are seen in absorption against the diffuse
background emission. 
The IRDCs are optically thick even at 24$\mu$m. Therefore, their apparent surface 
brightnesses are more of an indication of the emitting column density on the line of 
sight to the cloud (i.e., the cloud's distance) than a sign of the optical depth of 
the IRDCs. For example,   we note that the IRDCs at high and low Galactic plane latitudes, such as
G0.2+0.47 and G-0.67+0.18 are the darkest clouds, suggesting that they are
distributed locally along the line of sight.

As was shown in the ratio maps, the  large scale distribution  
of IRDCs forming a dust lane  
 implies  an early phase of massive star or star cluster 
formation in the Galactic center 
region.
IRDCs  are  dense ($>10^5$ cm$^{-3}$), cold ($<25$K) and attain high column 
densities ($10^{23}$ cm$^{-2}$) in 
the Galactic disk \citep{egan98,carey00,simon06}. 
The extreme physical properties of the clouds in
the Galactic center region should be even 
more suitable for young, massive star cluster  formation.   
We note the string of IRDCs is coincident with  the ``Dust Ridge''
\citep{lis01}, as shown in Figure 1c.
What is remarkable is 
a  large-scale east-west ridge of IRDCs tracing dense molecular gas  between
G0.25+0.01 and Sgr B2.  
This continuous east-west IRDC ridge   
shows a sinusoidal appearance on a scale of $\sim5'$ corresponding to 12
pc. The western half of this large-scale ridge of IRDCs  is quiescent 
\citep{lis01}
in its star
formation activity whereas the eastern half shows spectacular
on-going star formation.
The string of submillimeter emitting molecular clouds includes G0.25+0.01 and
the Sgr B complex and is darker than the well-known 
20 and 50 \kms molecular clouds
 M-0.13-0.08 and M-0.02-0.07 as well as clouds in the negative longitude
side of the Galactic center. This behavior  suggests that the 
 positive-longitude IRDCs are closer  than the   20 and 50 
\kms\
and Sgr C clouds. 
The large-scale distribution of IRDCs traces coherent  molecular
lanes  running parallel to the nuclear disk
\citep{bania77,bally88,oka05,tsuboi99, 
martin04}, as  shown in ratio maps of Figures 2 and 3. 
The color contrast of the coherent, large scale dust lane suggests  that 
the molecular gas in the central molecular zone is oriented in the same direction 
as that of  the  well-known stellar  bar  \citep{binney91,stark91}.


\subsection{Distribution of Compact  Sources}

The blue color (4.5$\mu$m emission) in Figure \ref{fig:13}a represents the stellar 
density 
distribution in Sgr A
where the evolved stellar core of the bulge peaks. This is mostly unresolved stellar 
sources, not intrinsically diffuse emission. 
A number of red (24$\mu$m) point sources
are also seen in this figure  that appear to be embedded within IRDCs suggesting
that they are protostellar candidates, as will be discussed in section 5. The 
red color
reveals embedded stellar sources that heat surrounding dust grains to high 
temperatures.
Numerous compact HII regions with radio continuum counterparts are distributed
along the Galactic plane near  the Galactic center.

To reveal the distribution of
intrinsically red compact sources against the strong extended emission, we
constructed a 24$\mu$m to 8$\mu$m intensity ratio image, which is presented in 
Figure \ref{fig:14}a. Regions with relatively strong 24$\mu$m emission are 
light 
whereas regions with strong 8$\mu$m (PAH) emission are dark. Most of the 
emission from the ISM (including that on the line of sight to IRDCs) disappears 
into a flat gray background. The disappearance of IRDCs is because they are 
opaque at both 8$\mu$m and 24$\mu$m. The extinction ratio (A$_{8\mu \rm 
m}$/A$_{24\mu \rm m}\sim1$) \citep{LiDraine01} implies a lack of reddening even 
for clouds that are not opaque (e.g., their edges).  The ``removal'' of the 
typical ISM emission and dark clouds accentuates a number of structures and 
helps reveal some fainter, compact and extended sources with excess 24$\mu$m 
emission. The most striking feature in the ratio map is the distribution of 
compact emission from a high concentration of strong IR-excess objects, 
candidate YSOs or UCHII regions distributed between $l=0\degr$ and 
$l=-1.2\degr$.

Figure \ref{fig:14}b shows 
the relationship between the compact 24$\mu$m point 
sources and the molecular gas. 
It is a composite image indicating  24$\mu$m 
in green and 450$\mu$m in red.  Sub-millimetre emission 
at
450$\mu$m is optically thin and is known to be an excellent tracer of molecular gas
distributed in the central molecular zone. The comparison of the distribution of
24$\mu$m emission and 
the ratio map shows two features. One is the distribution of
extended sub-millimetre emission being strongest in the positive longitude side 
the
Galactic center. It is clear that molecular gas distribution at $l>0\arcdeg$ is
more uniform and more intense at 450$\mu$m than that at negative longitudes.
Similar asymmetric structure  is also noted in CS and CO line observations
\citep{bally88,martin04,oka05,tsuboi99}.  This asymmetric distribution shows 2/3
of the mass of molecular gas on the positive longitude side of the Galactic center
region \citep{bally88,oka05,martin04}. The presence of such an
asymmetry is mainly due to  massive clouds associated with Sgr B2 and the dust ridge and 
a lack of massive, dense
clouds at $l<0\arcdeg$ with the exception of Sgr C.

The second feature is the distribution of 450$\mu$m emission at negative longitudes
showing that 24$\mu$m sources anti-correlate with the distribution of dust clouds.
Remarkably, the distribution of 24$\mu$m point sources appears to be sandwiched by two
layers of  dust and gas clouds running parallel to the Galactic plane
and symmetrically distributed with
respect to the Galactic center.
 The ``bow-tie'' dust
layers, as schematically drawn in Figure \ref{fig:1}c, stretch along 
both  positive and 
negative longitudes and 
 coincide with IRDCs at 8$\mu$m and 24$\mu$m. 
Submillimeter emission is also noted from a cloud at  negative 
latitudes  b=-5$'$. This feature is the negative latitude counterpart to 
the layer of submillimeter emitting cloud 
in the region between G0.25+0.01 
and Sgr B1 at positive longitudes. 
 The negative latitude submillimeter layer does not have an  IRDC counterpart,  
implying that  it 
is most likely located on the far side of Sgr B2.
Thus, it is likely
that these molecular layers are closer to us at positive longitudes than at negative ones. 
What is remarkable from the mid-IR and submillimeter images are that
the "bow-tie" structure  in fact consists of two layers  of 
molecular and dust clouds that   
are symmetrically distributed with respect to the Galactic center.  However, 
it is most likely that the positive longitude 
side of these molecular layers is  closer to us than those 
at negative latitudes. 

We also note that unlike the gas and dust
distributions, the 24$\mu$m point sources are  more uniformly distributed on the 
negative
longitudes than at positive ones.
 Given that the chain of  IRDCs that include Sgr B2 are 
thought to 
be closer than those at negative longitude, it could be that they hide a 
distant population of 24$\mu$m sources and create an apparent asymmetry, as discussed 
in more detail below. 

\section{Candidate YSOs}

To investigate the nature of the 24$\mu$m compact sources, 8--24$\mu$m ([8]-[24]$\mu$m) color 
magnitude 
diagrams (CMDs) are constructed for sources  within $|b|=10'$  and  $|l|$=1.4 
degrees, corresponding to the central 0.4$\times$ 0.05 kpc region. The 
CMD of these sources identified in
the positive and negative longitudes is shown in Figure \ref{fig:15}. 
These data used the MIPS catalog at 24$\mu$m \citep{hinz09}
and IRAC catalog at 8$\mu$m \citep{ramirez08}.  A total of 172,670 sources is 
detected,
70,794 and 101,876 of which are found in the region restricted to positive and
negative longitudes, respectively. 
There is an excess 
of
31082 sources identified in negative longitudes.  This asymmetry could be
explained by the high extinction experienced by mid-IR sources in the region where
dense molecular clouds are highly concentrated.  Alternatively, the distribution
of compact dusty sources could be due to an intrinsic excess of 
YSOs  or evolved AGB stars. 
The dashed line in Figure 6 is chosen  empirically, as discussed below,  
to separate the reddest YSOs from AGBs. 
Each of the two  possibilities is discussed
below.

OH/IR stars are oxygen-rich, mass-losing cool giants that 
 represent evolved 
AGB stars at 
the end of their lives \citep{Habing96}. These post AGB stars are known to be dusty with 
excess emission at infrared wavelengths. YSOs are also known to have excess emission in 
infrared wavelengths and are characterized by 
 dusty envelopes and disks that absorb the radiation from the central protostar 
\citep{whitney03b,adams87}. To explore the color of evolved stars in the 
Galactic center, we examined the spatial distribution of known OH/IR stars
in the region shown in  
Figure \ref{fig:15} 
using the OH 1612 MHz catalogs \citep{lindqvist92,sevenster97a,sevenster97b,sjouwerman98}.
 OH/IR 
surveys are not affected by visual extinction. These surveys  have uniformly sampled 
 the region  covered by  IRAC observations. 
The shallow but uniformly sampled survey by \citet{sevenster97a} finds 
a total of 7 and 9 
OH/IR stars in the restricted region, for positive and negative longitudes, respectively. 
Similarly, the spatial distribution of OH/IR stars on a large scale is symmetric with 
respect to the Galactic center. Although, there are not many OH/IR masers that we found to 
have infrared counterparts in the restricted region of our survey, we find no evidence of 
asymmetry in the distribution of OH/IR stars with respect to the Galactic center.

To examine whether the observed color of the infrared sources 
(Ortiz et al. 2002; Ojha et al. 
2007)  can be due to the 
excess of YSOs, we first identify the color of OH/IR stars. The reddening 
toward these Galactic center sources is minimally affected by the differential 
extinction between 8 and 24$\mu$m. \citet{Flaherty07} have recently derived 
A(8$\mu$m) - A(24$\mu$m) $\sim 0.0 \pm0.05 \times$ A(K). 
The color variation 
between 8 and 24$\mu$m should be $<$0.2 
mag  assuming  A(K) $\sim 3.2$ 
or $\sim$30 magnitudes of visual extinction. 
Thus, the observed color of the sources is 
intrinsic. Here, we make an empirical assumption that the very red sources are 
likely to be candidate YSOs.  Figure \ref{fig:16}a shows the CMD of all the 
24$\mu$m sources found in the surveyed region by IRAC. The colors of OH/IR 
stars that are found using the survey by \citet{sevenster97a} are shown as 
triangles in Figure \ref{fig:16}a and are distributed to the left of the dashed 
line that was selected empirically.  The color of OH/IR masers is clearly 
distinguished from the reddest and faintest sources (by a few magnitudes). The 
dashed line in Figure \ref{fig:16}a is the slope that separates most of the YSO 
candidates to the right from the rest of the sources that have 24 and 8$\mu$m 
counterparts. The left side of the slope in Figure \ref{fig:16}a indicates a 
trend of IR excess sources with luminosity. The CMD of the LMC shows a similar 
trend  among luminous AGB stars due
to the relationship between increasing radiation pressure resulting in
more mass loss in  dust, and 
therefore more IR excess (see Fig. 3a of Whitney et al. 2008). Thus, we believe 
the sources to the right of the slope in Figure \ref{fig:16}a are dominated by 
candidate YSOs. The color-color diagram between [8]-[24]$\mu$m is also shown in 
Figure \ref{fig:16}b where the candidate YSOs are shown as crosses. In the 
restricted $l=\pm1.4^0$, $b=\pm10'$ region the total number of YSO candidates 
is found to be 347 and 212 distributed on the negative and positive longitudes, 
respectively. The spatial distribution of candidate YSOs with [8]-[24] greater 
than 4 is shown in Figure \ref{fig:16}c revealing that the reddest infrared 
sources are distributed mostly along the Galactic plane. The numbers of YSO 
candidates are lower limits because the colors of stars in the saturated 
regions 
near $l\sim 0^0$ and $l\sim0.2^0$ could not be examined (see Figure 
\ref{fig:16}c). In spite of these restrictions, we note that most of the 
candidate YSOs are distributed in the negative latitudes, very close to the 
Galactic equator.

The saturated sources  at 24$\mu$m as well as the 
contaminating  sources limit our ability to  
address the issue of completeness. 
However, we believe that the distribution 
of OH/IR sources validates our empirical choice for the slope of the dashed 
line used to identify most of the YSO candidates. An additional step is taken 
in 
$\S5.3$ to 
identify the fraction of YSO candidates  via SED fitting of the sources selected 
from  the right region  
of the CMD of Figure 7. As discussed below, 
$\sim$60-65\% of the total number of sources from the CMDs are SED fitted.   
 Ultimately, future spectroscopic 
measurements  of YSO candidates should be able to address the completeness 
issue.



\subsection{Excess  Candidate YSOs in l$ < 0\degr$?}

To test the possibility that there is an asymmetric distribution of
candidate YSOs with respect to the Galactic center, we examined two tracers that
are not affected by visual extinction.  First, we examined compact radio continuum
sources \citep{fyz04} and found 44 and 58 sources in the
positive and negative longitude sides of the Galactic center, restricted to the same 
regions where candidate YSOs are found. It 
is known that Sgr
B2 is one of the most active star-forming sites in the Galaxy at $l\sim0.7^0$
\citep{mehringer98,depree05}. We note that 16 of the 44 compact radio
sources at  the positive longitudes arise only from ultracompact HII regions in 
Sgr
B2. On the other hand, compact radio continuum sources in the $l<0^{\arcdeg}$ 
region
are distributed more  uniformly than in the region at $l>0$ degrees. The excess
number of 14 sources at negative longitudes  combined with non-uniform
distribution of compact radio continuum sources at  positive longitudes 
imply
an asymmetry of radio continuum sources that can not be explained by excess
extinction at positive latitudes. This is consistent with the 
excess of infrared sources in the negative longitudes of the Galactic center. Additional support for this picture comes 
from H$_2$O maser study of the same
region showing a similar distribution to that of compact 20cm radio continuum
sources. This survey used the 
$Infrared\ Astronomical\ Satellite$ (IRAS) Galactic 
center catalog \citep{gtaylor93} to search for H$_2$ masers.  The largest concentration of star 
forming water masers is found in Sgr B2
whereas the star forming H$_2$O masers are populated uniformly in
the negative longitude sides of the Galactic center \citep{gtaylor93}. 
For example, the targeted survey by Taylor et al. 
which excluded the Sgr B2 region, detected a total of 13 and one star forming
H$_2$O masers at the negative and positive longitudes, respectively.  
Although water maser 
survey observations  have not uniformly sampled the Galactic center region, the available data 
are consistent with the picture that the largest concentration of quiescent
molecular clouds  such as  the string of IRDCs distributed in $l> $0\degr\ 
do not have a high density  of 24$\mu$m
sources, H$_2$O masers and ultracompact radio continuum sources \citep{lis94}.

Since most of the  candidate YSOs are distributed asymmetrically on the 
negative
longitude side of the Galactic center,  we constructed another histogram of
candidate YSOs for  $l< 0^0$. This histogram should be  a better representation 
of
the   scale height of the uniformly distributed YSO candidates. Gaussian fits to
the narrow and broad  components give FWHM$\sim5.17'\pm0.2'$ centered at b=-2.4$'$
and FWHM$\sim28.6'\pm2'$ centered at $b=-0.02'$, respectively. The  scale height
of  the YSO candidates in l$<0^0$ is $h\sim$6.3 pc  is similar to $h\sim7$ pc 
when all the sources at both positive and negative longitudes are included.


\subsection{Distribution of 24$\mu$m Sources}

To examine the distribution of YSO candidates as a function of latitude, a 
histogram was made using the data from the central 2.6$^0 \times 2^0 (l\times 
b)$. Figure \ref{fig:17}a shows a broad and narrow component. The broad and 
narrow components are fitted with two Gaussians. The broad component is 
centered near $b\sim0'$ with  a FWHM=$64.9'\pm2'$ whereas the narrow component is 
displaced with respect to the Galactic plane at $b=-3'$ with a FWHM$\sim 
5.6'\pm0.2'$. Figure \ref{fig:17}b shows the distribution of the candidate YSOs 
when they are restricted to $|b| < 10'$. This distribution is consistent with 
the narrow latitude distribution of YSO candidates with an offset with respect 
to the Galactic equator near $b=-2.4'$. This value is likely to be close to the 
true latitude of the Galactic plane since Sgr A*, the massive black hole at the 
Galactic center, lies at $b=-2.8'$. Unlike the broad distribution that is 
likely due to foreground sources toward the Galactic center, the narrow 
distribution of the candidate YSOs suggests a uniform sample of sources located 
along the midplane of the Galactic center. The scale height of the candidate 
YSOs is estimated to be $h\sim$7 pc assuming the distance to the Galactic 
center is 8.5 kpc.  Molecular line observations of this region based on CS 
(1-0) transition also finds a narrow scale height $h\sim10$ pc \citep{bally88}.  
The narrow scale height of both candidate YSOs and molecular clouds can be used 
as a support for the association of infrared sources and the disk population of 
molecular gas distributed in the Galactic center region.  To test that the 
reddest sources belong to a population of YSOs distributed in the midplane of 
the Galactic center region, 
the YSO candidates from the color-magnitude diagrams of Figures 6 and 7 
are presented in two
dimensional histograms in Figures 8c and 8d to emphasize the higher fraction of 
brighter, 
redder YSO 
candidates
within $|b| <$ 10\arcmin\ and  the exterior region $|b| >$
10\arcmin\ , respectively. 
These CMDs are selected from the point source catalogs. The 
fraction of redder YSO candidates with [8]-[24] $>$ 3, 
relative to the total number of sources  in the inner 10$'$ 
is found to be  a factor of five to ten  higher than the region 
beyond the inner 
10\arcmin. This suggests that there are more red YSOs in proportion to the total 
YSOs detected for the innermost region of the Galactic center than those found 
in the outermost region. There are 1720 YSO candidates, 599 of which are in the 
region restricted to $-1.4\arcdeg < l < 1.4\arcdeg$ and $-10' < b < 10'$; there 
are 347 and 252 sources distributed in the negative and positive longitudes, 
respectively.

To compare the scale heights of  candidate YSOs and the bright 24$\mu$m point 
sources, we constructed Figure \ref{fig:18}a which shows the distribution of bright
unsaturated  24$\mu$m  sources between 0 and 5 magnitudes as a function of
latitude within
$|l|<1.3^0$. The narrow and wide components reflect the scale
height of all bright stars which include both the evolved dusty stars and YSO
candidates.  The fitted values give two components with  FWM$\sim15'$ and 96$'$, both
of which are broader than those of the candidate YSO's. To measure the scale
height  of the bright dusty stars without being contaminated by the YSO
candidates, we subtracted the candidate YSOs and determined   the latitude
distribution of dusty sources. Two components were fitted with FWHM$\sim16.2'\pm0.3'$
and 105$'\pm3'$. The scale height of the narrow and bright component corresponds to
a value of $\sim20$ pc. The higher values of the scale heights of bright dusty
sources  compared to YSO candidates 
suggest they are more evolved dynamically than the  YSOs.
An important implication of these scale heights is the  support for 
a new  population of  YSOs distributed in the Galactic center region.

Using the mosaic image based on the combined MSX and MIPS data, we also 
constructed  
the
brightness distribution of 24$\mu$m emission     as a function of latitude and
longitude. Figure \ref{fig:18}b shows the latitude distribution  where two broad and narrow
components are detected. The broad background emission is  subtracted followed by
a Gaussian fit to the narrow component, giving a FWHM$\sim8'$ which corresponds to
a scale height of $\sim10$ pc. This value is slightly thicker  than the scale
height of candidate YSOs having  $h=7$ pc but  close to the scale height of
molecular gas $h$=10 pc. The  fit to the broad component gives FWHM=$61'$,
representing  the scale height of bright sources in the disk distributed along 
the
line of sight. 


To examine the brightness distribution of point sources for a given flux,  we
plot  the brightness distribution of 24$\mu$m sources as a function of longitude
 and present the results  in three different panels in  Figure \ref{fig:19}. The flux  
of selected
 YSO candidates  range between   0-5, 5-7 and 7-10 magnitudes corresponding to 
bright, 
intermediate and faint sources in the entire region of the 24$\mu$m survey but
restricted only to within $b=\pm10'$. Such a tight latitude restriction most
likely selects the  sources that populate the Galactic plane. It is clear
that the brightest 24$\mu$m  sources show a stronger  asymmetry in longitude, as
shown  in Figure \ref{fig:19}a,  when compared to the distribution of fainter sources,
as shown in  Figure \ref{fig:19}b,c. The faintest stars dominate at greater  
longitudes but
are not  detected in the innermost region of the Galactic center. This is because
of the confusion and the saturation caused by  the extremely strong  background. These
distributions indicate  that the brightest 
24$\mu$m point sources  exhibit an excess at negative longitudes similar to that seen in 
the candidate YSOs. The fainter 24$\mu$m sources seem to be uniformly distributed.

\section{The SEDs of  candidate YSOs}

One of the difficulties in identifying YSOs by selecting them from 8--24$\mu$m color 
magnitude diagrams is the contamination by dusty and bright AGB stars.
In addition, YSOs can populate left of the CMD line as
well (as seen in Figs. 6 \& 7).  
 In order to 
have more confidence in identifying YSO candidates and to characterize their 
properties, we 
obtained the Spectral 
Energy Distributions (SEDs) of selected number of YSO candidates using measurements 
at eight wavelengths from 1.24$\mu$m to 24$\mu$m in order to determine the 
evolutionary phases. The SED fitting technique 
is motivated by physical models and determines
the physical properties of the  sources.
This technique 
is used for further refinement of  the list of YSO 
candidates based on  the CMD technique with  its 
ad hoc distinctions in color. 
The SEDs  
of the sources are  analyzed by comparing to a set of SEDs produced by a large
grid of YSO models \citep{Robitaille06,whitney03a,whitney03b}.  We use a linear
regression fitter \citep{Robitaille07} to find all SEDs from the grid of models
that are fit within a specified $\chi^2$ range to the data.  The $\chi^2$ of the
fit depends on the errors assumed for the data.  We set lower limits on the
fractional flux error (dF/F) to be 20\% to account for variability between the
2MASS,  IRAC and MIPS observation dates. The minimum reduced $\chi^2$ 
 for each  source is determined.  For all the models with a reduced $\chi^2$
within 1 of the minimum value, physical parameters corresponding to those models are
averaged along with their standard deviations. The models are not always
well-constrained due to the incomplete SED information (many lower and upper
limits and relatively few overall data points). However, this averaging is constrained by 
the finite extent of the grid of models. 

We carried out our search for candidate YSOs 
toward two regions, 
using SED fitting of
sources  selected from CMDs. 
 One is a small cluster of  24$\mu$m
sources at G359.43+0.02 and the other is  the candidate YSOs  distributed  within
$|b|=10'$ and $|l|=1.3^0$.
We caution that 
 the CMD selection  biases us to
redder sources. In addition,  we have employed 
conservative errors in the SED fitting.  
Robitaille et al. (2006) defined three evolutionary stages based on
their derived model properties, analogous to the classification based
on the observed SED slope \citep{lada87}:
Stage~I objects have envelope infall rate
$\mdote/\mstar>10^{-6}$\,yr$^{-1}$, Stage~II objects have
$\mdote/\mstar<10^{-6}$\,yr$^{-1}$ and $\mdisk/\mstar>10^{-6}$, and
Stage~III objects have $\mdote/\mstar<10^{-6}$\,yr$^{-1}$ and
$\mdisk/\mstar<10^{-6}$.  
That is, Stage I objects are young protostars embedded in an opaque
infalling envelope.  In Stage II objects
the envelope has mostly dispersed, and the central star is surrounded
by an opaque disk.  In Stage III the disk is optically thin. 
It is not clear if there is a long-lived disk stage (Stage
II)  in high-mass sources, or if the disk is dispersed with the
envelope.
There are good reasons for the conservative (20\%) error bars, but it means 
that Stage
II and III YSOs with smaller IR excesses may fail to be distinguished  
from reddened stellar atmospheres.
It is clear that the YSO list will be incomplete and
biased towards younger evolutionary stages, which are readily
distinguishable from even dusty evolved stars.
We discuss these two groups  separately below and then  
estimate  the  star formation rate.

\subsection{G359.43+0.02 Cluster}

A striking cluster of 24$\mu$m sources lies  about 8$'$ north of the 
Sgr C complex G359.4-0.1
with its thermal and nonthermal radio continuum components.
Figure \ref{fig:20}a,b show grayscale images of
G359.43+0.02 at 24$\mu$m and 20cm, respectively. This cluster of 24$\mu$m 
source   (G359.43+0.02)  
is distributed in the region where the northern 
extension
of the nonthermal radio  filament of Sgr C dominates the radio continuum 
emission. In
spite of  strong extended background emission from  the nonthermal filaments,
there are several compact radio sources that  are detected toward  this cluster.
The positions,  the flux densities at 20 and 6cm and the sizes of individual 
radio sources are listed in Table 1; 
column 1 identifies the source number, columns 2 and 3 show the Galactic coordinates, 
and columns 4  to 7 give the flux densities in mJy and  the source sizes at 6 
and 20cm, respectively.   These values are estimated by 
 two-dimensional Gaussian fits to individual compact radio sources at 6 and 20cm.  
The finding chart for the  radio and 24$\mu$m sources is shown  in Figure 
\ref{fig:20}c. 
  The crosses drawn on Figure \ref{fig:20}a,c show the positions
of radio continuum point sources 
 whereas the 
circles drawn on Figure \ref{fig:20}b,c show the
position of compact 24$\mu$m  sources.  Because of the saturation of several 24$\mu$m
sources  in this cluster, there are only two radio sources, as listed in Table 1, 
that have cataloged 24$\mu$m counterparts. 


The CMD covered by the region shown in Figure \ref{fig:20}a,b is shown
in Figure \ref{fig:20}d. A total of 28   candidate YSOs  is detected 
in the YSO candidate region of 
Figure \ref{fig:20}d. We selected these  candidate YSOs and fitted their SEDs. 
The  SED fitting leads to the rejection of 10  sources as   YSO candidates. 
Thus 
the total 
number of YSO fitted sources is about 64\% of  
the sources that are selected from CMDs. The fraction of Stage I objects is
78\% of the total number of SED fitted sources.  

The rejected sources are likely to be evolved stellar sources, for 
which the model YSO SEDs give poor fits.
Other reasons for  rejecting sources 
could be i) variability between IRAC and MIPS, ii) errors in
photometry,  iii) source confusion
where multiple sources contribute and give an SED that is not  well fit by a
single YSO model, iv) 
bad extraction of data or v) there are no correct models in the grid. 
 
Figure \ref{fig:21}a 
shows the fitted SEDs for each individual YSO candidate source and 
Table 2 shows  the 
physical
characteristics derived from their SED fitting. 
These sources are assumed to be 8.5 kpc
away with Av values rangig between 5 and 50 magnitudes. 
Given that 
9 out of 18 of these have fewer than 5 data points,  
we note from  from Figure \ref{fig:21}a  that the flux of most cluster members  is
elevated  at 24$\mu$m when compared to IRAC and 2MASS data points. This supports 
the identification of these  sources as YSOs. The parameters of the fits to these
18 sources suggest that the masses of the 
individual protostars range between 5 
and 11.7
\msol\ with  corresponding luminosities ranging between 6.5$\times10^2$ and 
10$^4$ \lsol. 

Columns 1-10 of Tables 2, 3 and 4 show (1) the
the SST (Spitzer Space Telescope) Galactic center (GC) number based on  the IRAC catalogs 
\citep{ramirez08}, 
(2-3) Galactic coordinates,  (4) $\chi^2$ values, 
(5) number of acceptably fitting models ``nfits'',  (6) the visual extinction 
Av, (7-8) the average  mass and luminosity  of the protostar, (9) the 
average mass accretion 
rate which includes  their corresponding standard deviations,  and
(10) the evolutionary stage of the fitted sources. 
The error bars are 1$\sigma$ standard deviations of the well-fit values.
The standard deviation of the fitted values reflect the limited number od data points that went 
into a fit. Thus, additional data points in other wavelength bands should
improve  the reliability  in a fit. 
We  examined  $\chi^2$ with
 values  ranging between 0.5 to 2. 
There were many  fits that did not  go through the data points when a value of 2 was 
selected. 
On the other hand, when we used a value of  0.5, there were not enough fits. Thus, we 
visually selected fits with $\chi^2\sim1$. 
 Av values rangig between  5 and 50 magnitudes. 
 Av restriction is unlikely to biase  the results.  It is restricting some of 
the less
likely models from being included in the fitting.  
For example, if the Av to a source is 10, and 
the fitter
finds a model with Av=0, it is going to find a more embedded model than it should in order to 
fit  the high
extinction.  By selecting a reasonable range of Av, we are getting more appropriate YSO models 
to  fit the source.

\subsection{Sources within $|b|=10'$}

The number of candidate YSOs, as shown to the right in the CMDs of Figure \ref{fig:16}a
 is 347 and 212 distributed at  negative and positive longitudes, respectively. 
These numbers are obviously lower limits due to the bright saturated background 
at 24$\mu$m. SED fitting showed a total of 213, 112 and 35 for Stage I, II, and 
III 
sources. We note that $\sim60$\% of YSO candidates are identified as 
Stage I sources.
When grouped by longitude,  the total numbers of Stage I, II, and III 
sources are 
131, 69, 23 at l$<0\degr$ whereas 82, 43, 12 sources are distributed at 
l$>0\degr$, respectively.
The total number of acceptable sources that are SED fitted  within the restricted 
region is 360. 
This number is  $\sim$64\% of  the total 
number of sources that are selected from CMDs. A similar fraction of 
SED fitted stars is  found in the G359.43+0.02 cluster.
The total fraction of the SED fitted  YSOs in Stage I is $\sim$60\%. 
Assuming that these sources are within the central region and hence located  at 
distances of 7.5 and 8.5 kpc, 
Tables 3 and 4 list 
the physical parameters of individual sources distributed at  
negative and 
positive longitudes, respectively.

\subsection{Star Formation Rate} 
\subsubsection{G359.43+0.02 Cluster}

We now estimate the star formation rate (SFR) from the sum of average 
masses and luminosities derived from fitting of the SEDs of the YSOs. 
We first 
determine the SFR in the G359.43+0.02 
cluster which is a subset of the detected YSOs found 
throughout the nuclear disk. 
The characteristics of this cluster are described in $\S5.1$.    
The total  stellar mass in the cluster is 
estimated to be $\sim148$ \msol. Assuming a distance  $\sim8.5$ kpc, SED fitting 
to a 
sample of 18 sources yields  masses ranging from 8 to 20 \msol\ and classifies 
14 candidates in Stage I, three  in Stage II and one in Stage III.  This 
result suggests that the majority  
(nominally 78\%)   are Stage I YSOs, which have stellar ages $\sim 
10^5$ years 
and  masses in excess of 8 \msol.  
or low-mass stars, 
The canonical age for a Class I source for low-mass stars is thought to be 1-5$\times10^5$ 
years.    This
is based on the ratio of number of Class I to Class II sources in nearby star forming regions.  The 
ages of
the Class II sources are known from their stellar parameters (temperature, radius) and PMS stellar
evolutionary tracks, so a combination of theory and observations.  and Class 0 sources are 
estimated to be 
$\sim10^4$ yrs, from ratio of  the numbers and dynamical timescales of their outflows.  For higher 
mass stars which presumably form much faster, we believe $10^5$ years for a Stage  I source is 
reasonable \citep{lada99,beuther07}.

To compute the total cluster mass of only Stage I YSOs, we adopt a standard broken 
power-law form of the IMF \citep{whitney08,kroupa01}, with the number of stars per unit 
mass scaling as $M^{-1.3}$ between 0.08 and 0.5 \msol\ and as $M^{-2.3}$ between 0.5 
\msol\ and 50 M$_\odot$ (designated the Kroupa IMF hereafter).  Figure 12b shows the 
broken power-law form, as a dashed line, 
which is normalized to pass through the peak in the histogram of the stellar mass
distribution derived from SED fitting of the YSOs in the G359.43+0.02 cluster.
 The stellar mass associated with the 
Stage I YSOs in this cluster is $\sim$ 1400 \msol.

The lack of observed high mass stars is most likely due to saturated sources and 
confusion.  Another possibility is that the SED models of massive stars with a large 
inner hole in the surrounding material were not included in the grid, and thus are not 
represented in this study. Lastly, most massive stars embedded in the dense molecular 
material, which is on average 10$^2$ higher than that in the Galactic disk, may evolve 
rapidly. The residence time of an expanding remnant in dense molecular clouds such as 
found in the Galactic center is unknown. Numerical simulations are needed to examine 
whether the evolution of supernova remnants  give shorter residence time in high 
density 
medium than in low density medium \citep{tilley06}.  This 
may be relevant to the 
fact that the Galactic center region is known to host  a high cosmic ray flux 
\citep{law08}. The advantage of using only Stage I YSOs is that the average lifetime of 
these sources is better constrained than when all YSOs are included. Assuming that the 
typical age of Stage I YSOs is 10$^5$ years, the SFR in this cluster is $\sim 
1.4\times10^{-2} \msol\, \rm {yr}^{-1}$.


\subsubsection{Nuclear Disk}

\subsubsubsection{SFR Using YSO Candidates}

We now estimate the SFR in the nuclear disk restricted to a region between 
$|l|<1.3\degr$ and $|b|<10'$. The characteristics of the YSOs are described in $\S5.2$. 
The stellar mass associated with only Stage I YSOs gives a total mass of $\sim 
1.4\times10^4$ \msol. The star formation rate is estimated to be $\sim$ 0.14 
\msol\,yr$^{-1}$. Figure 12c shows the histogram of the mass distribution for both 
longitudes as well as the broken power law of the chosen IMF. SFRs for YSOs at negative 
and positive longitudes are estimated to be 0.1 and 0.05 \msol\,yr$^{-1}$, 
respectively.  The SED fitting of YSOs confirms a 10$^5$ year duration of star 
formation because the number of Stage I YSO candidates dominates over Stage II and III 
YSOs. This implies an asymmetric star formation rate with respect to the Galactic 
center. The cause of this asymmetry could be due to a short-lived random event causing 
a lop-sided star formation rate or it might be due to a more stable structure, possibly 
related to the bar.

The mass of molecular gas in the central region allows us to determine the star
formation rate per unit mass and compare it with the efficiency of star formation
elsewhere in the Galaxy.  The observed anti-correlation of molecular clouds and the
population of YSO candidates suggest that molecular clouds have already been consumed
in the formation of YSOs.  If we make an assumption that a fraction of the molecular
mass presently distributed in the central region has gone into formation of YSO
candidates, and adopt an initial molecular gas mass of 10$^6 - 10^7$ \msol, then the
star formation efficiency is estimated to be $\times10^{-7} - 10^{-8} \rm yr^{-1}$.
The total mass of molecular gas is estimated to be $\sim5\times10^{7}$ \msol
\citep{pierceprice00}.  The estimated star formation efficiency of the Galactic disk
is 5$\times10^{-9}$ yr$^{-1}$ \citep{gusten04}.  The dominance of Stage I YSO
candidates relative to Stage II suggests that a burst of star formation must have
occurred $\sim10^5$ years ago, consistent with our estimate above being higher than the
average star formation efficiency. The global properties of interstellar  and stellar 
components of the Galactic center region also suggest star formation occurs in burst 
\citep{tutukov78,loose82}.


\subsubsubsection{SFR Using Thermal Radio Measurements}
 
If the  candidate YSOs are massive and are sufficiently evolved, they are likely
to produce HII regions that can be detected at radio wavelengths if their 
spectral
type is   earlier than B3 stars \citep{felli00}. To examine further the 
nature of infrared sources that have radio continuum counterparts, the masses
inferred from radio data are compared  with those inferred from SED fitting. 
Assuming the fluxes given in Table 1 are HII regions and are produced by  
optically thin
bremsstrahlung radiation, then we can estimate the lower limit to the number of
ionizing photons to sustain the ionization. As given by \citep{condon92},
$$ N_{Ly} =6.3 \times 10^{52} s^{-1} (T_e /10^4)^{-0.45} \times (\nu /GHz)^{0.1}
(L_{thermal} / \\ 10^{27}\rm  {ergs^{-1} Hz^{-1}})  $$
where T$_e$ is the electron temperature, $\nu$ is the observed frequency and
L$_{thermal}$ is the observed luminosity. The number of Lyman continuum photons
per second calculated from  5 GHz data for radio  sources 5 and 6  in Table 1 are
3.6 and 3.4 mJy corresponding to N$_{Ly} \sim 1.5\times10^{46}$. These rates 
are consistent with  B2 ZAMS stars
\citep{vacca96,panagia73}. The average masses estimated from SED fitting for sources 5 
and 6  are 8.8 and 9.6 \msol, respectively. The discrepancy between the inferred mass from 
free-free continuum flux and from SED fitting may come from the fact these YSOs are more 
evolved and SED modeling  is not accounting for  the Lyman 
continuum flux and therefore, mass determination   may not be accurate. 
An additional uncertainty is that some of the YSO candidates may not be YSOs.
These sources are likely to  be found  in high galactic latitudes
as well as in the regions where  there is no evidence for dense  molecular clouds
or IRDCs.  

The resulting population of high-mass stars
produces ionizing photons and associated free-free emission. 
For our adopted IMF, 
this population produces ionizing
photons at a rate $Q \approx 3\times10^{50}$ s$^{-1}$ and a brehmsstrahlung
flux of $\sim 50$ Jy at 5GHz. Recent single dish radio observations of 
this region show a flux density of $\sim 10^3$ Jy at  5 GHz which includes 
both 
thermal and nonthermal emission \citep{law08}. The large scale 
4\degr $\times$ 1 \degr radio study of this 
region shows that only $\sim$ 20\%
of the flux at 5GHz is likely to be thermal. Thus, the YSO population may 
contribute up to a quarter of the observed thermal emission 
from the Galactic center region assuming that the fraction of thermal to 
nonthermal emission is the same in the central 400$\times$50 pc
($|l|<1.3\degr$ and $|b|<10'$) 
as it is in the central 4\degr $\times$ 1 \degr. 
The remaining  thermal flux of $\sim$150 Jy  may be associated with highly evolved HII 
regions resulting from  earlier  generations of star formation 
activity. This point will be discussed further in section 7.





\section{4.5$\mu$m  Excess Sources}

In previous sections, we have identified YSO candidates in the Galactic center 
region by studying the CMDs of the point sources in the surveyed region. We 
then refined the list of selected point sources from the CMD. The SED fitting 
of these selected sources used IRAC, MIPS and 2MASS data. These YSOs are 
identified to be mainly of Stage I with an age of $\sim10^5$ years.  We now 
identify YSOs that may be in a phase of star formation younger than 
$\sim10^5$ years by examining the color of their emission in the IRAC 
bands. Recent studies have identified 
``green fuzzies'' on the basis of excess emission at 4.5$\mu$m (Chambers 
et al. 2009).  
(the 4.5$\mu$m sources are also known as 'extended green objects (EGSs)' 
Cyganbowski
et al. (2008)). Active cores show a correlation between 24$\mu$m point sources
and the "green fuzzies" whereas a quiescent core shows no IR emission. The 4.5
$\mu$m excess emission is considered to be due to shock 
excited 
H$_2$ 0-0 S(9) line emission \citep{noriega04} or the CO $\nu$=1--0 rovibrational 
bandhead \citep{Marston04}. Using this color selection criterion, a recent IRAC 
survey of the Galactic plane has identified a number of interacting supernova 
remnants showing excess emission at 4.5$\mu$m, indicating shocked, excited CO 
and H$_2$ molecular line emission \citep{reach06}. A strong correlation between 
shocked molecular H$_2$ gas and methanol masers has also been noted 
\citep{lee01}, suggesting outflow is taking place at a very early stage of 
massive star formation, prior to the formation of a HII region. Also, more 
recently, the 4.5$\mu$m excess emission has been detected in DR21, a massive 
star formation site, in which the excess 4.5$\mu$m color is 
considered to be 
due a shocked molecular outflow \citep{Marston04,Smith06,cyganowski08} or 
associated with high-mass protostars within the cores of IRDCs 
\citep{Rathborne05,Beuther05}.

\subsection{Correlation with IRDCs}

Since there is a great deal of molecular material as traced by IRDCs in the 
nuclear disk, we searched for excess emission at 4.5$\mu$m (``green'' in 
3-color images combining 3.6, 4.5 and 8 $\mu$m emission) that traces molecular 
line emission. A quantified indication of excess 4.5 $\mu$m emission is 
obtained by constructing the ratio 
$I(4.5) / [I(3.6)^{1.2}*I(5.8)]^{0.5}$.
 This 
essentially is the ratio of the actual 4.5 $\mu$m intensity to that determined 
by a power--law interpolation of the 3.6 and 5.8 $\mu$m intensities. It was 
found empirically that a slight ($\sim$10\%) modification to the interpolation 
coefficients helps emphasize sources with unusually strong 4.5 $\mu$m emission. 
In a map of this ratio across the entire IRAC survey of the Galactic center, 
most point sources exhibit a uniform ratio that is not very different from the 
background. In regions of high extinction (e.g. IR dark clouds, and Sgr B2), 
this ratio rises because the reddening reduces the 3.6 $\mu$m intensity. 
However, there were of order 100 sources, located in regions of both high and 
low extinction that exhibit ratios distinctly higher than those of other 
sources in their vicinity (within several arc minutes). These include 33 
sources that were visually selected as ``green" sources in 3-color images. 
Their individual color images are shown in Figure \ref{fig:22}; the 4.5$\mu$m 
excess sources show both extended and compact emission. The ratio also clearly 
identifies a number of interesting sources that do not stand out in the 3-color 
images. Our search should be compared with two recent studies. Using an 
algorithm that  finds ``green fuzzies,'' Chambers et al. (2009)  identified 
excess 4.5$\mu$m emission by confining to the cores of IRDCs.  Another study by 
Cyganowski et al. (2008) visually identify ``EGOs''. Our technique is sensitive 
to both compact and extended sources with 4.5$\mu$m excess emission, unlike 
other searches.


The spatial distribution of the 4.5$\mu$m excess sources is marked on Figure 
\ref{fig:23}a which displays the ratio image of the region surveyed  by IRAC.
 There are 33 sources that are characterized to have 
4.5$\mu$m excess emission, three of which (g1, g2 and g4) 
coincide with known planetary 
nebulae \citep{jacoby04}.  
Thus, there are 30 sources that are
candidate YSOs. 
 It is quite possible that the list of 4.5$\mu$m excess sources 
 is contaminated by 
additional PNe, distributed in the crowded region of the Galactic center. 
Half of the 4.5$\mu$m 
excess sources are near Sgr B2, G0.25+0.01, M-0.02-0.07,  and Sgr C (G359.5-0.0). 
The rest appear to be distributed away from the Galactic plane, most 
likely associated with local objects distributed at negative latitudes. Although there 
is no bias in the selection of of 4.5$\mu$m excess sources toward 
IRDCs, Figure 
\ref{fig:23}a shows that most of the 4.5$\mu$m excess
 sources are distributed in the vicinity of IRDCs. 
The highest density of 4.5$\mu$m 
excess sources is near Sgr B2. 

\subsection{Correlation with Methanol Masers}

To correlate the position of highly embedded YSO candidates with early 
sites of massive star formation, as traced by methanol masers, Figure 
\ref{fig:23}b presents the spatial distribution of the known methanol masers on 
an 8$\mu$m image. A survey of the inner 2 degrees of the Galactic center showed a 
total of 23 class II methanol masers at 6.7 GHz \citep{caswell96}. 
We used limited targeted surveys of class I methanol sources. 
 The crosses show the position of class II methanol 
masers \citep{caswell96}. The 6.7 GHz methanol maser observations searched the 
region along the plane between $|l|<0.9^0$ and $|b|<0.5^0$. Table 5 lists the 
position of the sources that are found within our IRAC survey. Table 6 shows the 
positions and fluxes of a subset of 4.5$\mu$m excess sources that were observed 
with MIPS at 70$\mu$m.

It is now well established that methanol masers are signposts of ongoing 
massive star formation throughout the Galaxy. Class II methanol masers are 
recognized to be radiatively pumped, unlike class I methanol masers, which  known to 
be collisionally pumped \citep{Menten91}. Tables 7 and 8 show the positions of 
class I and II methanol masers and their corresponding velocities, 
respectively. The correlation of methanol masers with the 4.5$\mu$m excess 
sources in the surveyed region was first reported by Yusef-Zadeh 
et al. (2007a). The number 
of 4.5$\mu$m excess sources distributed in the Galactic center region is 14 
but we detected 6 methanol maser counterparts. Given the limited sensitivity 
of 1 Jy in the maser survey by \citet{caswell96}, it is possible that many of 
the 4.5$\mu$m excess sources have weak maser counterparts or that they signify 
different evolutionary phases of massive protostars when compared with the 
onset of methanol maser emission. A more detailed study of the correlation of 
6.7 GHz methanol masers and EGOs with IRDCs in the GLIMPSE data was recently 
reported by Cyganowski et al. (20008).


\subsection{The SEDs of 4.5$\mu$m Excess Sources}

Using 2MASS, IRAC and MIPS point source catalogs as well as 850 and 450 $\mu$m data 
\citep{pierceprice00}, we constructed the SEDs of individual 4.5$\mu$m excess sources 
to be used in  model fitting. Due to the lower resolution of the submillimeter data 
and the 
excess flux of molecular emission in the 4.5$\mu$m IRAC band, we show upper limits to the 
peak flux densities at 4.5, 450 and 850 $\mu$m \citep{pierceprice00,ramirez08}. The 
distances to many of the 4.5$\mu$m excess sources are unknown due to the difficulty in 
estimating the kinematic distances of sources in the direction toward the Galactic 
center. 
We assumed that the sources that are distributed at high galactic latitudes are most 
likely foreground sources whereas the range of distances for low 
latitude sources galactic latitudes is 
restricted  to $|b| < 10'$ and they are near the  Galactic center. 
The model fitting for foreground sources was  
restricted to distances 
ranging between 2 and 6 kpc whereas the the range of distances for low galactic 
latitudes 
is between 7.5 and 8.5 kpc. The derived masses and luminosities 
 of the 4.5$\mu$m 
excess 
sources based on the well-fitted SED models are shown in Tables 9 and 10 for Galactic 
center and foreground sources, respectively. 
It should be pointed out that if the fitter finds a low-Av model to fit a source, that does 
not mean that the source is not at the Galactic center. 
The fitter will find models that can fit a  range of
Avs even if the source is at the Galactic center  by restricting this variable to reasonable 
values. This 
restriction allows the underyling YSO models to more appropriate models for fitting that source.
We note that for the very 
young sources 
that we believe are associated with 4.5$\mu$m excesses,  the YSO mass estimate 
is the current mass.  

Assuming that the sources that are not SED fitted are YSOs, there are currently 
no 
available models in the grid to fit these sources.  Also, the parameter ``nfits'' in 
Table 9 depends on how many datapoints are available.  The fewer the datapoints, the 
more models can fit the data.  If there are too many fits to the datapoints, it 
suggests the source is poorly constrained (which will show up in the standard 
deviations). On the other hand, if there are too few fits, it suggests the models are 
wrong (e.g., single source YSO trying to fit a cluster) or the data are bad. We note a 
a strong correlation between the number of data points and $\chi^2$, as well as an 
inverse correlation between the number of data points and the number of "acceptable" 
models. And as the number of acceptable models increases, the (relative)  
uncertainties also increase, especially for $< \lstar > $ and $ < \mdote >$. 
Nevertheless, it is clear that the number of fits and standard deviations allow us to 
identify well-constrained sources, unlike the CMD technique. We identify ten 4.5$\mu$m 
sources in Stage I and four in Stage II and III evolutionary phases. The fraction of 
Stage I sources is $\sim$70\% which is slightly lower than that found in the 
G359.43+0.02 cluster (78\%) and higher than the fraction of 60\% found in the region 
between $|b|=10'$ and $|l|<1.3^0$. This is consistent with the idea that the 4.5$\mu$m 
excess sources are in their early phase of evolution. The Tables presented here give 
sufficient information for use of the fits in future studies.



The visual extinction levels derived from fitting the SEDs of the sources 
which are most likely located near the Galactic center
range between 
5 and 47 magnitudes. Figure \ref{fig:galcen_models}a shows the fitted  SEDs of 
selected sources with high extinction. The number of good fits, as seen in Figure 15 
and Table 9, depends both on the quality of the data, and the number of models that 
happen to have an SED that looks like the one fitted.  We note from Figure 15 that g0, 
g5, g27, g31 are not well fit by the models as they seem to have narrow peaks at about 
$4-5\mu$m, dropping down on either side. Thus, unlike the rest, these sources are 
unlikely to be accreting YSOs. The correlation of 4.5$\mu$m sources and IRDCs is strong 
and is consistent with recent analysis by Chambers et al. (2009);  we found 11 of the 
4.5$\mu$m sources are correlated with IRDCs, half of them are not extended and six of 
the 14 sources have methanol maser counterparts. The correlation of the majority of 
4.5$\mu$m excess sources with IRDCs, which are known to lie near the Galactic center, 
supports our initial choice for placing the low latitude 4.5$\mu$m excess sources at 
the distance of the Galactic center.

Appendix 1 describes each of the Galactic center and foreground 4.5$\mu$m excess 
sources.


\bigskip
\section{Discussion}
\subsection{Star Formation History of the Nuclear Disk}

To better understand the nature of star formation history in the complex region of the 
nuclear disk, we have used several different measurements. These measurements include 
4.5$\mu$m excess sources, Stage I YSOs, thermal HII regions and diffuse nonthermal radio 
flux, each of which is described below.

We adopt, as in section 5.3, a standard broken power-law form of the IMF 
\citep{whitney08,kroupa01} to estimate on going star formation rate based on SED fitted 
4.5$\mu$m excess sources. We believe these excess sources trace the earliest phase of star 
formation.  Given a large uncertainty in the completeness of 4.5$\mu$m sources, the 
integrated 
mass of Stage I sources using Kroupa IMF, as before, is estimated to be $\sim$ 522 
\msol. 
Assuming that typical age of 4.5$\mu$m excess sources is $\sim5\times10^4$ to 
$10^5$ years, we 
estimate SFR 0.017 to 0.009 \msol yr$^{-1}$, respectively. Given the vast amount of 
dense 
molecular gas 
corresponding to $\sim(5)\times10^7$ \msol\ residing in the Galactic center region, the 
efficiency of star formation is estimated to be $\sim1-3.4\times10^{-10}$ yr$^{-1}$.

A more robust approach to estimate SFR, as discussed earlier in section 5.3.2, is to use 
the evidence for a population of Stage I YSO candidates in the nuclear disk and make an 
estimate of 
the SFR$\sim$0.14 \msol yr$^{-1}$ in the last 10$^5$ years. The star formation rate 
during 
this period appears to be an order of magnitude higher that that estimated 
from 4.5$\mu$m excess sources. These excess sources are in an earlier phase of 
their evolution and are younger than 10$^5$ years. 

To estimate the SFR  from young stellar objects older than 10$^5$ years, we used 
24$\mu$m flux density. This flux is representative of the ionizing flux and hence of ages 
of a few million years. We find a total 24$\mu$m flux density of $1.17 \times 10^5$ Jy 
within the central region, $|l| < 1.3\degr$, $|b| < 10'$ (i.e., a radius of 182 pc). The 
K-band extinction over this region appears to average slightly more than two magnitudes 
(Dutra et al. 2003) and the observed ratio of A$_{24}$/A$_K$ $\sim$ 0.5 \citep{chapman09}. 
We therefore corrected the observed flux density for 1.1 magnitudes of extinction.  For a 
distance of 8.5 kpc, the resulting 24$\mu$m luminosity is $\sim9 \times 10^7$ L$_\odot$. In the 
formulation of Rieke et al. (2009), this luminosity implies a SFR of 0.07 M$_\odot$ 
yr$^{-1}$, assuming a Kroupa IMF. This estimate implicitly assumes that the SED fitted 
YSOs did not contribute significantly to the total measured flux at 24$\mu$m. 

Lastly, nonthermal radio emission can be used as another proxy to determine SFR. 
This SFR probably applies to conditions   some tens of
millions of years ago (Helou \& Bicay 1993).
The flux 
density of nonthermal emission is estimated to be 800 Jy at 6cm. If we extrapolate this 
flux to 21cm flux assuming a power law spectral index of 0.8, we derive a luminosity at 
this wavelength of 61.2 L$_\odot$. This then yields a SFR of 0.007 M$_\odot$ yr$^{-1}$ 
\citep{rieke09}.

Taken together the above estimates provide a reasonable picture of large-scale  star 
formation activity in the nuclear disk.  We 
suggest that the region within $\sim200$ pc of the 
Galactic center was in a very quiescent state $\sim10^7$  years  ago, and that 
the SFR has 
been rising slowly, peaking to a value of 0.14 M$_\odot$ yr$^{-1}$
around 10$^5$ years ago and then perhaps falling  
since then to the current value of $\sim$0.01 M$_\odot$ yr$^{-1}$.


\subsection{The Schmidt-Kennicutt Law in the Galactic Center}

A motivation for this work has been to test the influence of the extreme ISM
conditions in the central 400 pc of the Milky Way on the process of star
formation.  We have found that the behavior of various SFR indicators is
consistent with expectations from the global properties of other galaxies.
The value of q$_{24}$ = log(f$_{21cm}$/f$_{24\mu}$m) $\sim$  2.1, where 
$f_{21cm}$ and $f_{24\micron}$ represent flux densities  at 21cm and 
24$\mu$m, respectively,   is at the
extreme high end of the range for external galaxies and is larger than is
typically observed in the centers of nearby galaxies (Murphy et al. 2006)
but can be explained in terms of a relatively low rate of star formation in
the past.  The properties of the individual forming massive stars also are
consistent with those of similar objects elsewhere in the Galaxy.

A less ambiguous test for changes in star forming properties is based on the
Schmidt-Kennicutt Law.  This law is an empirical power-law relation between
the surface densities of the interstellar gas and of star formation (see,
e.g., Kennicutt 1998). If this law holds between the disk and center of the
Milky Way, it would be strong evidence that the high interstellar gas
temperatures in the latter region do not fundamentally alter the manner in
which stars form there.
Fuchs et al. (2009) have determined the relevant surface densities for the
Milky Way disk near the sun and show that this region does indeed behave as
expected according to the Schmidt-Kennicutt Law.  We now estimate the same
parameters for the central 0.8 kpc of the Galaxy. We take the mass in gas
from Pierce-Price et al. (2000) to be $5.3 \pm 1.0 \times 10^7$ M$_\odot$.
Our estimates of the SFR have a range: 0.007 M$_\odot$ yr$^{-1}$ from the
non thermal radio, 0.03 M$_\odot$ yr$^{-1}$  from the youngest forming
massive stars; 0.062 M$_\odot$ yr$^{-1}$ from the 24$\mu$m luminosity; and a
peak of 0.14 M$_\odot$ yr$^{-1}$  from the $\sim$ 10$^5$ year old YSO
candidates. A long term time-averaged value can also be estimated. The
enclosed mass in this region is $1.5 \times 10^9$ M$_\odot$ (Haller et al.
1996). Therefore, a strict upper limit on the average SFR over the past
10$^{10}$ years is 0.15 M$_\odot$ yr$^{-1}$; a more plausible average would
be 25\% to 50\% of this value.  This average is in good agreement with our
measurements of the recent SFR and suggests that the variations represent
fluctuations around this average. 
 We take the typical recent SFR to lie
between 0.04 and 0.08 M$_\odot$ yr$^{-1}$. 
 In Figure 16, we plot the
Galactic center values on Figure 4 of Fuchs et al. (2009). 
In this figure,
different types of Sa, Sb, and Sc  galaxies are shown as open triangles, open circles, 
and 
filled circles. The 
red star is the value for the solar
neighborhood and the blue cross is that for the Galactic center. The black
line has a slope of 1.4, corresponding to the Schmidt-Kennicutt relationship
found for external galaxies (Kennicutt 1998). 
The agreement is excellent,
suggesting that to first order the star formation in the Galactic center
region is not strongly affected by the environmental differences compared 
with the disk of the Galaxy.  

\section{Conclusions}

In this paper, we have studied the distribution of interstellar dust clouds and the
nature of star formation in the Galactic center region using the 24$\mu$m {\it
Spitzer} MIPS survey augmented with additional mid-IR, near-IR, submillimeter
and radio data. By making use of MSX data in regions where the MIPS data are
saturated, we constructed a complete 24 $\micron$ image of the Galactic center
region with unprecedented sensitivity and resolution.

The comparison between the 24$\mu$m and 8$\mu$m images showed numerous examples
of compact regions of 24 $\micron$ emission surrounded by partial shells of 8
$\micron$ emission. These structures are characteristic of H II regions, as seen
throughout the Galactic plane, where luminous stars have heated large dust
grains thus enhancing 24 $\micron$ emission, and have destroyed PAHs thus
suppressing 8 $\micron$ emission.

We noted a number of elongated IRDCs distributed over a wide range of angular
scales toward the Galactic center. Many of the IRDCs lie within a ``bow-tie''
structure, consisting of two layers parallel to the Galactic equator and at both
positive and negative longitudes. Molecular line survey studies show
counterparts to a large fraction of the IRDC of the bow-tie structure. The
darkness of the two coherent IRDC layers provides relative locations of
individual clouds with respect to each other. In particular, we note that clouds
distributed on the positive longitudes (e.g., Sgr B2 and G0.25+0.01) are closer
to us than those on the negative longitudes.

A result of the 24 $\micron$ imaging is the recognition of a population
of compact and point sources distributed between the two layers of IRDCs. This
population is predominantly found at negative galactic longitudes, and includes a
distinct cluster of sources at G359.43+0.02. Many of these sources are YSO
candidates. Candidates are selected as particularly red objects in the 24 - 8
$\micron$ color magnitude diagram, and characterized by modeling their SEDs.
This led us to estimate the star formation rate of 0.14 \msol yr$^{-1}$ based on
SED fitted Stage I YSO candidates. The preponderance of Stage I YSO candidates
implies that massive star formation took place in the nuclear disk about $10^5$
years ago. The derived scale heights of the population of YSO candidates
($h\sim7$ pc) as well as that of the molecular layer ($h\sim10$ pc) suggest
their being dynamically young. These measurements are consistent with a highly
efficient burst of high mass star formation in the nuclear disk about 10$^5$
years ago. This aspect of the study characterizes the unique nature of the
Galactic center region, apart from the general Galactic disk. 
Evidence for large-scale IRDCs throughout the Galactic center 
is consistent with the idea that  massive star formation and 
massive cluster formation takes  place in the central region of the Galaxy. 
Future study of this
region with its steep gravitational potential provides a testing ground for
large-scale formation of massive stars and/or star clusters.

Finally, we explored an independent means of identifying YSO candidates by
finding sources with excess 4.5 $\micron$ emission, indicating the presence of
shocked molecular emission from protostellar sources. This color distinction was
used as a proxy to probe the sites of star formation even earlier than 10$^5$
yrs old. We found 33 sources that show excess 4.5$\mu$m emission. Those at the
lowest latitudes were presumed to be at the Galactic Center, while those at
higher latitudes often shown indication of being more local sources. SED fitting
as well as the correlation of 4.5$\mu$m excess sources with methanol masers and
IRDCs indicated the majority of these YSO candidates in Stage I and are likely
to be associated with molecular outflows in sites of on-going star formation.
Assuming that the 4.5$\mu$m excess sources provide us with all the sites of
on-going star formation and that the SED fitted mass determination of the
4.5$\mu$m excess sources is applicable to these protostars, then we conclude
that the efficiency of on-going star formation is lower than that estimated for
the Galactic disk and is consistent with earlier studies of this region
\citep{lis94}. The estimated values  of SFR over different time scales up to $10^7$ 
years ago  suggest   that  large 
scale star formation in the nuclear disk is likely to 
be continuous over a long period. However,  
mild burst-like activity could increase  SFR by an order of magnitude 
on a $10^5$ year time scale.


We acknowledge the use of SCUBA data taken by D. Pierce-Price.
This work is partially supported by the NSF under award number AST-0807400 and 
JPL/Caltech contract 1255094.

\section{Appendix}

We assumed that the sources that are distributed at high galactic latitudes are most 
likely foreground sources whereas 
the  low 
latitude sources ($|b| < 10'$)   are near the  Galactic center. 
We first describe 4.5$\mu$m excesses 
sources which are most 
likely located in the Galactic center region followed by foreground sources. 
In all the 
figures presented below, the circles mark the position of the 4.5$\mu$m excess sources. 
For extended 4.5$\mu$m sources, the peak of the excess emission is used to 
identify the closest point sources in the IRAC and MIPS catalogs by Ramirez et al. 
(2008) and Hinz et al. (2009), respectively.

The squares represent sources that were observed at 70$\mu$m. The distribution of class 
I and II methanol masers are also drawn on these figures as plus(+)  and cross (x) 
signs, respectively.

\subsection{Galactic Center 4.5$\mu$m Excess Sources}

\bigskip
{ \bf G1.041-0.072 (g0)}\\
This source lies at the eastern edge of the  ridge of IRDCs and 
is not well fit as it 
shows a narrow peak in its SED, thus it is unlikely
to be at an early phase of a  YSO.    
The fitted 
SED gives a protostellar source with  a mass of $\sim10$ \msol.
This 4.5$\mu$m excess 
source has a 24$\mu$m counterpart, as shown in Figure \ref{fig:26}a  and lies 
within   an arc-minute of an extended 8$\mu$m  and 24$\mu$m HII 
source G1.05-0.071  at l=1$\degr\ 3'\ 4''$, b=$-4'\ 15''$. This extended source is also 
shown in the 3-color image, as presented in Figure \ref{fig:22}.  
Figure \ref{fig:26}b shows contours of 850$\mu$m emission that  trace 
the elongated east-west 
IRDC. The strongest 850$\mu$m emission coincides 
with G1.05-0.071.


\bigskip
{ \bf G0.826-0.211(g3)}\\
This 4.5$\mu$m excess source is embedded in the IRDC associated with Sgr B2. The distribution of 
the IRDC ridge at the location of g3 widens in a plume-like structure pointed southward of the 
Galactic plane. Figure \ref{fig:26}c,d show 
contours of 450 and 850$\mu$m emission from this 
prominent IRDC. A 24$\mu$m and 8$\mu$m source coincides with g3,  which is shown as a 
circle in 
Figure \ref{fig:26}.  This 4.5$\mu$m excess source was also observed at 70$\mu$m, as shown by a 
square sign in Figures \ref{fig:26}c,d, but was not detected (see Table 6). Most of the 
well-known ultracompact HII regions associated with Sgr B2 are 
located to the northwest of g3, 
as shown in Figure \ref{fig:23}a. The SED fit to this source gives a mass of 13.7 \msol. This 
4.5$\mu$m excess source is located within 20$''$ of SiO 17452-2819 with a radial velocity of 90 
\kms\ \citep{shiki97}. The location of this maser source with respect to g3 supports the idea 
that this SiO maser is associated with the Sgr B2 star forming region and is a counterpart to a 
young stellar object.  The correlation of submillimeter emission, IRDC, 24$\mu$m point source 
and excess 4.5$\mu$m emission is consistent with g3 being a massive YSO associated with an active 
core.

\bigskip
{\bf Sgr B2 (g6-g10)}\\
The largest concentration of 4.5$\mu$m excess sources is located in Sgr B2, 
 which is one of the most massive star forming sites in the Galaxy.  SED fits to the five 
4.5$\mu$m excess sources g6 to g10 give the highest protostellar masses among the 
4.5$\mu$m excess sources, ranging between 8 and 22 \msol.  Figure \ref{fig:27}a,b shows 
contours of 450 and 850$\mu$m emission that are superimposed on 8$\mu$m and 20cm images 
of Sgr 
B2, respectively. 
It is clear that the IRDC associated with Sgr B2 is traced remarkably 
well by submillimeter and molecular line emission
from several active molecular 
cores  \citep{jones08}. 
The distribution of compact HII 
regions, as shown best at 
20cm wavelength, indicates that they lie  at the southern edge of  
the dark cloud. 
The distribution of UC HII regions  as well as  24$\mu$m sources
suggests 
a later phase of massive star formation at the southern edge of IRDC where UC HII regions 
are  concentrated.  However, the distribution of methanol sources 
are seen silhouetted against the middle  of the infrared dark cloud.  
Figure 
\ref{fig:27}c makes this point even clearer
 by showing that most of the methanol sources 
are displaced with respect to 
24$\mu$m sources heated by UC HII regions lie   at the edge of the IRDC.
The high column density of Sgr B2 and the saturation of the brighter
 24$\mu$m sources 
are likely 
responsible for a lack of additional detection of 4.5$\mu$m excess sources and YSO 
candidates in Sgr B2. 
Nevertheless, the projected displacement between 24$\mu$m sources and 
methanol sources
 is consistent 
with a picture  that  star formation in the IRDC  associated with Sgr B2 has 
taken place from outside in.

 Figure \ref{fig:27}d shows contours of emission from one of the 4.5$\mu$m 
excess sources, g6, that shows extended emission at 4.5$\mu$m.  Three of the 
five 4.5$\mu$m excess sources, g6, g8 and g9, coincide with ultracompact HII 
regions Sgr B2 R, I and F \citep{gaume95}. We note that  Sgr B2-F itself 
breaks up into multiple radio components that  are considered to be members of 
a cluster of massive ultracompact HII regions \citep{depree05}. The parameters 
of the SED fit to g9 should be considered with caution, given that multiple 
sources could contribute to the SED. Fortunately, the flux from a cluster of 
massive stars is likely to be dominated by the most massive star as luminosity 
is a steep power of mass and thus the assumption of all the luminosity arising 
from a single cluster member is reasonable.


\bigskip
{\bf G0.376+0.04 (g16)}\\
The 4.5$\mu$m excess 
source associated with g16  is detected against the IRDC ridge 
at  G0.376+0.04. This source coincides with one  of the string of 
submillimeter emitting clouds, known as the dust ridge in Figure 1c \citep{lis98}.
g16 coincides with a
class II methanol maser  at a velocity of 37 \kms\ that  is 
listed as source 8 in Table 8.
Molecular line studies of the ridge of IRDCs are dominated by
velocities ranging between  20 and 30 \kms\ \citep{oka05}. 
Figures \ref{fig:28}a,b show contours of 450$\mu$m and 850$\mu$m emission 
and are superimposed on 
grayscale  24$\mu$m and 20cm images of this cloud. This is the first 
evidence that 
this cloud shows  a signature of massive star formation as the fitted SED 
indicates  10 \msol\ for this YSO, the first  evidence that this cloud is a site 
of massive star formation \citep{argon2000}. 
Following the definition of active and quiescent cores by Chambers et al. (2009),  
the eastern half of the IRDC ridge is in its quiescent phase of
star formation whereas the western half is forming massive YSOs. 
The 4.5$\mu$m excess source also lies in the vicinity 
of a nonthermal radio filament that  appears to terminate in the IRDC and 
about 1$'$ east of g16, as shown in Figure \ref{fig:28}b.  
Figure \ref{fig:28}c shows contours of 4.5$\mu$m emission from G0.376+0.04. 

\bigskip
{ \bf G359.932-0.063 (g23) }\\
This 4.5$\mu$m excess 
 source lies at the edge of 
the 50 \kms molecular cloud M-0.13-0.08 \citep{HerrnsteinHo05,ArmstrongBarrett85}. 
Figures \ref{fig:29}a,b show contours of submillimeter emission 
from two  prominent 50 \kms\ and 20 \kms\ molecular clouds, known to 
be the closest clouds to the Galactic center. 
Contours of  450$\mu$m and 850$\mu$m emission from 
these clouds are superimposed on the 
 8$\mu$m and 20cm images of the Sgr A region, respectively. 
The 50 \kms\  cloud is known to be interacting with the Sgr A E East SNR G0.0+0.0 
\citep{zylka90}  
and several OH masers at 1720 MHz have trace the interaction site \citep{fyz99}. 
The shell-like Sgr A East supernova remnant  is shown  in Figure \ref{fig:29}b. 
The position of 
g23 lies within five  arcseconds of OH(1720 MHz)   maser  H at a velocity of 49 \kms. 
The excess 4.5$\mu$m emission at this position could be associated with the 
supernova shock interaction. 
4.5$\mu$m excess emission  has been  seen toward a number of 
supernova remnants \citep{reach06}.
However, the parameters of the SED of this 
source suggest that  G359.23-0.063 
is associated with a  11 \msol\  protostar (see Table 9).
If so, the protostar presents the earliest  phase of massive star 
formation in the 50 \kms\ cloud, perhaps triggered by the interaction 
with the expanding Sgr A East supernova remnant. 
A  chain of ultracompact HII regions 
$\sim2'$ northeast of g23 may trace  massive star formation at 
a later phase than  that shown by g23. Figure 
\ref{fig:29}c shows a 24$\mu$m image 
of IRDC associated with the 50 \kms and clearly shows  the IRDC toward  the 
region where g23 is detected.  The evidence for collisionally excited 
methanol emission 
 has also been reported toward the 50 \kms\  
(\citep{haschick90}, see source 2 in Table 7). This 
result supports  supports
the presence of  massive star formation in this region. 
Figure \ref{fig:29}d shows contours of 4.5$\mu$m emission from g23. 

The 20 \kms\ and 50 \kms\ IRDCs are also located in a region where 19 
radio filaments (A1-A19)   have been detected at 20cm  \citep{fyz04}. 
Two of these nonthermal radio filaments RF-A3 and RF-A9 are labeled in Figure 20b. 
 The nonthermal filaments (e.g., G359.90-0.07 and  G359.88-0.07) 
may be  interacting with 
the 20 \kms molecular cloud \citep{fyzpound05}.


\bigskip
{ \bf  G359.841-0.080 (g25)}\\
G359.841-0.08 is another 4.5$\mu$m excess 
source that is found at the western edge of 
the 20 \kms\ molecular cloud 
M-0.02-0.07 \citep{HerrnsteinHo05,ArmstrongBarrett85}, 
which is traced by  the prominent 
IRDC at 24$\mu$m, as presented in Figure \ref{fig:29}c. Evidence for a signature of star 
formation is supported by the presence of star formation  by  a class I 
methanol maser and a  HII region 
toward the  eastern half  
of the cloud. The methanol maser source 
is detected 
at a velocity of 21 \kms\ \citep{haschick90,valtts00}. 
Additional observations are  needed to confirm the association of 
this 4.5$\mu$m excess 
 source with the 20 \kms cloud. Contours of 4.5$\mu$m emission are 
shown in Figure \ref{fig:29}d. The peak of submillimeter emission from the 
20 \kms\ IRDC  shows no signs of star formation, thus,  is considered to be 
a quiescent core.  

\bigskip
{ \bf G359.599-0.032 (g27)}\\
This 4.5$\mu$m excess source is projected against the IRDC ridge, 
on  the negative longitude 
side of the Galactic center.
Figure \ref{fig:30}a,b shows a 24$\mu$m and 6cm 
continuum images of  this 
source, respectively, and the contours of 4.5$\mu$m emission are  shown in 
Figure \ref{fig:30}c. The parameters of 
the fit to this source suggests a  
14.5$\pm3.5$ \msol\ protostar. 
However, g27 shows  a narrow peak in its SED and 
is not  fitted well  by the 
models.  
This source is located about 35$''$ north of an HII region
and a shell-like structure with a diameter of 30$''$ centered at G359.61-0.045. 
The shell-like structure is best seen at 24$\mu$m in Figure \ref{fig:30}a.

\bigskip
{\bf G359.437-0.102 (g29)}\\
Figure \ref{fig:31}a,b shows  contours of 450$\mu$m and 850$\mu$m emission superimposed 
on 8$\mu$m and 20cm continuum images of the Sgr C HII region, respectively. This 
well-known HII region is one of the brightest extended HII regions in the Galactic 
center region. 
 The Sgr C HII 
region was saturated at 24$\mu$m, so we combined the image with the MSX image at the 
same wavelength band. 
The Sgr C IRDC is elongated and  runs roughly along the Galactic plane. It is 
 detected at both 8$\mu$m 
and  24$\mu$m, as  shown in 
Figure \ref{fig:31}c.  This IRDC is traced by submillimeter emission showing several 
cores  along its elongated structure.
The peak emission at 24$\mu$m coincides with the peak HII 
emission at 20cm and suggests that warm grains are being  heated by the central O4 
star.

The 
4.5$\mu$m excess
 source g29 lies  at the peak  of  the submillimeter emission  and is 
shown as a circle in Figure \ref{fig:31}a,b,c. The crosses show the 
positions of class II methanol 
masers with velocities of -52 and  -53 \kms.  The masers 
are listed as sources number 1 and 2  in Table  8.
Radio recombination line measurements 
 of the Sgr C HII region show a radial velocity of 
$\sim-65$  \kms (Liszt 1992; Liszt and Spiker 1995).
Figure 
\ref{fig:31}d shows contours of the extended 4.5$\mu$m excess emission. The 
corresponding color image of 
the 4.5$\mu$m excess source can be seen in Figure 13. 
The modeled SED of the 4.5$\mu$m 
excess source 
G359.436-0.102 in Sgr C indicates that the central protostellar candidate coincides 
with the northern maser source. Figure 15a  shows the best 
fitted models to the data using 2MASS, IRAC, MIPS and SCUBA data accounting for 22 
magnitudes of visual extinction.  This SED fit is consistent with the identification of 
the 4.5$\mu$m excess source with a massive 14 \msol\ protostellar source (see Table 9) 
associated with the methanol masers.
 
The region surrounding 4.5$\mu$m excess source 
(g29) is complex. The emission at 4.5 and 5.8$\mu$m is extended whereas the emission at 
8$\mu$m has a  point-like structure and does not appear to have a counterpart at shorter 
wavelengths. 
The 8$\mu$m source is $\sim6''$ away from the center of the 4.5$\mu$m emission. There 
are three other nearby compact  sources;  higher resolution study is needed to 
clarify the nature of these 8$\mu$m sources. However, it is 
most likely that the extended emission from the region close to the 4.5$\mu$m excess
source represents  shocked molecular outflow from the YSO candidate. 
This is consistent with  the presence  of submillimeter core, methanol masers,  IRDC and 
4.5$\mu$m excess, all of which  suggest
an active phase of massive star formation at the northwestern  edge  of Sgr C.
The southeastern half of the 
IRDC shows no signs of active star formation.

There is  a puzzling feature at the northeastern boundary of  Sgr C IRDC, a lack 
of strong free-free radio continuum and submillimeter 
emission   l=-$34' 10''$, b=$-5' 45''$.  This implies that 
there is a cavity in the   ionized gas and 
emission by dust   at the 
interface between  the submillimeter  core and the Sgr C  HII region. 
However, it is possible that the appearance of the cavity results from 
poor  sensitivity in the submillimeter.  

At 20cm, there is a  compact bright continuum source at 
l= $-34' 29''.19$, b=$-6' 38''.15$ 
which is  about 1$'$ or 2.4 pc (the Galactic center distance of 
8.5 kpc) from the positions of the methanol masers. The flux density of this source is 
20$\pm1.8$ mJy. 
The most prominent nonthermal radio filaments of Sgr C, as shown in 
Figure 22b,  becomes quite faint  
inside the HII complex. 
The long nonthermal filaments  appear to lie at eastern edge of
the Sgr C cavity and the HII region.  Like other filaments described 
in the Sgr A and Sgr B clouds, the prominent nonthermal filament appears to terminate 
at 
the Sgr C IRDC.

\bigskip
{ \bf G359.30+0.033 (g30)}\\
This 4.5$\mu$m excess 
source is detected at the edge of an elongated and narrow IRDC at l=40$'$, b=2$'$. 
This feature appears to widen  in the direction away from the Galactic plane, 
resembling the vertical molecular Clumps 1 and 2  
that are found at l=355$^0$ and l=3$^0$ and b=0$^0$ \citep{bania77}. 
Figures 
23a,b show an 8$\mu$m and 24$\mu$m images of this remarkable filamentary 
dark cloud. The  cloud is  drawn schematically in  Figure 1c. The  brightness
of this cloud in the ratio map shown in Figure 3b suggests that it is likely to be 
on the front side of the Galactic center. 
Contours of  4.5$\mu$m emission from the region where excess emission is detected 
at this wavelength are  shown in Figure \ref{fig:32}c. 
The fit to the SED of this source suggests a  8.9 \msol. 
protostar (see Table 9).  

\bigskip
{ \bf G359.199+0.041 (g31)}\\
This 4.5$\mu$m excess 
source is  projected against an east-west elongated IRDC. 
Figures \ref{fig:33}a,b show  24$\mu$m  and 
20cm continuum images of this source. 
A methanol maser source  and a 20cm continuum 
source lie  in its vicinity, 4$'$ away from the 4.5$\mu$m excess 
 source. 
If the methanol source is associated with the same IRDC toward which the  4.5$\mu$m 
excess  
source is detected, then the IRDC is a site of massive star formation. 
Contours of 4.5$\mu$m emission from this cloud are  shown in Figure \ref{fig:33}c.
The fit to the SED of this source suggests  a  
14.4 \msol protostar (see Table 9). However, the SED is not fitted well  
by  the models because of the narrow peak at mid-IR wavelengths.

\bigskip
{ \bf G358.980+0.084 (g32)}\\
The 4.5$\mu$m excess 
source 32 is projected at the center of an IRDC which has a
spidery  appearance.  Figures \ref{fig:34}a,b show  24$\mu$m and 8$\mu$m 
images of this source.  There is no evidence of a compact radio 
counterpart  at 20cm to the 4.5$\mu$m excess 
source, though, a compact source is 
detected about 1$'$  southwest of it.  The
methanol maser coincident with  the position of the 4.5$\mu$m excess 
 source supports the presence of  on-going  massive star formation in this dark 
cloud. 
The SED fit to the source identified by its excess 4.5$\mu$m emission 
gives a 7.6 \msol\. protostar. Figure \ref{fig:34}d presents contours of 4.5$\mu$m 
emission 
from this source. 

\subsection{Foreground 4.5$\mu$m Excess Sources}

Figure \ref{fig:foreground_models}b and Table 10 show the SEDs of individual 
foreground sources and list 
the physical characteristics derived from the fits, respectively. 
The estimated 
A$_v$ range between 4 and 33  
magnitudes.  Due to the distance uncertainty, the mass estimate is not well constrained. 
We note that with the exception of three sources, all the 4.5$\mu$m excess sources 
appear to be extended. In addition, with the exception of one,  all the SED fitted 
sources appear to be in 
Stage I suggesting their early evolutionary phase.

\bigskip
{ \bf 0.955-0.786 (g1), 0.868-0.697 (g2) and  0.780-0.740 (g4)}\\
All the three 4.5$\mu$m excess 
sources g1, g2 and g4 are 
located at the edge of the region surveyed  with IRAC. 
Source g1, g4  and g2 coincide with the  positions of
three 
planetary nebulae JaST 76 and JaST 70 in \citep{steene01} and 
PN G0.8-0.6 in  \citep{kerber03}, respectively.  The SEDs of these sources
are shown in  Figure \ref{fig:25}b. The successful fitting of these PNe  
shows that  SED fitting  can not be used 
to distinguish them from other source types. 
Additional compact 
 4.5$\mu$m excess sources may be 
PNe that are misidentified as YSOs.

\bigskip
{\bf G0.708+0.408 (g5)}\\   
This source is projected against an IRDC observed 
with the MIPS and IRAC data. 
Figures \ref{fig:35}a,b show a 24$\mu$m image and 
a 4.5$\mu$m contour of the 4.5$\mu$m excess 
 source. 
The SED fit to this source suggests  a 6 \msol\ protostar (see Table 10) as 
shown in Figure 
\ref{fig:25}b. However,  the SED is not 
fit fit well because of its   narrow peak.


\bigskip
{\bf 4.5$\mu$m Excess 
 Sources Associated with Sh20 (g11-g15, g12-g14)}\\
Several 4.5$\mu$m excess 
 sources associated with Sh20, are distributed at   negative latitudes 
where several prominent optical emission  nebulae have been 
identified. The 4.5$\mu$m excess sources 
g11, g12, g13, g14 and  g15
are  most likely  associated Sh20 (Sharpless 20)
 centered at l=0.5$\degr$, 
b=-0.3$\degr$  with 
an  extent of 10$'$
\citep{marsalkova74}. 
The SED fitting to these 4.5$\mu$m excess 
 sources give masses in the range between 2 and 14 \msol\ (see 
Table 10).

Figures \ref{fig:36}a,b show 24$\mu$m and 8$\mu$m images of G0.542-0.476 where
4.5$\mu$m excess 
 source g11 is detected.
The 4.5$\mu$m excess 
 source appears to 
be projected against a dearth of emission at 8$\mu$m in the vicinity of an elongated IRDC 
with 
an extent of 2$'$. 
A sharp linear feature with an extent of 
$\sim1'$ lies  to the north. The lack of a 
counterpart at 24$\mu$m and/or at  radio wavelengths  suggests that this 
linear feature is likely to  arise from  PAH emission.  Figure \ref{fig:36}c shows 
a
radio continuum image of the region where g11 is detected and Figure 27d shows 
a 4.5$\mu$m contour map of g11. 

Other 4.5$\mu$m excess 
 sources that appear to be associated with Sharpless 20 are
G0.517-0.657 (g12), G0.483-0.701 (g13) and G0.477-0.727 (g14). 
Figures 28a,b,c show 24$\mu$m, 8$\mu$m and 20cm grayscale images of the 
region where these 4.5$\mu$m excess 
 sources are detected. The 4.5$\mu$m emission contours 
for  g12, g13 and g14 are shown in 
Figures 28d,e,f, 
respectively. It is clear that this is a region of massive star formation 
where sharp extended shell-like structures and a compact radio continuum 
source coincident with g14 are seen at 24$\mu$m and 20cm. 
No 8$\mu$m counterparts are found. 


Another HII   structure with the appearance of a ``question mark'' 
is a  component of  Sharpless  20. 
The 4.5$\mu$m excess 
 source g15 lies at the center of this HII region. 
Figures \ref{fig:38}a,b,c show grayscale images
at 24$\mu$m, 8$\mu$m, and 20cm, respectively, where 
as a contour map of  g15 at 4.5$\mu$m is shown in Figure \ref{fig:38}d.

\bigskip
{\bf G0.315-0.21 (g17)}\\
This 4.5$\mu$m excess 
 source lies in the vicinity of two stellar cluster candidates 
that  are located within  one  arc-minute of each other and within the Sh2-20 
HII region \citep{sharpless59,dutra03}.
This cluster also has variable X-ray emission, indicating very young stars
 \citep{law04}. Figures \ref{fig:39}a,b show contours at 
450$\mu$m  and 850$\mu$m superimposed on a grayscale 24$\mu$m and 
20cm images of this HII region.   
Two sources in the  24$\mu$m and 20cm 
images lie  in the vicinity of a submillimeter peak. 
Contours of 4.5$\mu$m emission 
from both the northern and southern sources are  shown in 
Figure \ref{fig:39}c. The 4.5$\mu$m excess 
 source coincides with the southern component 
of the emission shown in these figures.  
The presence of methanol masers 
coincident with the 4.5$\mu$m excess 
 source and a bright submillimeter peak
provide  strong  
evidence for massive star formation  in this region. 
The fit to the SED of the 4.5$\mu$m excess 
 source is also consistent with 
12.8 \msol\ protostar. 

\bigskip
{\bf G0.167-0.445 (g18), G0.091-0.663 (g19) \& G0.084-0.642 (g20)}\\
The 4.5$\mu$m excess 
sources g18, g19 and g20   
are projected toward the HII region RCW 141,
which is centered 
near  l=0.3$\degr$, b=-0.2$\degr$ 
 with an  extent of  6$'\times4'$. 
Figures \ref{fig:40}a,b show 8$\mu$m and 20cm grayscale images of g18, respectively. 
The two 8$\mu$m sources appear to be embedded within  an IRDC. However, the one 
indicated by a cross shows  strong excess emission at 4.5$\mu$m.  
A nonthermal radio filament that  runs perpendicular to the
Galactic plane is projected 
against the 4.5$\mu$m excess 
 source. 
Figures 32a,b show grayscale 24$\mu$m and 8$\mu$m images 
of g19 and g20, respectively. It is clear that the southern source 
g19 is embedded in an IRDC.  Figure  \ref{fig:41}c shows that neither of these 
4.5$\mu$m excess 
sources has  20cm continuum  counterparts. 
Figures 
\ref{fig:41}d,e show contours of 
4.5$\mu$m emission from g19 and g20, respectively.

\bigskip
{\bf G359.939+0.170 (g22)}\\
The 4.5$\mu$m excess 
 source g22 is located in an HII complex with an IRDC and 
a prominent 8$\mu$m nebulosity. Figures \ref{fig:42}a,b show
contours of 850$\mu$m emission superimposed on 
grayscale images at 24$\mu$m and 20cm, respectively.
The 850$\mu$m emission and the IRDC at 24$\mu$m 
appear to lie at the edge of the nonthermal filament,  as seen 
in Figure \ref{fig:42}b.   
A large-scale view of the dark cloud and 
the  extensive 8$\mu$m emission to its eastern edge are  shown 
in Figure \ref{fig:42}c. Contours of the source responsible for excess
4.5$\mu$m emission are  shown in Figure \ref{fig:42}d. 

\bigskip
{\bf G359.972-0.459 (g21), G359.907-0.303 (g24) \& G359.618-0.245 (g26)}\\
The 4.5$\mu$m excess 
sources g21, g24 and g26   are 
distributed in the vicinity of  RCW 137 which is roughly 
 centered close to l=0.1$\degr$, b=-0.3$\degr$ 
with an extent of 2$'\times2'$. The most interesting of these 4.5$\mu$m excess  
sources is g26,  which  is embedded in a  prominent submillimeter 
core tracing an IRDC. 
Figures \ref{fig:43}a,b show contours of 450$\mu$m and 850$\mu$m emission 
superimposed on 
8$\mu$m and 20cm grayscale images of the IRDC associated with g26. 
The submillimeter cone-shaped structure peaks where 
methanol maser emission is  detected. The closeness 
of methanol masers,  the 4.5$\mu$m excess 
 source and the submillimeter core support the evidence for 
massive star formation in this cloud. Figure \ref{fig:43}c shows 
contours of 4.5$\mu$m emission from
g26, which has an  estimated   7 \msol\ protostar (see Table 10). 

\bigskip
{\bf G359.57-0.27 (g28)}\\
The 4.5$\mu$m excess 
 source g28  lies at the edge of  an IRDC 
that extends for 
2$'$.   Images of the  region at 24 and 8$\mu$m  are 
shown in 
Figures \ref{fig:44}a,b.
   g28 is located in the darkest area of the IR dark cloud, which shows 
an  east-west elongation. 
The source was also detected at  70$\mu$m,  which allows 
us  to constrain the mass of the YSO candidate to be 
4 \msol\ assuming an $<$A$_{v}$$>$=4.   Contours of 4.5$\mu$m emission 
are  shown 
Figure \ref{fig:44}c.


\clearpage
\begin{figure}
\centering
\includegraphics[scale=0.8,angle=-270]{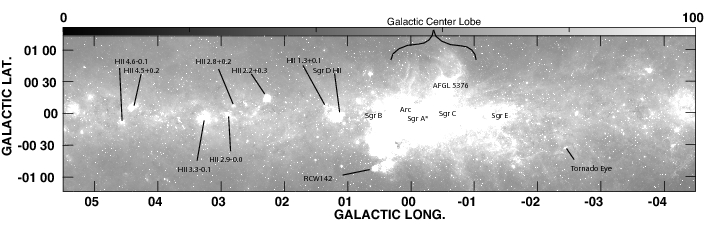}\\
 \caption{
(\textit{a}) A 24$\mu$m image of the inner $\sim10^0\times2.5^0 (l \times b$) 
of the central region of the Galaxy based on combining MIPS and MSX data. 
}
\label{fig:1}
\end{figure}

\begin{figure}
\ContinuedFloat
\centering
\includegraphics[scale=0.6,angle=0]{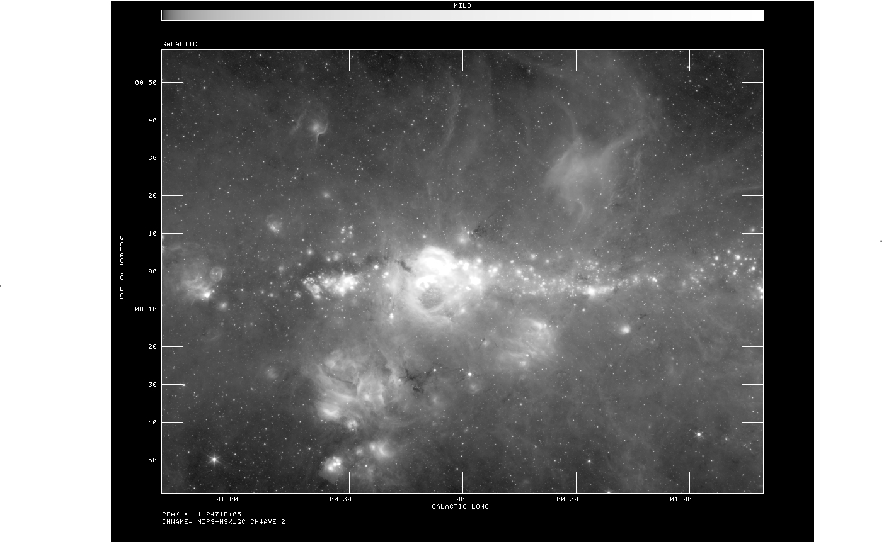}\\
  \caption{
    (\textit{b}) 
 A close-up view of the central 2.5$^0\times2^0$ 
showing  the  prominent HII complexes  (Sgr A -- E) at 24$\mu$m, 
}
\label{fig:1}
\end{figure}

\begin{figure}
\ContinuedFloat
\centering
\includegraphics[scale=0.6,angle=0]{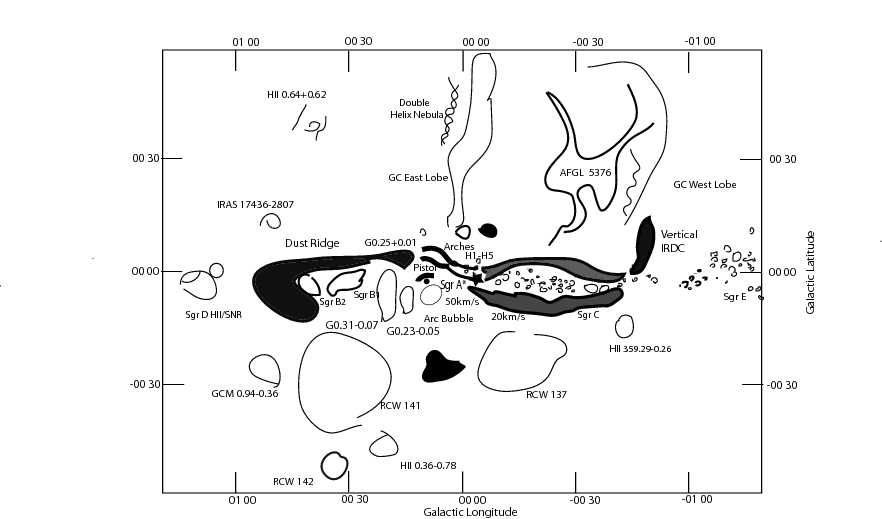}
  \caption{
    (\textit{c}) A schematic diagram of prominent features 
detected at 24$\mu$m from the region shown in (b).
}
\label{fig:1}
\end{figure}

\clearpage
\begin{figure}
\centering
\includegraphics[scale=0.6,angle=-90]{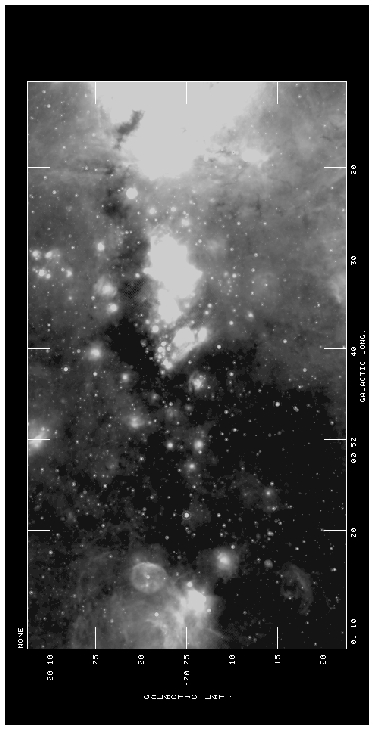}\\
\includegraphics[scale=0.6,angle=-90]{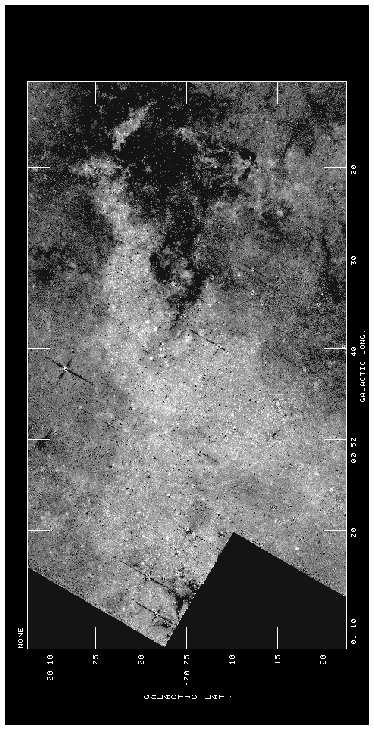}
  \caption{
    (\textit{a - Top})  24$\mu$m emission from  prominent chain of 
IRDCs on the positive longitude side of the Galactic center. The thickest cloud  
surrounds Sgr B2 to the left near l=0.7\deg.
    (\textit{b - Bottom}) Identical to (a) except that the ratio of 
IRAC images
 is displayed. Positive values of this ratio are shown in 
light 
color which  reveals  low contrast features of the IRDCs.  The ratio image is 
constructed by 
 $I(4.5) / [I(3.6)^{1.4}*I(5.8)]^{0.5}$. 
}
\label{fig:2}
\end{figure}

\clearpage
\begin{figure}
\centering
\includegraphics[scale=0.6,angle=-90]{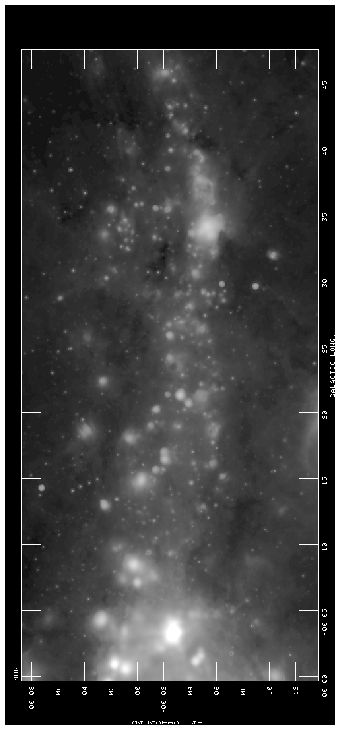}\\
\includegraphics[scale=0.6,angle=-90]{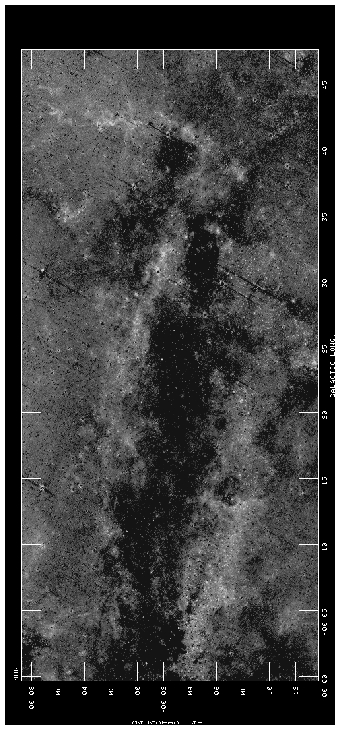}
  \caption{
    (\textit{a - Top}) Similar to Figure \ref{fig:2}a except that 24$\mu$m 
distribution from  
the negative  longitude side  is selected. Note the distribution of stellar 
objects along the Galactic plane. 
   (\textit{b - Bottom}) As in  Figure \ref{fig:2}b 
regions with higher  intensity ratios reveal details in the 
structure of IRDCs.  
Most of the stellar sources seen in (a) are mainly distributed between the 
two extended IRDCs running parallel to the Galactic plane. 
}
\label{fig:3}
\end{figure}

\begin{figure}
\centering
\includegraphics[width=4in, angle=90]{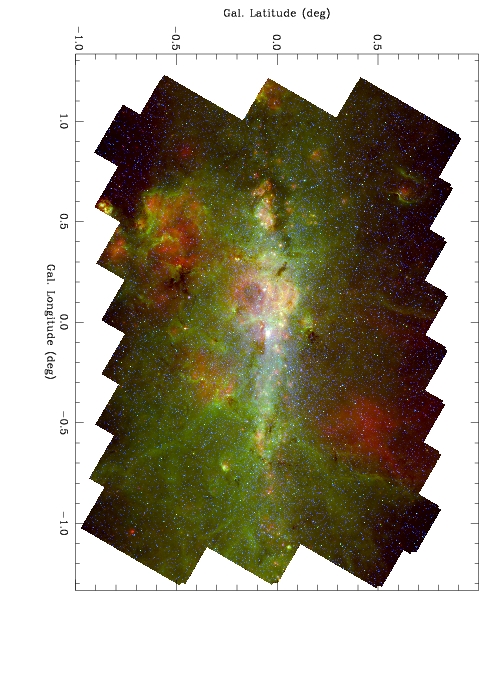}\\
\includegraphics[width=4in, angle=90]{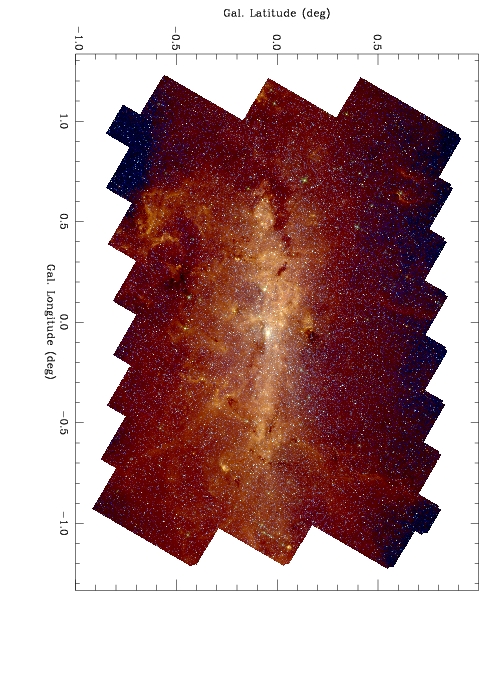}
\caption{ 
(\textit{a - Top})  Color images of the 24,  8 and 3.6$\mu$m   
emission from the central 2$^0\times1.4^0$ are shown in red,  green and blue
colors, respectively. The diffuse clouds peaking at 24$\mu$m are  centrally heated
whereas the  emission in green color is mainly due to PAHs in photodissociation 
regions. 
(\textit{b - Bottom})  Similar to (a) except that the 
4.5, 5.8 and 8$\mu$m images are shown  in blue, green and red colors, respectively. }
\label{fig:13}
\end{figure}

\begin{figure}
\centering
\includegraphics[scale=0.8, angle=90]{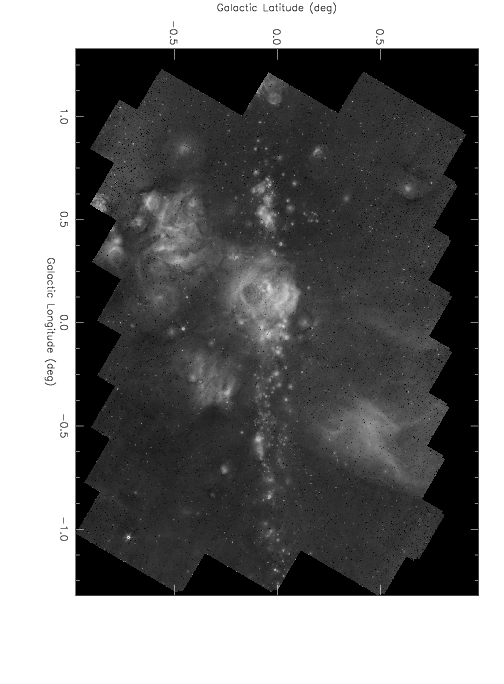}\\
\includegraphics[scale=0.6, angle=270]{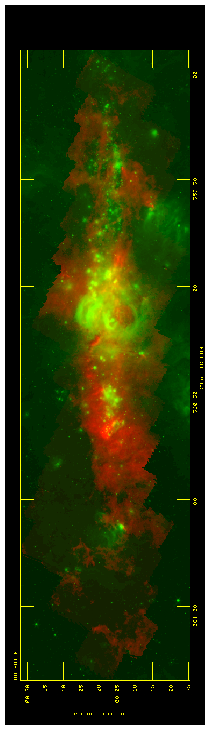}
\caption{
(\textit{a - Top}) 
A large-scale grayscale ratio  image  I(24$\mu$m)/I(8$\mu$m)  
shows the bright region to be mainly due to 24$\mu$m emission 
whereas dark region shows the region dominated by 8$\mu$m emission.
Note that the dark regions in this
ratio image do {\it not} represent the IRDCs.
(\textit{b - Bottom})
The distribution of 24$\mu$m emission in green color  is shown  against  
the distribution of 450$\mu$m emission \citep{pierceprice00} in red color.}
\label{fig:14}
\end{figure}


\begin{figure}
\centering
\includegraphics[scale=0.7, angle=0]{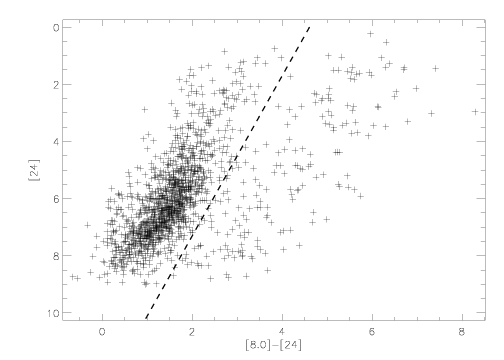}
\includegraphics[scale=0.7, angle=0]{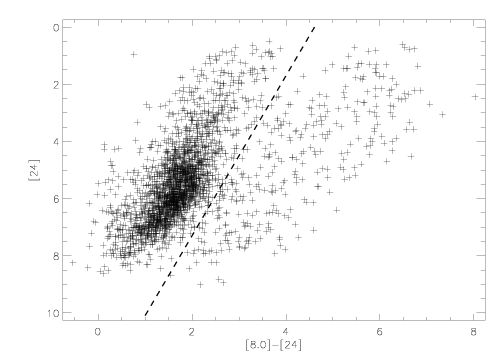}
  \caption{
    (\textit{a - Top}) The color magnitude diagram of 24$\mu$m 
sources covers the region b$\pm10'$ and 0 $<$ l $< 1.4^0$. 
The dashed line is an empirical separator between evolved AGB stars
(to the left) and candidate YSOs to the right. 
    (\textit{b - Bottom}) Similar to (a) except that the search area includes -1.4$^0 
<$ l $<$ 0. }
\label{fig:15}
\end{figure}

\begin{figure}
\centering
\includegraphics[scale=0.5, angle=0]{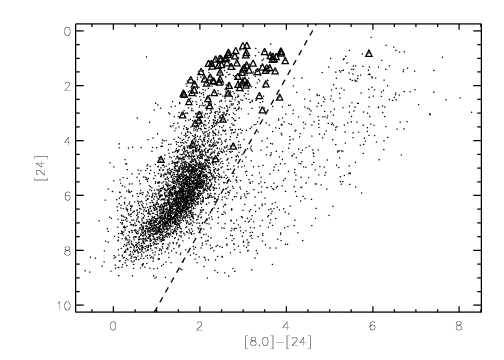}\\
\includegraphics[scale=0.5, angle=0]{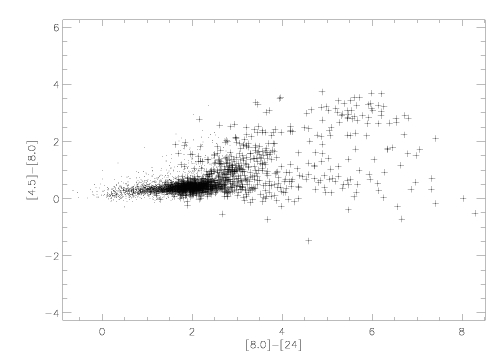}
\includegraphics[scale=0.5, angle=0]{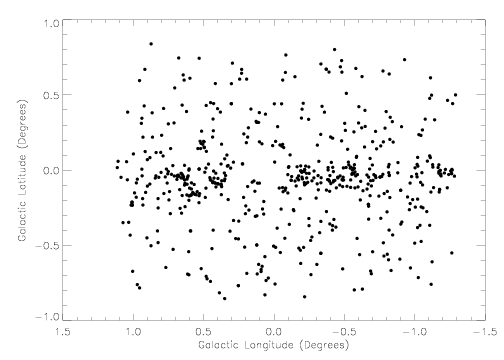}
  \caption{
    (\textit{a - Top}) 
The color magnitude diagram of the region covering 
$-10' < b < 10'$ and $-1.4\arcdeg < l < 1.4\arcdeg$. The triangles show the
locations of known OH/IR stars within the region. The dashed line is drawn
to separate the OH/IR stars from candidate YSOs with redder colors.
A total of 4,541 sources is plotted as dots in the CMD.
    (\textit{b - Middle}) Color color magnitude diagram of the region covered in (a). 
The crosses show the location of candidate YSOs identified in (a). 
    (\textit{c - Bottom}) The spatial distribution of 
24$\mu$m sources. The central 
hole 
centered on Sgr A is due to MIPS saturation at this wavelength. 
}
\label{fig:16}
\end{figure}

\begin{figure}
\centering
\includegraphics[scale=0.4, angle=0]{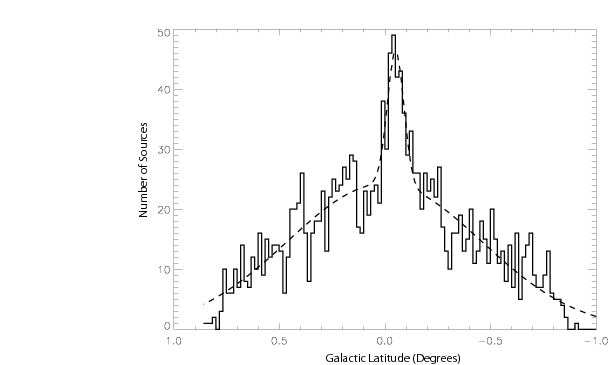} 
\includegraphics[scale=0.4,angle=0]{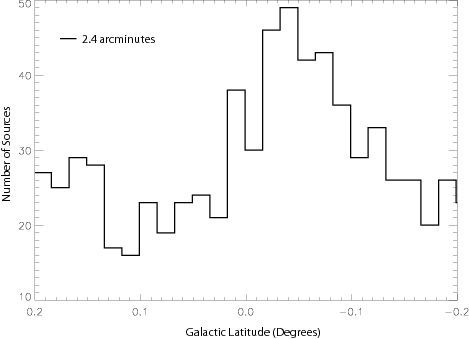}
\caption{
    (\textit{a - Left}) 
The histogram of YSO candidates as a function of
latitude for the central $|l|$ $<$ 1.3\degr\ shows two components,
which were  fit by  Gaussians. The broad component is centered near
$b\sim$0.0\degr\ with a FWHM of 64\arcmin\, whereas the narrow
component is centered at $b$=-3\arcmin\ with a FWHM of $\sim$5.5\arcmin .
Each bin is 1\arcmin .
    (\textit{b - Right}) 
The distribution of YSOs with Galactic latitude for the
same region as (a) but over the region $|b|$ $<$ 0.2\degr .
The units of latitude  are degrees. The latitude of Sgr A* is $b=-2' 46.2''$. }
\label{fig:17}
\end{figure}

\begin{figure}
\ContinuedFloat
\centering
\includegraphics[scale=0.5, angle=-90]{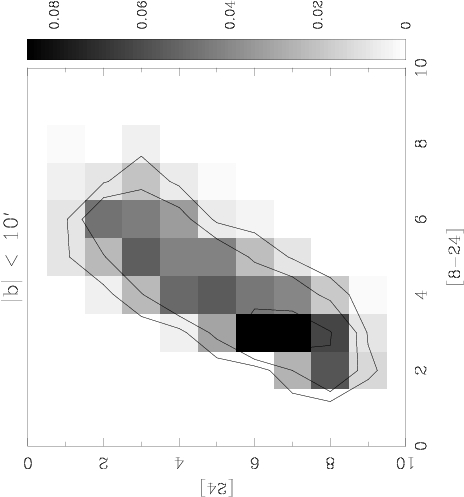} 
\includegraphics[scale=0.5, angle=-90]{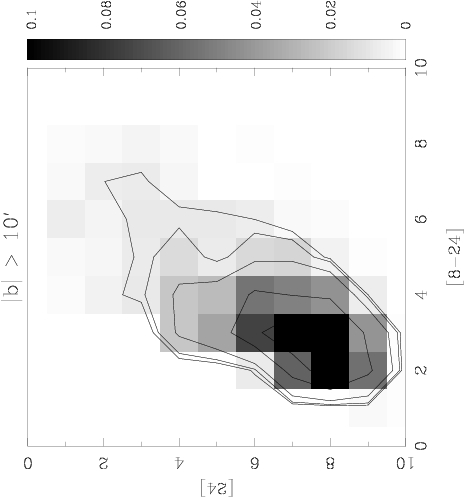} 
  \caption{
The YSO candidates for the inner region (\textit{c - Top}) 
and the outer region
     (\textit{d} Bottom) 
binned the sources 
in both [8]-[24] color and [24] magnitude with 
a bin size of 1 magnitude. The number of YSO candidates
 in each bin is then divided by 
the total
number of YSO candidates in the region to give a measure of the 
color-magnitude
properties of the YSO candidates. Contour levels of  
0.1, 0.25 and 0.6  for (c) and   
0.007, 0.01, 0.05 and 0.08 for (d) are 
also drawn.  It is clear that there is a higher fraction of brighter, redder YSO
candidates in the innermost region of the Galactic center when compared to outside 
sources.
}
\label{fig:17}
\end{figure}

\begin{figure}
\centering
\includegraphics[width=2.9in,angle=0]{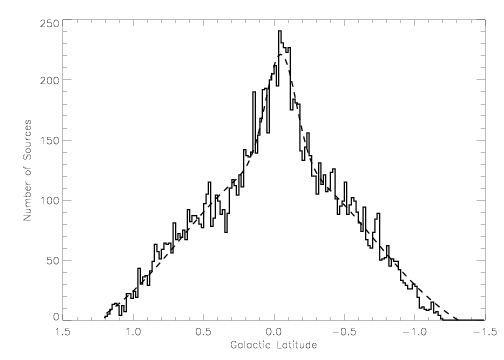}
\includegraphics[width=2.9in, angle=0]{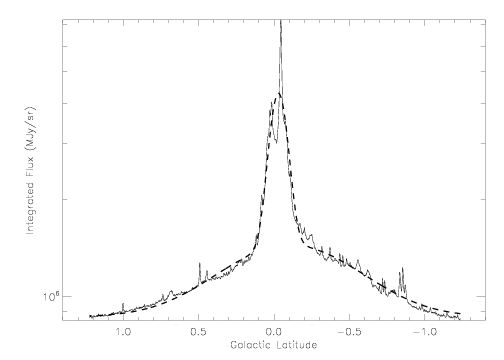}
\caption{
    (\textit{a - Left}) 
 The distribution of unsaturated MIPS 24$\mu$m point
sources brighter than 5 magnitudes, binned as a function of Galactic
latitude. Each bin is 0.1\arcmin . The distribution can be decomposed
into two Gaussian components, a narrow and a broad component with FWHM
of 15\arcmin\ and 96\arcmin , respectively.
    (\textit{b - Right}) 
The total brightness distribution of 24$\mu$m emission
as a function of latitude based on the combined MIPS and MSX image
using $|l| < 4\degr $. The units of latitude  are degrees.
}
\label{fig:18}
\end{figure}



\begin{figure}
\centering
\includegraphics[scale=0.5, angle=0]{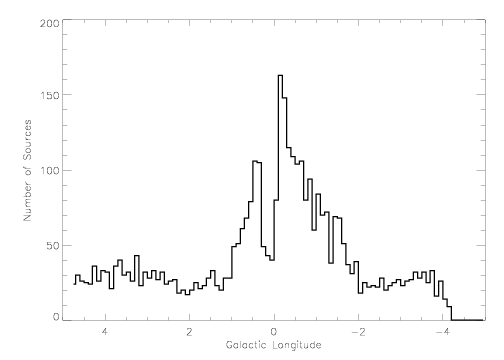}\\ 
\includegraphics[scale=0.5, 
angle=0]{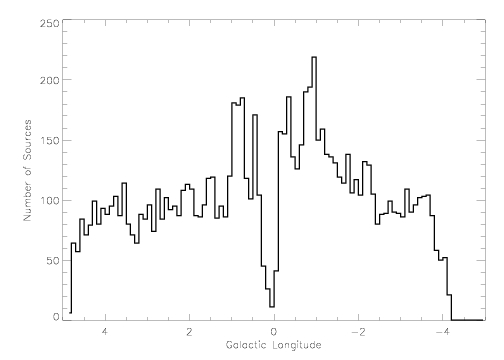}\\
\includegraphics[scale=0.5, 
angle=0]{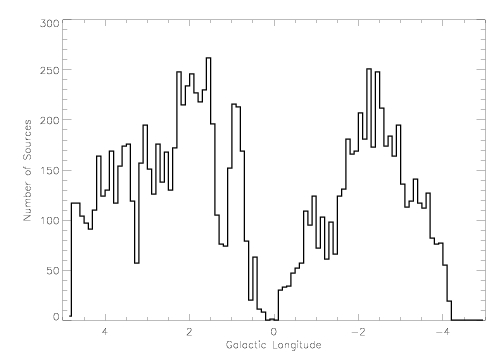}
  \caption{
    (\textit{a - Top}) 
The brightness distribution of 24$\mu$m point sources 
as a function of longitudes in shown for 
24$\mu$m sources ranging between  0 and 5 magnitudes. 
    (\textit{b - Middle}) The middle panel shows the distribution of  point sources 
with 
intermediate brightness between  5 and 7 magnitudes.  
    (\textit{c - Bottom}) Similar to the middle panel except for faint sources
between 7 and 10 magnitudes. All the plots have used 
the entire region of the 24$\mu$m  survey but restricted the latitudes to within 
b=$\pm10'$. The dips in the distribution near $l=0\degr$ are largely caused by the inability to 
detect faint sources in regions where 24$\mu$m  emission is extremely bright. 
The units of longitude are degrees.}
\label{fig:19}
\end{figure}

\begin{figure}
\centering
\includegraphics[scale=0.4, angle=0]{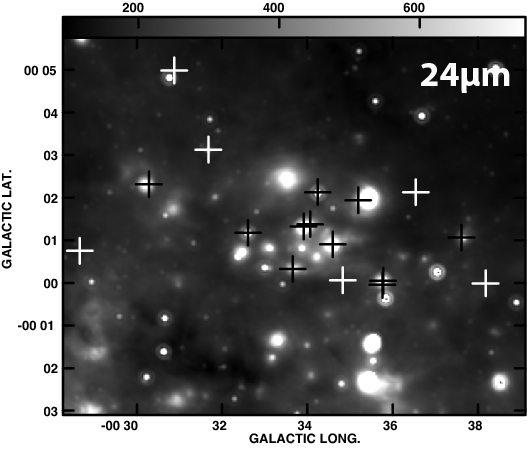}
\includegraphics[scale=0.35, angle=0]{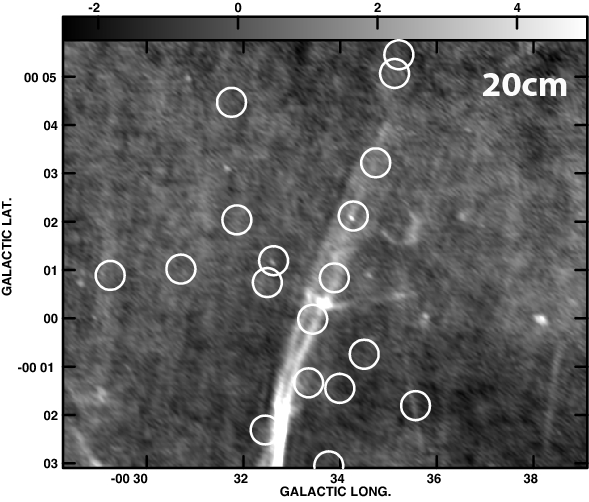}\\
\includegraphics[scale=0.35, angle=0]{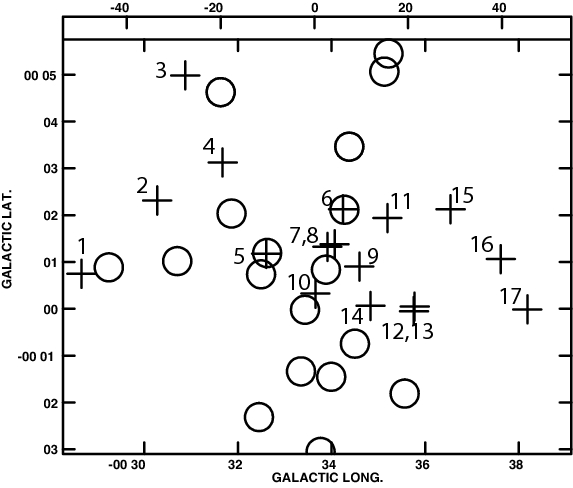}
\includegraphics[scale=0.4, angle=0]{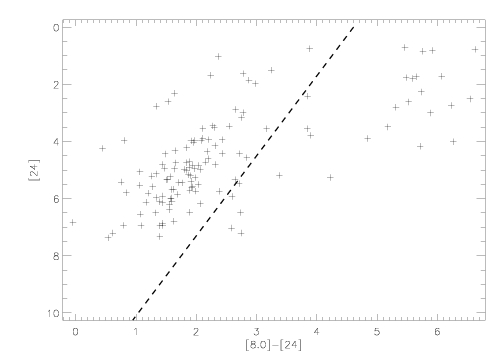}
  \caption{
    (\textit{a - Top Left})
A grayscale 24$\mu$m image of the G359.43+0.02 stellar cluster. 
    (\textit{b - Top Right}) A grayscale 20cm continuum image with a 
resolution of 
12$''\times12''$ with an identical size 
to that shown in (a). 
    (\textit{c - Bottom Left}) A finding chart of YSO candidate sources and 
radio continuum sources in the region shown in (a) and (b). 
    (\textit{d - Bottom Right}) 
A plot of the  CMD of the infrared sources  
found in  (a). A total of 18 candidate YSOs are selected from the area to 
the right  of the plot.  
The crosses present the position 
of the compact  radio sources as listed in Table 1 whereas   
the circles coincide with the position of 
unsaturated 24$\mu$m YSO candidate sources, as listed in Table 2.}
\label{fig:20}
\end{figure}

\begin{figure}
\centering
\includegraphics[scale=0.3]{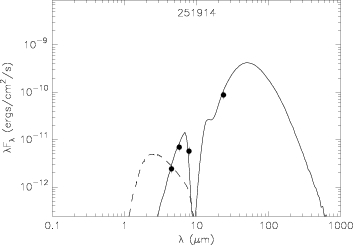}
\includegraphics[scale=0.3]{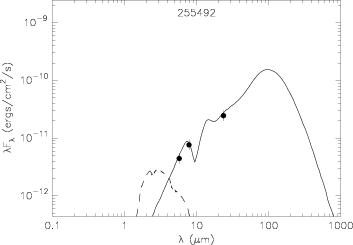}
\includegraphics[scale=0.3]{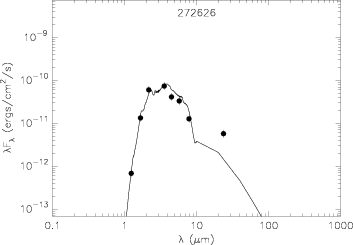}
\includegraphics[scale=0.3]{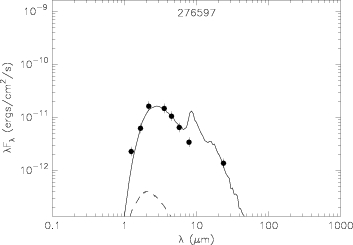}
\includegraphics[scale=0.3]{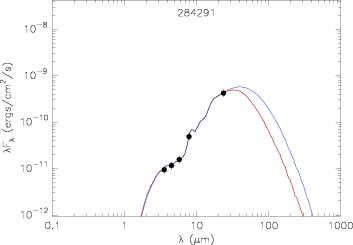}
\includegraphics[scale=0.3]{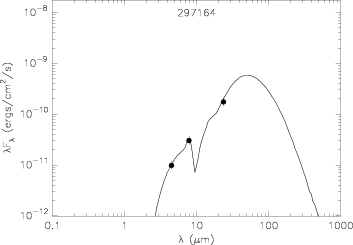}
\includegraphics[scale=0.3]{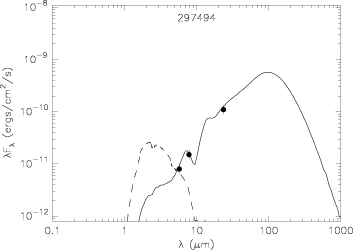}
\includegraphics[scale=0.3]{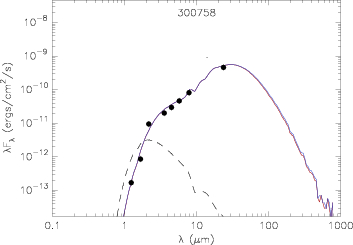}
\includegraphics[scale=0.3]{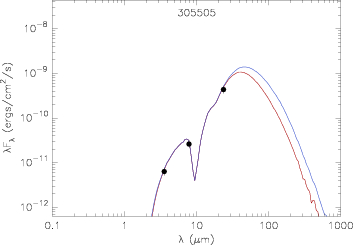}
\includegraphics[scale=0.3]{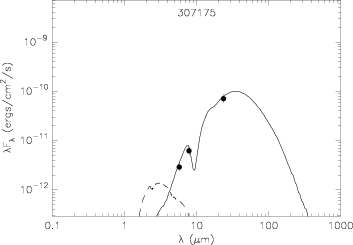}
\includegraphics[scale=0.3]{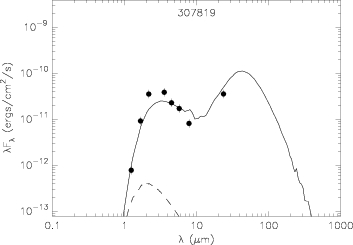}
\includegraphics[scale=0.3]{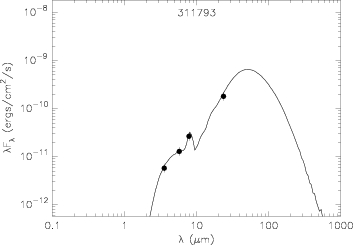}
\includegraphics[scale=0.3]{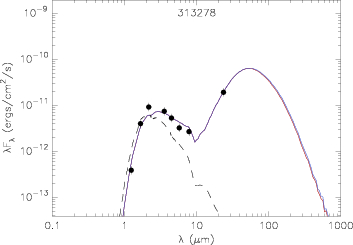}
\includegraphics[scale=0.3]{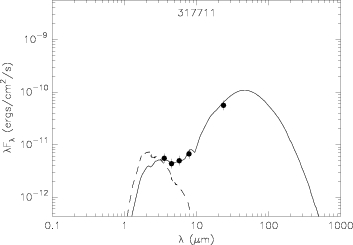}
\includegraphics[scale=0.3]{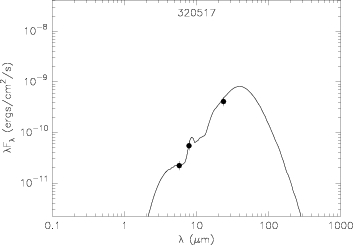}
\includegraphics[scale=0.3]{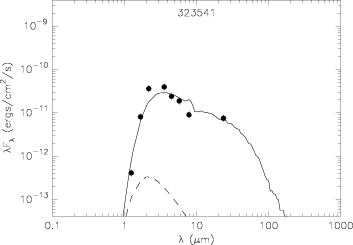}
\includegraphics[scale=0.3]{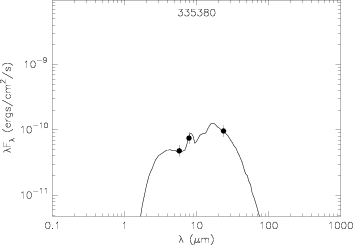}
\includegraphics[scale=0.3]{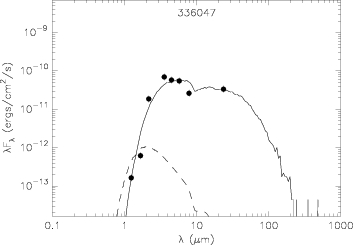}
\caption{
(\textit{a})
 The SED fitted plots of 18 candidate YSOs
found in the G359.43+0.02 cluster. 
The parameters of the fits are shown in Table 2.
}
\label{fig:21}
\end{figure}


\begin{figure}
\ContinuedFloat
\centering
\includegraphics[scale=0.5, angle=0]{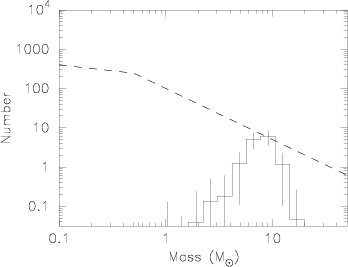}
\includegraphics[scale=0.5, angle=0]{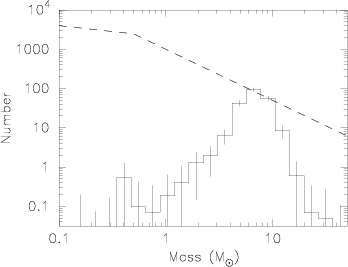}
\caption{
  (\textit{b Left}) The histogram of the mass of YSO candidates for the 
cluster G359.43+0.02. 
    (\textit{c Right}) Similar to (b) 
except for all the YSO candidates distributed within 
the Galactic disk restricted to 
$|b|=10'$ and 
$|l|<1.3^0$.
 Number" is both (a) and (b) represent all the good fits 
for a given source  in the 
distribution and is  normalized so that the  sum equals the average mass which  is
given in Tables 2 to 4.   Each source is  represented 
by a distribution of well-fit masses. 
The broken power-law form is fitted to the data and is 
shown  as dashed line in both (b) and (c). 
}

\end{figure}

\begin{figure}
\centering
\includegraphics[width=6in, angle=0]{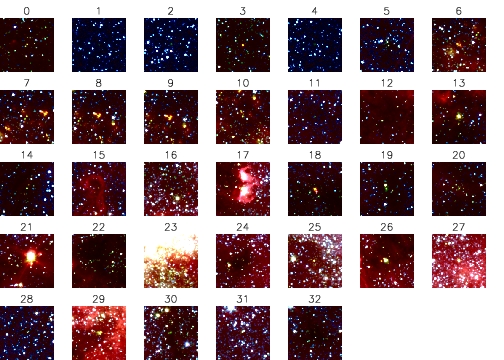}
\caption{
Color images of all the 4.5 $\micron$ sources found in the IRAC
survey region. Red, green, and blue colors denote 8, 4.5, and 3.6 $\micron$
emission respectively. Each image shows a region of 7.5$'\times7.5'$ centered
on the 4.5$\mu$m excess sources.   
\label{fig:22}}
\end{figure}

\begin{figure}
\centering
\includegraphics[scale=0.5, angle=0]{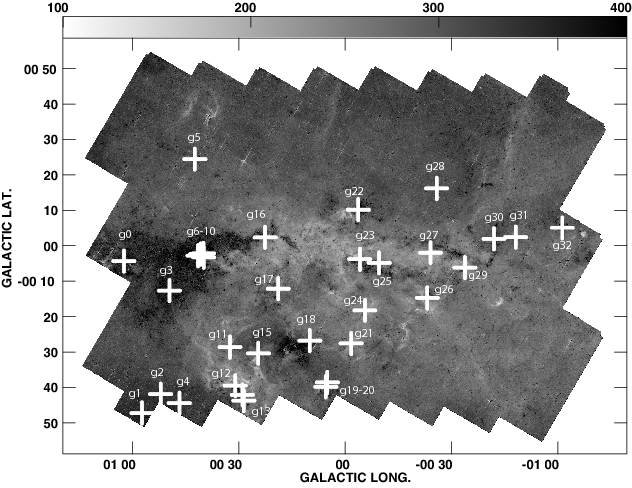}\\
\includegraphics[scale=0.5, angle=-90]{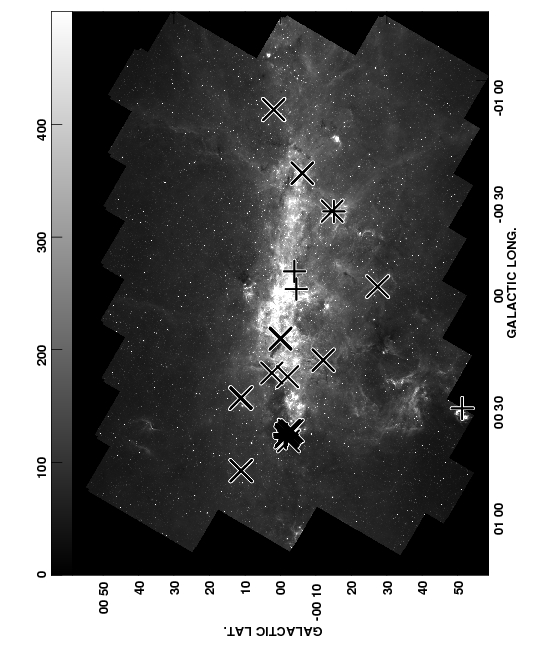}
  \caption{
    (\textit{a - Top}) 
The distribution of 4.5$\mu$m excess 
 sources, presented as crosses, is superimposed 
on a ratio image constructed from 
 $I(4.5) / [I(3.6)^{1.2}*I(5.8)]^{0.5}$. 
    (\textit{b -  Bottom}) The distribution of class I and II methanol masers,
 shown as "+" and "x" signs, respectively is superimposed 
on an 8$\mu$m image.   
\label{fig:23}}
\end{figure}

\begin{figure}
\centering
\includegraphics[scale=0.3, angle=0]{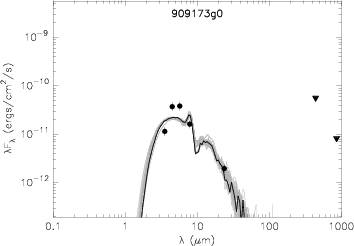}
\includegraphics[scale=0.3, angle=0]{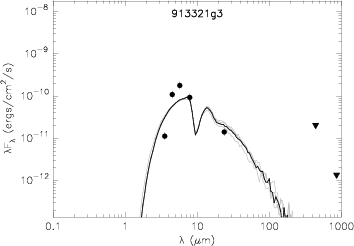}
\includegraphics[scale=0.3, angle=0]{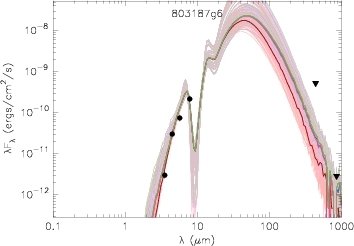}
\includegraphics[scale=0.3, angle=0]{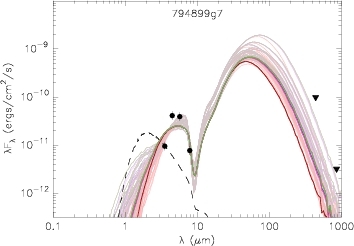}
\includegraphics[scale=0.3, angle=0]{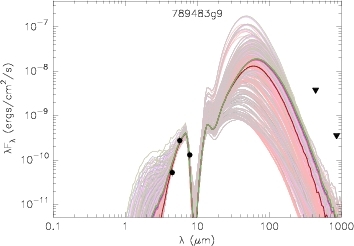}
\includegraphics[scale=0.3, angle=0]{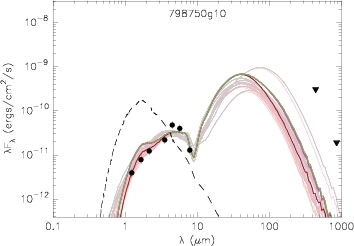}
\includegraphics[scale=0.3, angle=0]{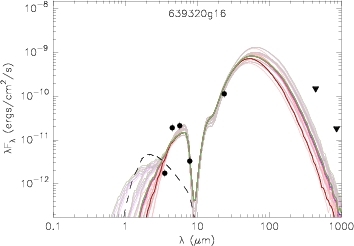}
\includegraphics[scale=0.3, angle=0]{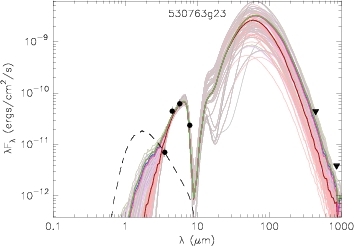}
\includegraphics[scale=0.3, angle=0]{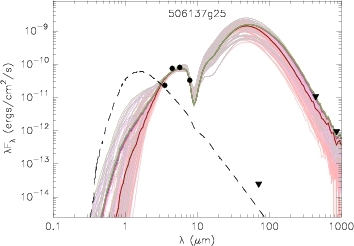}
\includegraphics[scale=0.3, angle=0]{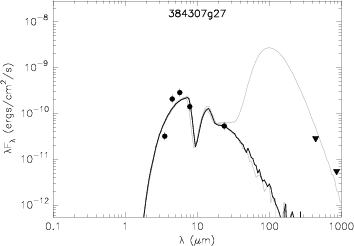}
\includegraphics[scale=0.3, angle=0]{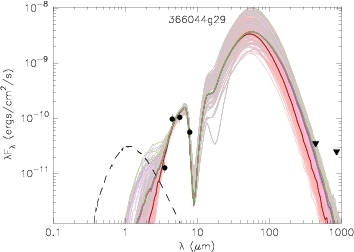}
\includegraphics[scale=0.3, angle=0]{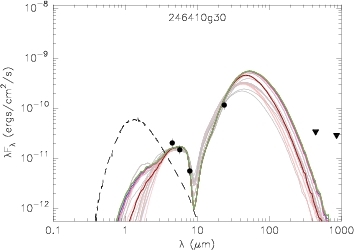}
\includegraphics[scale=0.3, angle=0]{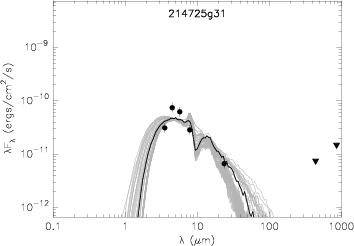}
\includegraphics[scale=0.3, angle=0]{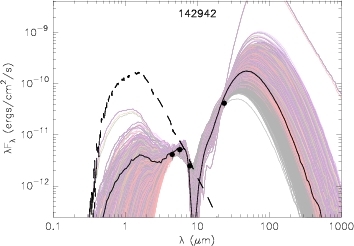}
\caption{
\label{fig:galcen_models}
(a) SED Fits to  4.5$\mu$m excess 
 sources toward the Galactic center.
The  colored lines (reddish)
correspond to fits made from different apertures.  The grey lines are for
IRAC sizes, and pink for 24$\mu$m. 
}
\end{figure}

\begin{figure}
\ContinuedFloat
\includegraphics[scale=0.3, angle=0]{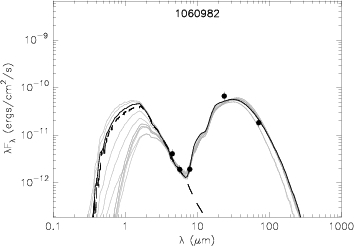}
\includegraphics[scale=0.3, angle=0]{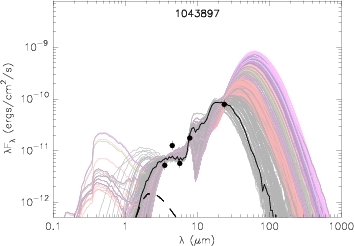}
\includegraphics[scale=0.3, angle=0]{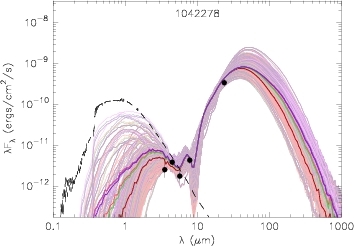}
\includegraphics[scale=0.3, angle=0]{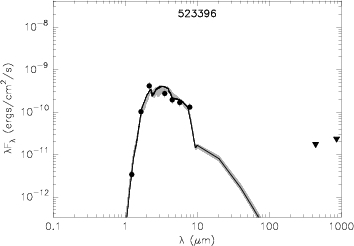} 
\includegraphics[scale=0.3, angle=0]{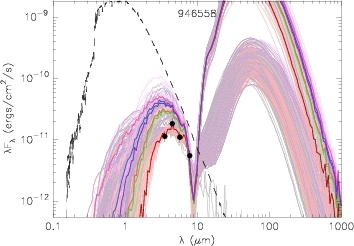}
\includegraphics[scale=0.3, angle=0]{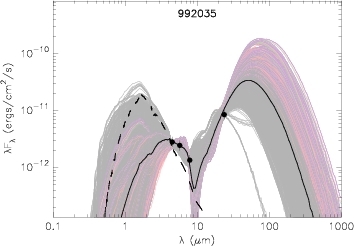} 
\includegraphics[scale=0.3, angle=0]{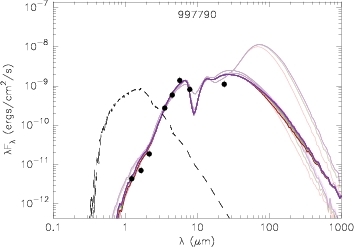}
\includegraphics[scale=0.3, angle=0]{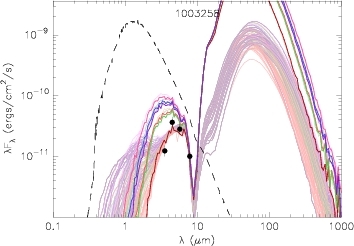}
\includegraphics[scale=0.3, angle=0]{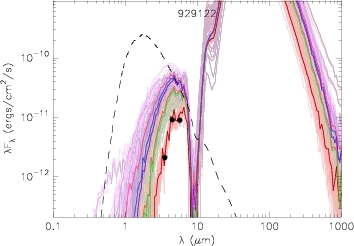} 
\includegraphics[scale=0.3, angle=0]{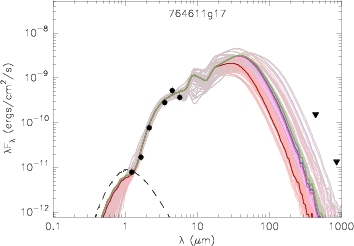}
\includegraphics[scale=0.3, angle=0]{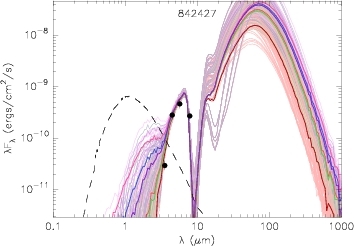}
\includegraphics[scale=0.3, angle=0]{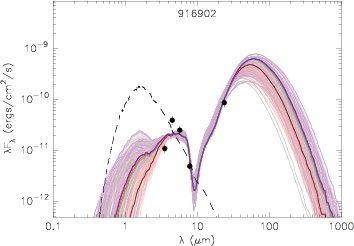}
\includegraphics[scale=0.3, angle=0]{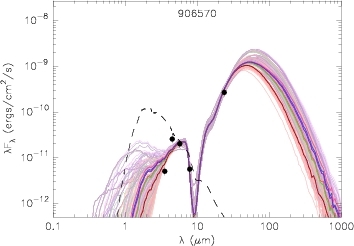}
\includegraphics[scale=0.3, angle=0]{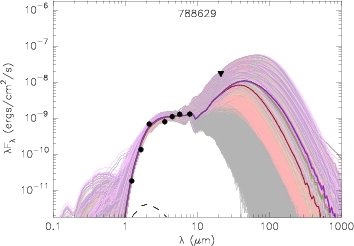}
\includegraphics[scale=0.3, angle=0]{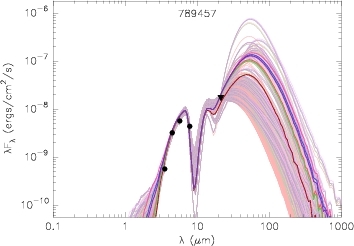}
\includegraphics[scale=0.3, angle=0]{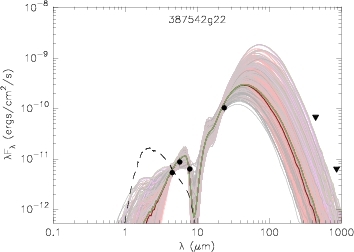}
\includegraphics[scale=0.3, angle=0]{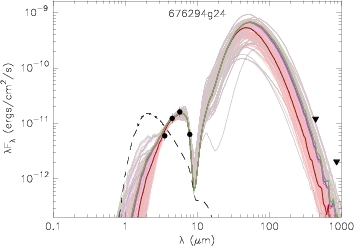}
\includegraphics[scale=0.3, angle=0]{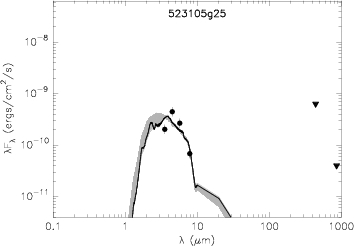}
\includegraphics[scale=0.3, angle=0]{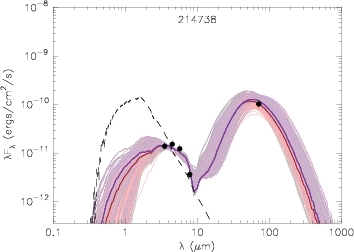}
\caption{
(b) SED Fits to foreground 4.5$\mu$m Excess 
 Sources
}
\label{fig:foreground_models}
\end{figure}

\clearpage

\begin{figure}
\centering
\includegraphics[origin=c,scale=0.7, angle=0]{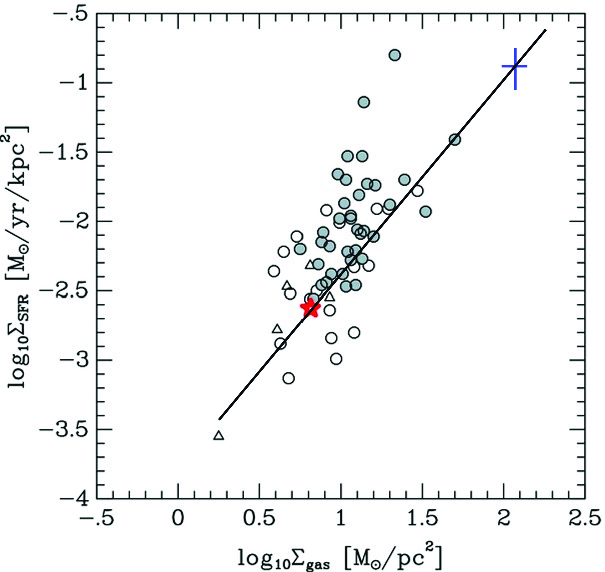}
  \caption{
 The correlation of the SFR with the gas surface density (Fuchs et al. 
2009; Kennicut 1998). 
Sa-Sb
galaxies are shown as open triangles, Sb-Sc as open circles, and Sc-Sd as filled 
circles. The
red star is the value for the solar neighborhood and the blue cross is that for the 
Galactic
center. The black line has a slope of 1.4, corresponding to the Schmidt-Kennicutt 
relationship
found for external galaxies (Kennicutt 1998). 
}
\label{fig:25}
\end{figure}

\begin{figure}
\centering
\includegraphics[origin=c,scale=0.3, angle=0]{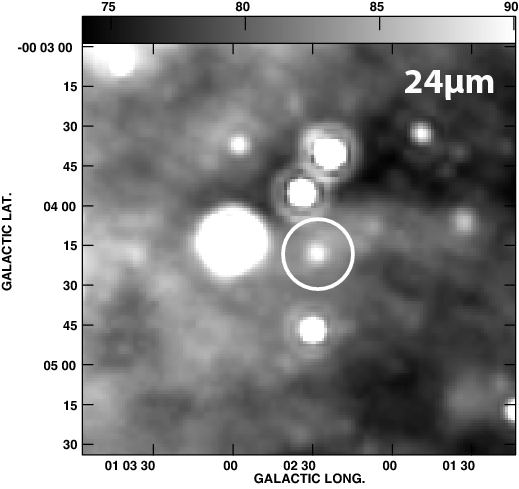}
\includegraphics[origin=c,scale=0.3, angle=-90]{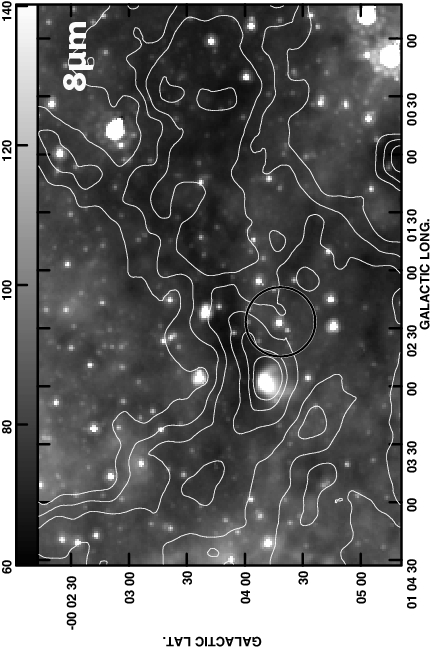}
  \caption{
    (\textit{a - Left}) A 24$\mu$m image of the 4.5$\mu$m excess 
 source G1.041-0.072 (g0). 
The circle represents the position of the 4.5$\mu$m excess 
 source which coincides with a
24$\mu$m source.  
    (\textit{b - Right}) Similar to (a) except that this  8$\mu$m image 
shows an segment of an IRDC ridge and g0. Contours of 850$\mu$m based on 
SCUBA observations are also superimposed on the 
figure.  Levels are at 1.750, 2, 2.25, 2.5, 2.75, 3
and 4 Jy/beam.
The circle
marks the  position of the 4.5$\mu$m excess 
sources whereas  the square represents sources that are  observed at 70$\mu$m.
The distribution of class I and II methanol masers are also 
drawn on these figures as plus(+)  and cross (X) signs, respectively. 
These symbols apply for  Figures 16-35. 
}
\label{fig:26}
\end{figure}

\begin{figure}
\ContinuedFloat
\centering
\includegraphics[origin=c,scale=0.3, angle=270]{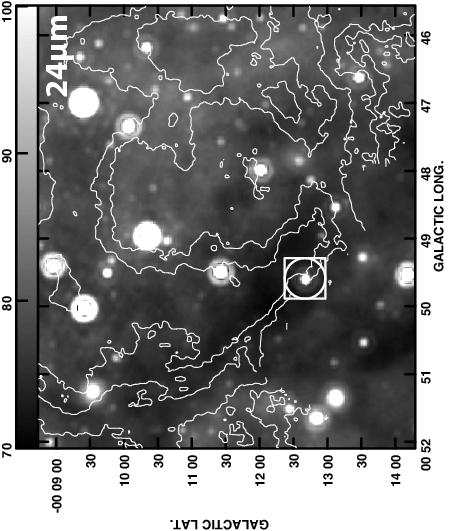}
\includegraphics[origin=c,scale=0.3, angle=270]{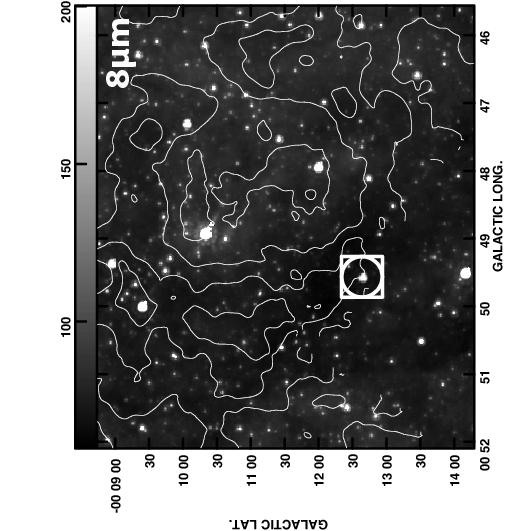}
  \caption{
    (\textit{c - Left}) Contours of 450$\mu$m emission are superimposed on 
a 24$\mu$m image of g3. The location of g3 is shown by a circle 
whereas the square sign shows the location at which 70$\mu$m observation was 
made. Levels at 6, 7, 9, 12, 14 and 20 Jy/beam.
    (\textit{d - Right}) Similar to (a) except that contours of  850$\mu$m emission
are superimposed on an 8$\mu$m image of g3.  Levels at 2, 2.5, 3, 3.5, 4, 4.5 
Jy/beam. 
}
\label{fig:26}
\end{figure}

\vfill\eject

\begin{figure}
\centering
\includegraphics[origin=c,scale=0.4, angle=270]{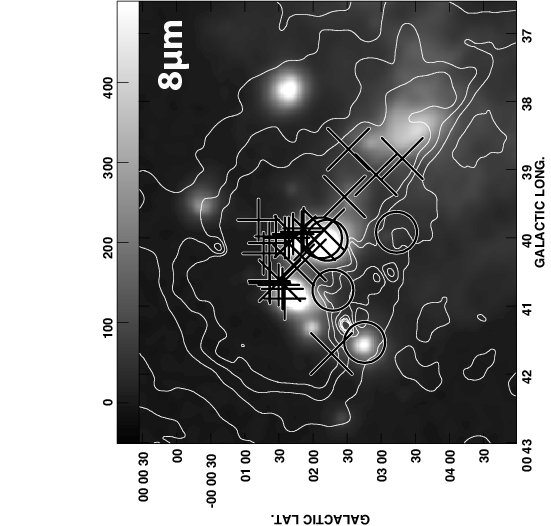}\\
\includegraphics[origin=c,scale=0.4, angle=0]{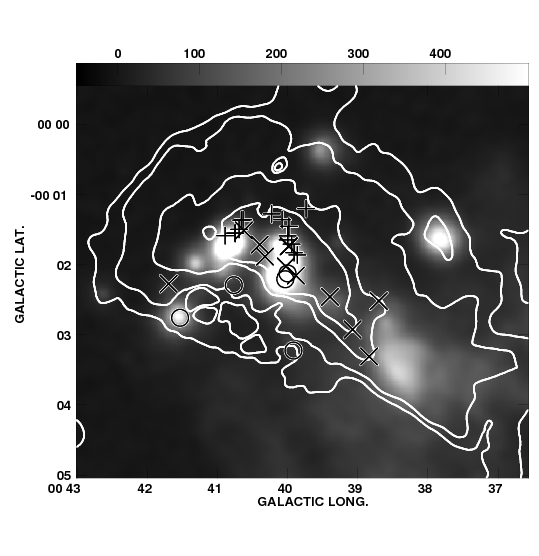}
  \caption{
    (\textit{a - Top }) Contours of 450$\mu$m emission superimposed on an 
8$\mu$m image of Sgr B2. 
 sources. Levels are at 35, 40, 50, 100, 200 and 400 Jy/beam.
    (\textit{b - Bottom}) Similar to (a) except that 850$\mu$m contours are 
superimposed on a 20cm radio continuum image. Levels are at 6, 7.5, 15 and 25 Jy/beam.
Symbols are as defined in Figure 17.
}
\label{fig:27}
\end{figure}

\begin{figure}
\ContinuedFloat
\centering
\includegraphics[origin=c,scale=0.4, angle=270]{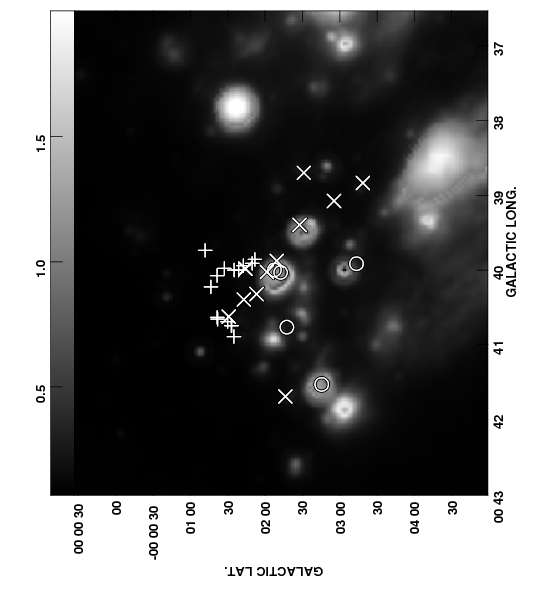}\\
\includegraphics[origin=c,scale=0.4, angle=0]{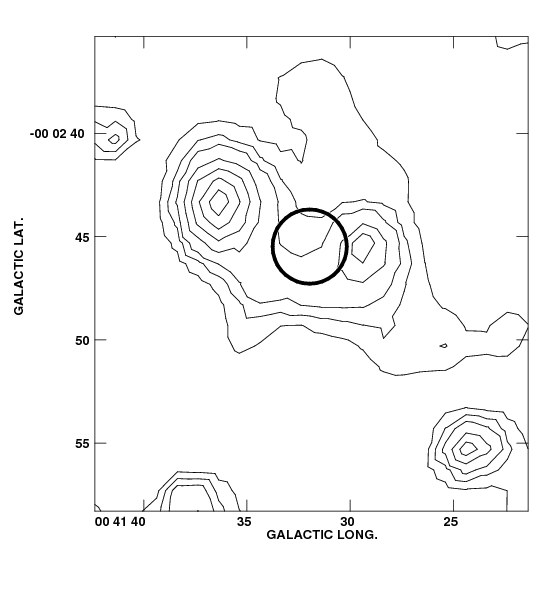}
  \caption{
    (\textit{c - Top}) A 24$\mu$m grayscale image of Sgr B2. 
    (\textit{d - Bottom}) Contours of 4.5$\mu$m emission from 
the 4.5$\mu$m excess 
source g6. Levels are at 25, 35, 50, 75, 100, 125 MJy sr$^{-1}$.
Symbols are as defined in Figure 17.
}
\label{fig:27}
\end{figure}

\clearpage

\begin{figure}
\centering
\includegraphics[origin=c,scale=0.35, angle=270]{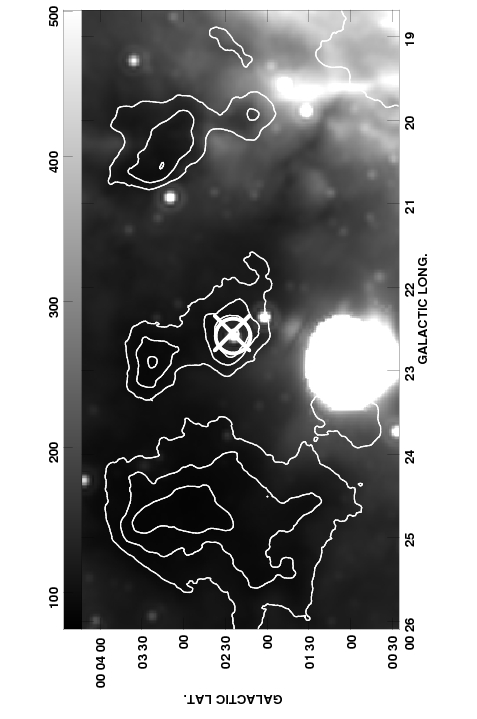}\\
\includegraphics[origin=c,scale=0.35, angle=270]{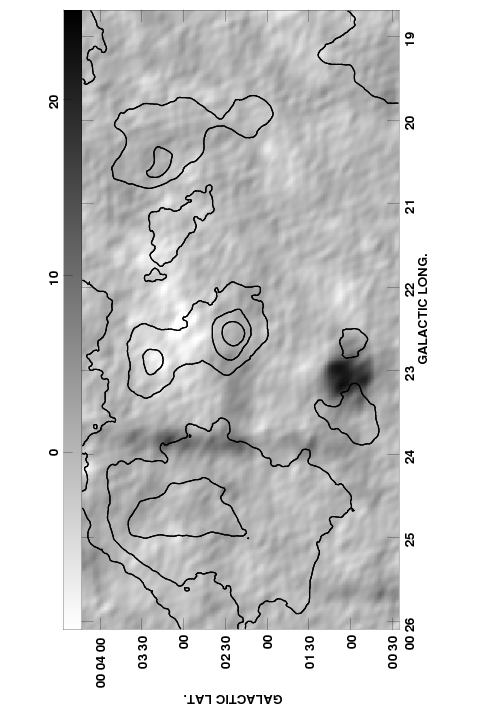}\\
\includegraphics[origin=c,scale=0.3, angle=0]{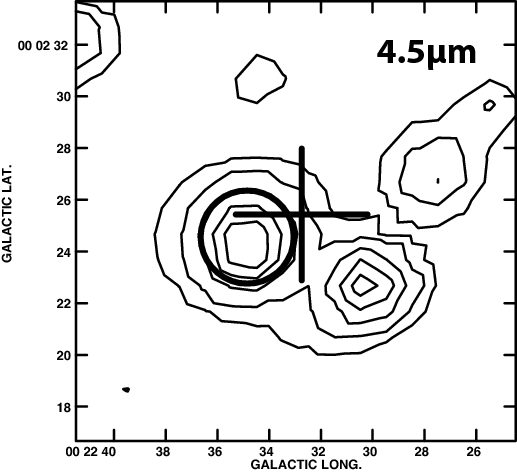}
  \caption{
    (\textit{a - Top}) Contours of 450$\mu$m emission around g16 are superimposed on
    a 24$\mu$m grayscale image. Levels are at 3, 4, 5.5 and 8 Jy/beam.
    (\textit{b - Middle}) Similar to (a) except that contours of 850$\mu$m emission
    are superimposed on a grayscale 90cm image. Levels are at 15, 20, 27.5 and 40 Jy/beam.
    (\textit{c - Bottom}) A 4.5$\mu$m contours of emission from the region where
    g16 is identified. Levels are at 20, 30, 50, 70, 90 and 110 MJy/sr.
Symbols are as defined in Figure 17.
}
\label{fig:28}
\end{figure}

\begin{figure}
\centering
\includegraphics[origin=c,scale=0.3, angle=270]{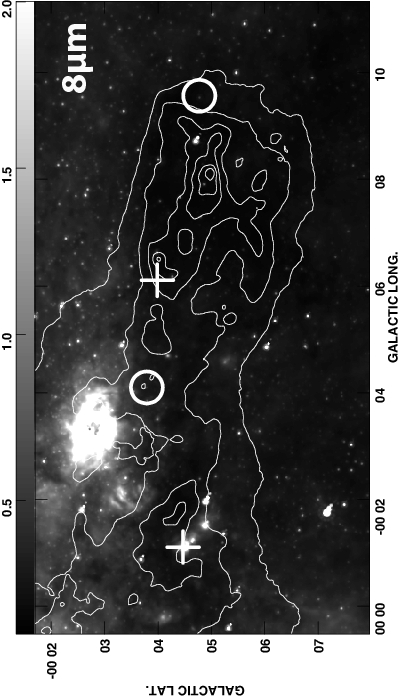}\\
\includegraphics[origin=c,scale=0.3, angle=270]{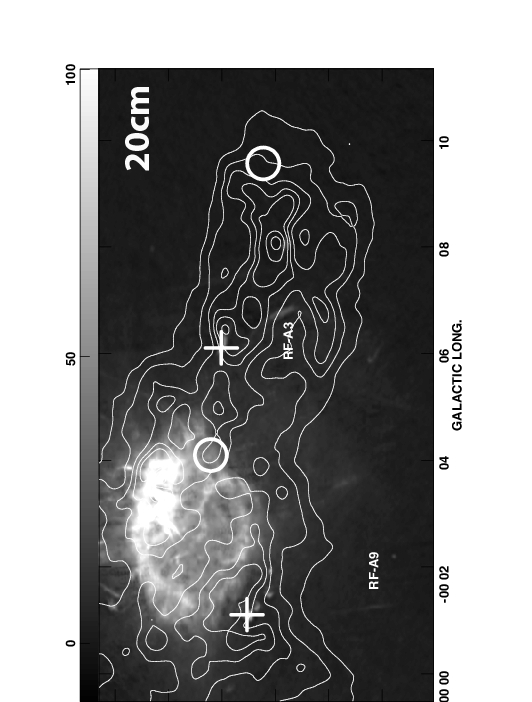}\\
\includegraphics[origin=c,scale=0.3, angle=270]{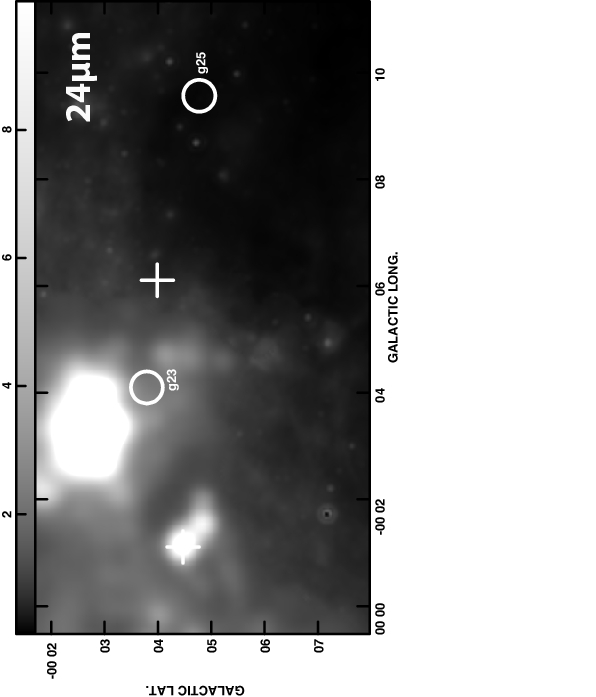}
  \caption{
    (\textit{a Top}) Contours of 450$\mu$m emission from the 
    50 and 20 \kms\ molecular clouds are superimposed on an 8$\mu$m image 
    of this region.  The circles show the positions of 
    4.5$\mu$m excess 
 sources g23 and g25, projected toward the 50 \kms and 20 \kms\ molecular 
    clouds, respectively.  The plus signs shown the position of class I methanol sources 
    in this region. 
    (\textit{b - Middle}) Similar to (a) except that contours of 850$\mu$m emission 
    are superimposed on a radio continuum image at 20cm. Levels are at 4, 5, 6, 7, 8, 10, 12 
    and 14 Jy/beam
    (\textit{c - Bottom}) A 24$\mu$m image of the Sgr A region based mainly on 
    the saturated MIPS data that are replaced by MSX data. The IRDCs associated with 
    the 50 and 20 \kms are best represented in this figure. 
Symbols are as defined in Figure 17.
}
\label{fig:29}
\end{figure}

\begin{figure}
\ContinuedFloat
\includegraphics[origin=c,scale=0.3, 
angle=270]{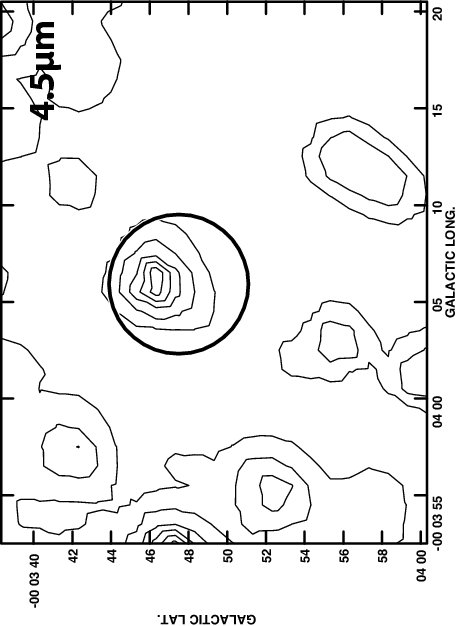}
\includegraphics[origin=c,scale=0.3, 
angle=0]{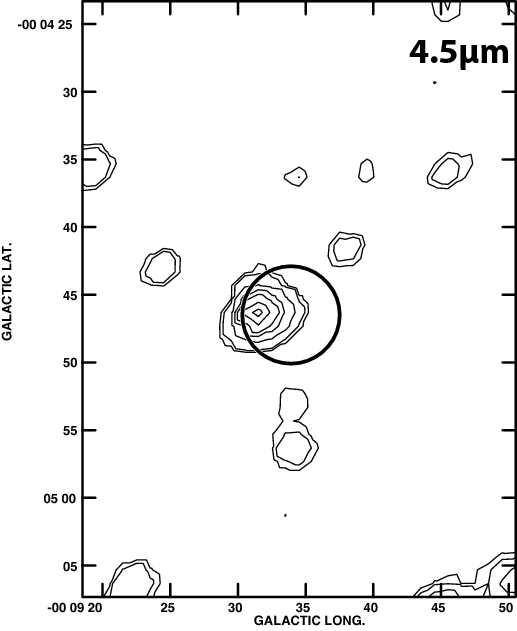}
  \caption{
    (\textit{d - Left}) Contours of 4.5$\mu$m emission from g23 at levels of 50, 
100, 200,
    300, 400, 500, 600 and 700 MJy/sr. 
    (\textit{e - Right})  Contours of 4.5$\mu$m emission from g25 at levels of 40, 
50, 100, 200,  300, 500, 700 MJy/sr.
Symbols are as defined in Figure 17.
}
\label{fig:29}
\end{figure}

\begin{figure}
\centering
\includegraphics[origin=c,scale=0.2 ,angle=0]{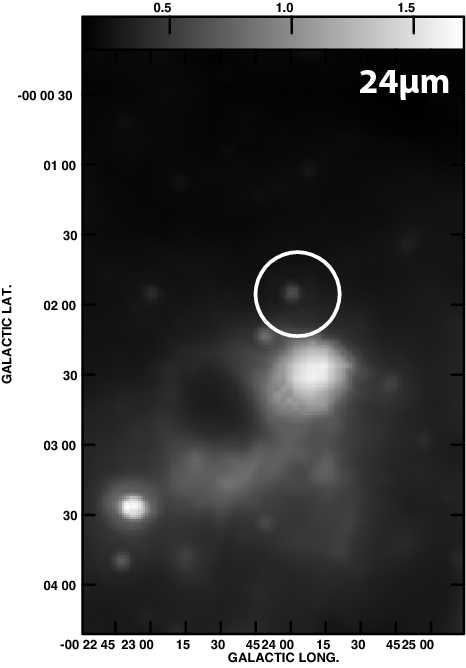}
\includegraphics[origin=c,scale=0.2, angle=0]{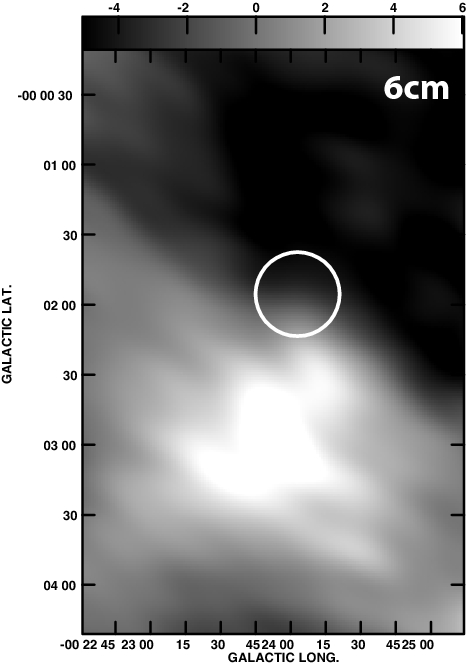}
\includegraphics[origin=c,scale=0.2, angle=-90]{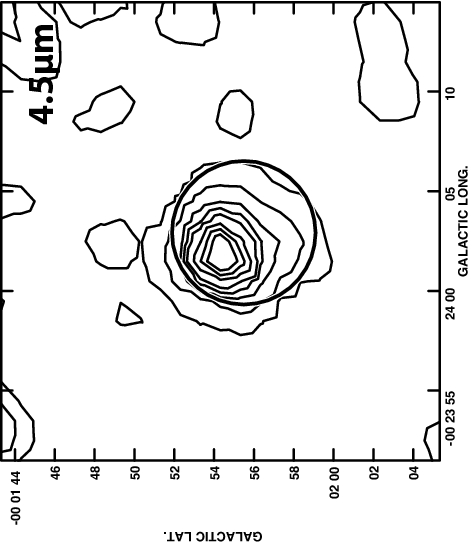}
  \caption{
    (\textit{a - Left}) A 24$\mu$m image of G359.6-0.03 or g27 as presented by a
circle where 4.5$\mu$m excess emission is detected.   
    (\textit{b - Middle}) A 20cm continuum image of (a). 
    (\textit{c - Right}) contours of emission from the source associated with g27 at 
4.5$\mu$m at 
levels of
    50, 100, 200, 300, 400, 600, 800 and 1000 MJy/sr.
Symbols are as defined in Figure 17.
}
\label{fig:30}
\end{figure}

\begin{figure}
\centering
\includegraphics[scale=0.3, angle=270]{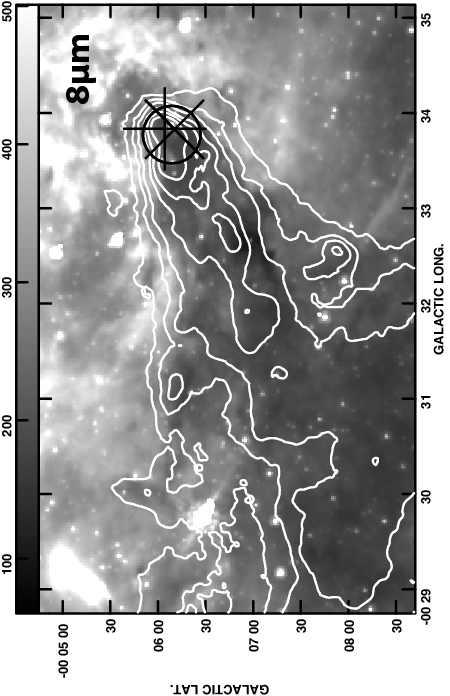}\\
\includegraphics[scale=0.3, angle=270]{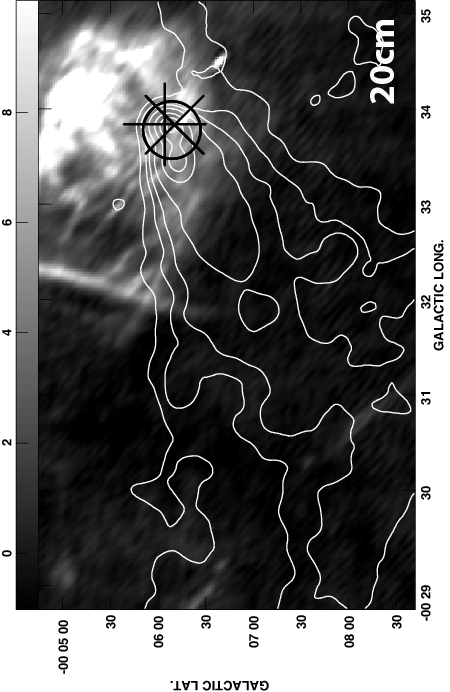}
  \caption{
    (\textit{a} Top) Contours of 450$\mu$m emission from Sgr C with a resolution 
8$''$ 
are superimposed on a grayscale 8$\mu$m image. 
Levels are at 5, 7.5, 10, 15, 20 
and 30 Jy/beam.
    (\textit{b} Bottom) Contours of 
450$\mu$m emission from Sgr C 
are superimposed  on a grayscale 20cm  image with a spatial resolution 
of 12.8$''\times12.8''$. 
Levels are at 2, 2.5, 3, 4, 6, 8, and 10 Jy/beam.
Symbols are as defined in Figure 17.
}
\label{fig:31}
\end{figure}

\begin{figure}
\ContinuedFloat
\centering
\includegraphics[scale=0.3, angle=270]{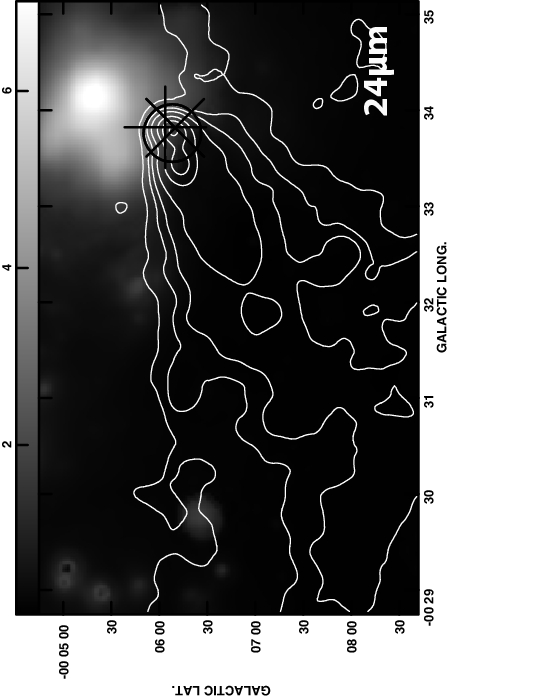}
\includegraphics[scale=0.3, angle=270]{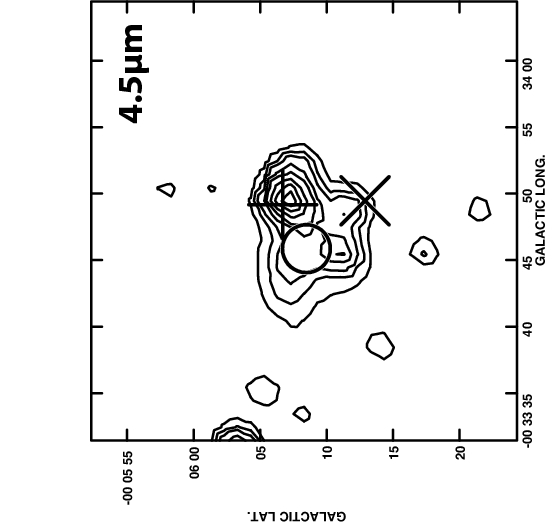}
  \caption{
   (\textit{c} Top) A MIPS image 
of the region shown in (a) at 24$\mu$m.  
    (\textit{d} Bottom) Contours of 4.5$\mu$m emission from the 4.5$\mu$m excess 
 source g29, as   identified by a circle. 
Levels are at  50, 75, 100, 150, 200 
and 300 MJy/sr. 
Symbols are as defined in Figure 17.
}
\label{fig:31}
\end{figure}

\begin{figure}
\centering
\includegraphics[height=2in,origin=c, angle=270]{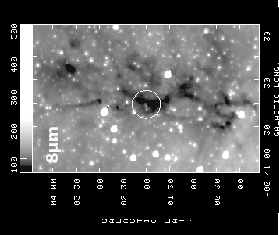}
\includegraphics[height=2in,origin=c, angle=270]{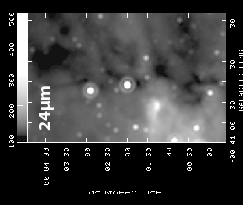}
\includegraphics[height=3in,origin=c, angle=0]{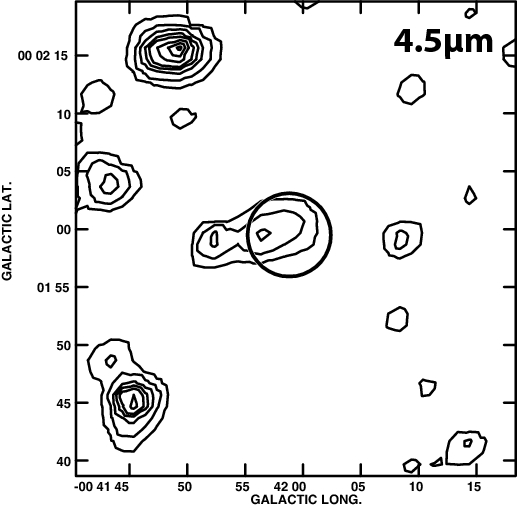}
  \caption{
    (\textit{a - Top Left}) A grayscale  8$\mu$m  image of the 4.5$\mu$m excess 
 source g30. 
The position of the 4.5$\mu$m excess 
 source is shown by a circle.  
    (\textit{b - Top Right}) Similar to (a) except at 24$\mu$m. 
    (\textit{c - Bottom}) Contours of 4.5$\mu$m emission from G359.30+0.033 at levels 
      of  25, 50, 100, 150, 200, 300, 400 and 500 MJy/sr.
Symbols are as defined in Figure 17.
}
\label{fig:32}
\end{figure}

\newpage

\begin{figure}
\centering
\includegraphics[origin=c,scale=0.3, angle=270]{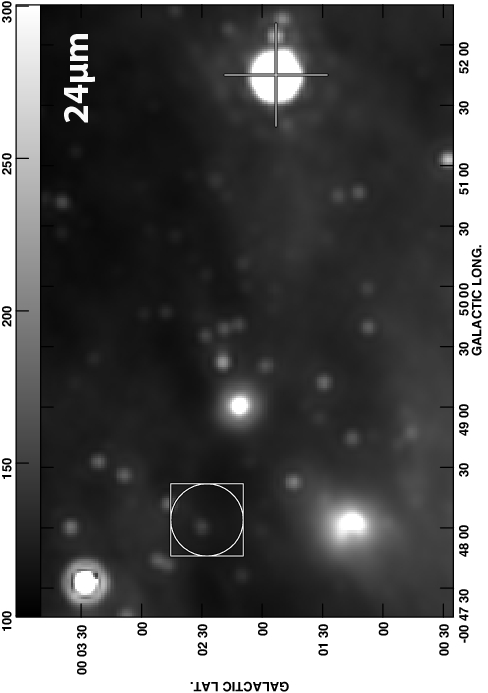}
\includegraphics[origin=c,scale=0.3, angle=270]{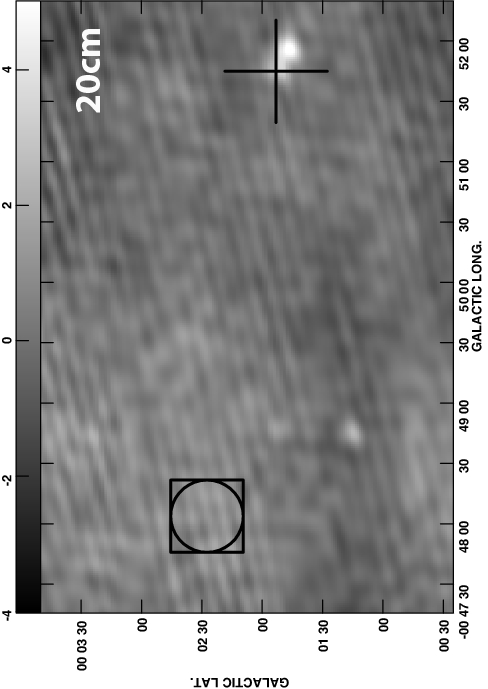}
\includegraphics[origin=c,scale=0.3, angle=270]{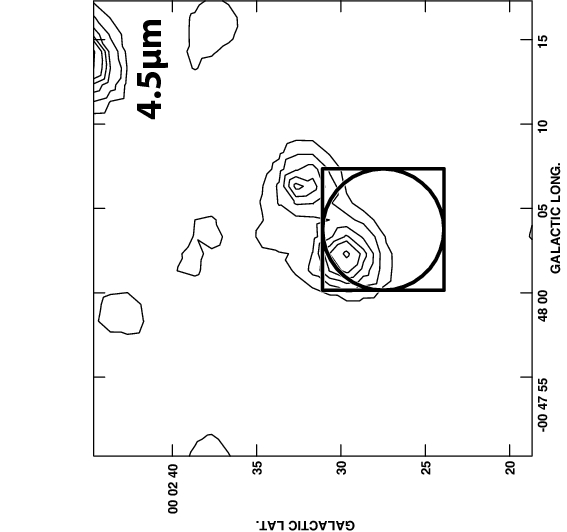}
  \caption{
    (\textit{a - Top Left}) A grayscale 24$\mu$m image of the 4.5$\mu$m excess 
 source g31.
    (\textit{b - Top Right}) Similar to (a) except that the image is shown at 20cm. 
    (\textit{c - Bottom}) Contours of 4.5$\mu$m emission from  
    G359.199+0.041 at levels of 50, 100, 200, 300, 400, 600, 800 and 1000 MJy/sr. 
Symbols are as defined in Figure 17.
}
\label{fig:33}
\end{figure}

\newpage

\begin{figure}
\centering
\includegraphics[origin=c,scale=0.3, angle=0]{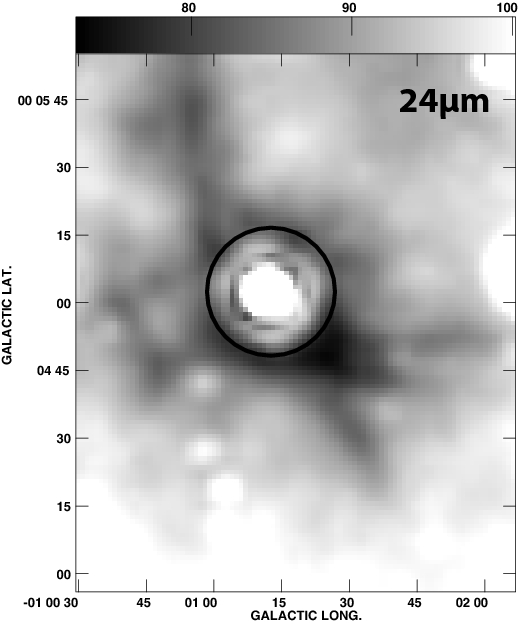}
\includegraphics[origin=c,scale=0.3, angle=0]{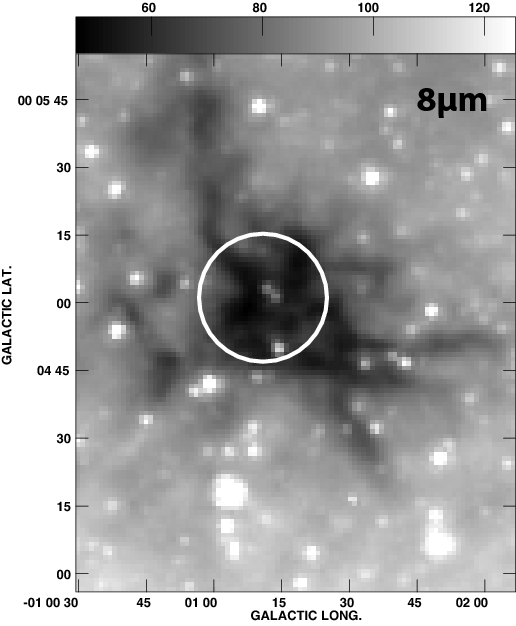}
\includegraphics[origin=c,scale=0.3, angle=0]{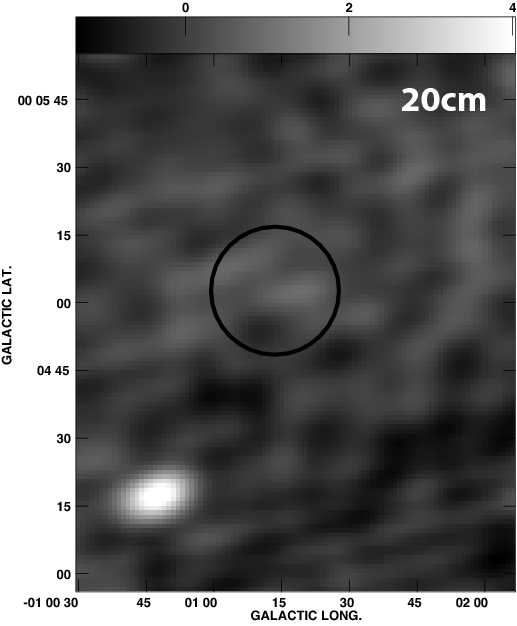}
\includegraphics[origin=c,scale=0.3, angle=0]{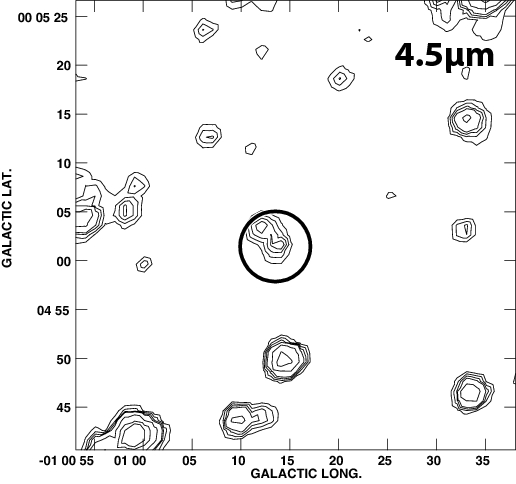}
  \caption{
    (\textit{a - Top Left}) A grayscale  24$\mu$m  image of the 4.5$\mu$m excess 
 source g32 from 
73 MJy/sr.
    (\textit{b - Top Right}) Similar to (a) except at 8$\mu$m from 47 to 125 MJy/sr.
    (\textit{c - Bottom Left}) Similar to (a) except at  20cm from range= -1.3 to 4 
mJy/beam.
    (\textit{d - Bottom Right}) 
Contours of 4.5$\mu$m emission from G359.30+0.033 at levels of 15, 20, 25, 30, 50, 100, 
150 and 300 MJy/sr. 
The position of the 4.5$\mu$m excess 
 source is shown by a circle. 
Symbols are as defined in Figure 17.
}
\label{fig:34}
\end{figure}

\begin{figure}
\includegraphics[origin=c,scale=0.3, angle=0]{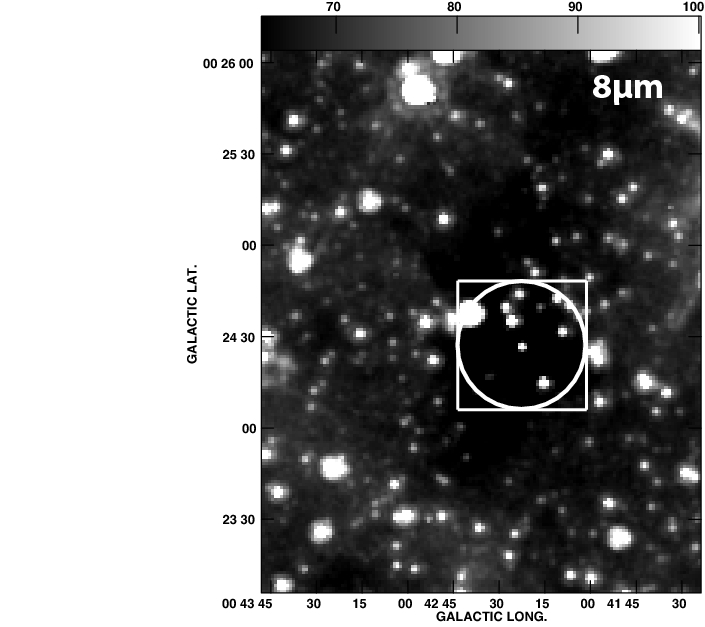}
\includegraphics[origin=c,scale=0.3, angle=270]{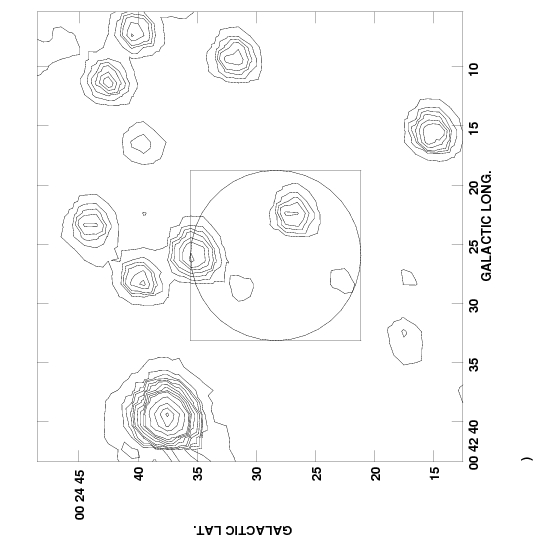}
  \caption{
    (\textit{a - Left}) An 8$\mu$m image from 64 to 100 MJy/sr of the 4.5$\mu$m 
excess  source g5,  G0.708+0.408,  
    projected in the  middle of an IR dark cloud.  
    (\textit{b - Right}) Contours of 4.5$\mu$m emission from 
    the 4.5$\mu$m excess 
   source g5 that shows excess emission at this wavelength. Levels are at 
      25 50 75 100 150 200 300 400 500 600 1000 2000 2500 MJy/sr.
Symbols are as defined in Figure 17.
}
\label{fig:35}
\end{figure}

\begin{figure}
\centering
\includegraphics[origin=c,scale=0.35, angle=270]{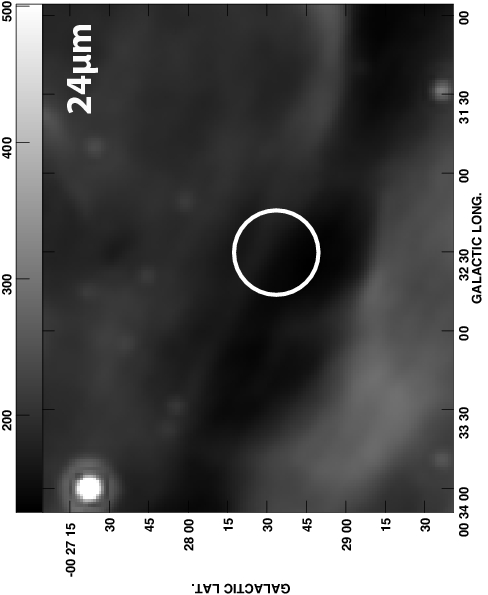}
\includegraphics[origin=c,scale=0.35, angle=270]{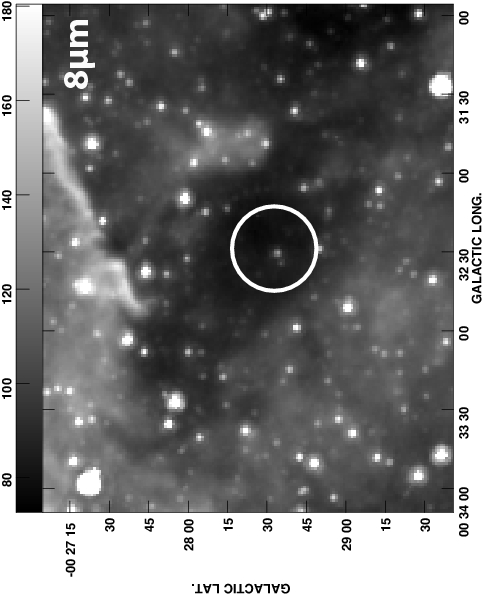}
\includegraphics[origin=c,scale=0.35, angle=270]{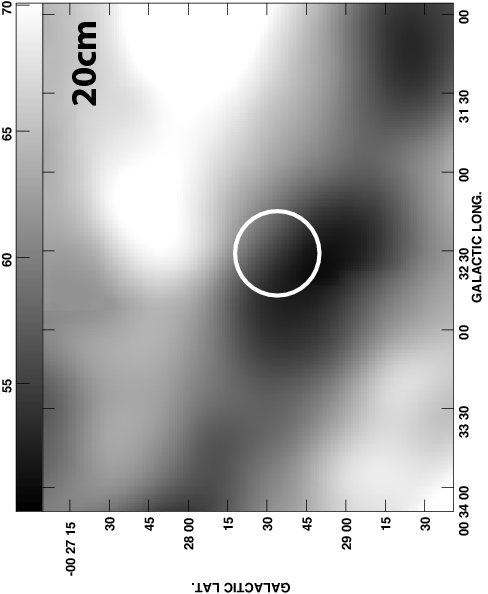}
\includegraphics[origin=c,scale=0.35, angle=270]{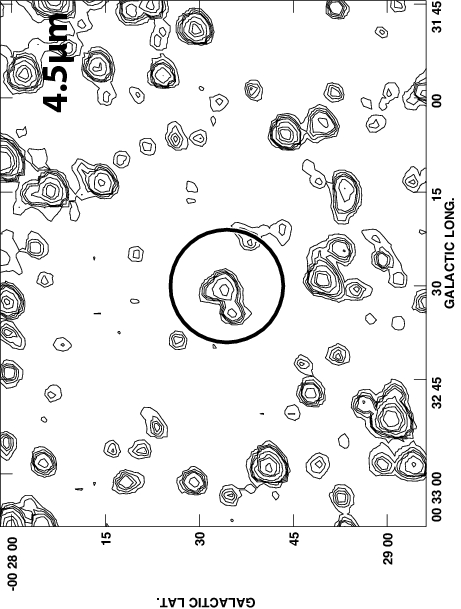}
  \caption{
    (\textit{a - Top Left}) A grayscale image of g11 at 24$\mu$m with a flux range 
between  130 and 500 MJy Sr$^{-1}$. 
    (\textit{b - Top Right}) Similar to (a) except at 8$\mu$m with a flux range 
between 73 and 180 MJy sr$^{-1}$. 
    (\textit{c - Bottom Left}) A 20cm continuum image with a flux range between  
     50 and 80 mJy beam$^{-1}$. 
    (\textit{d - Bottom Right}) Contours of 4.5$\mu$m emission from g11 with  
    contours at levels of 15, 20, 25, 30, 50, 100, 150 and 300 MJy sr$^{-1}$.
Symbols are as defined in Figure 17.
}
\label{fig:36}
\end{figure}

\begin{figure}
\centering
\includegraphics[origin=c,scale=0.4, angle=0]{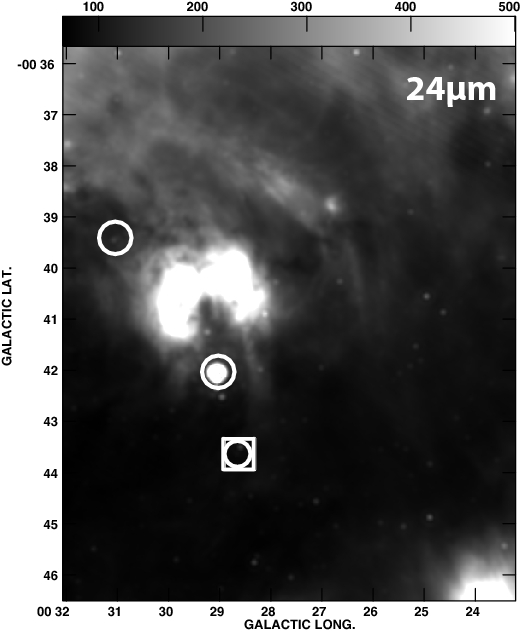}
\includegraphics[origin=c,scale=0.4, angle=0]{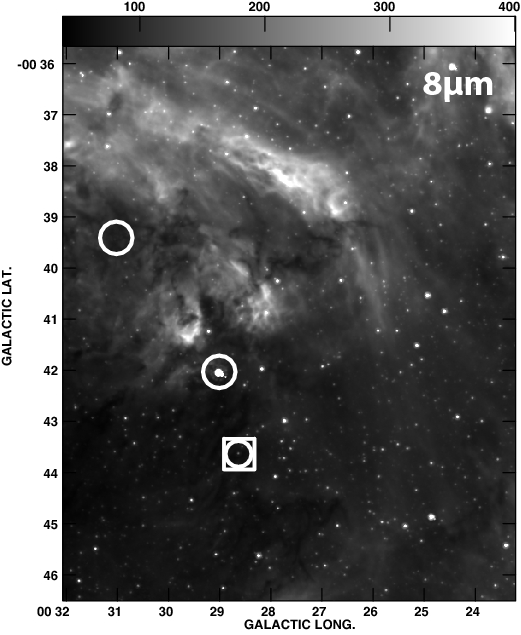}\\
\includegraphics[origin=c,scale=0.4, angle=0]{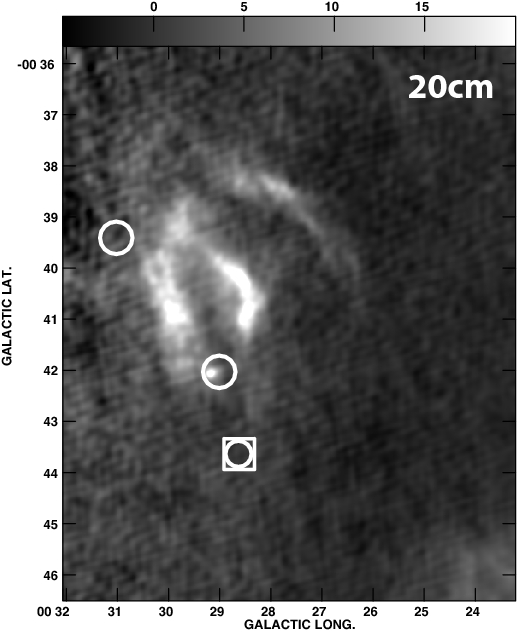}
\includegraphics[origin=c,scale=0.4, angle=0]{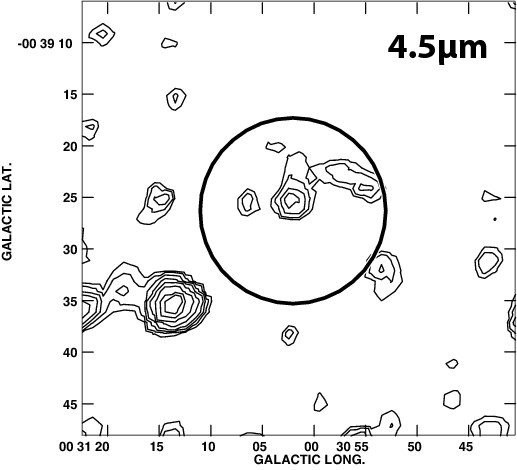}
\caption{
    (\textit{a - Top Left}) Greyscale image of g12, g13, g14 at 24$\mu$m with 
    a flux range between 66 and 500 MJy sr$^{-1}$. 
 (\textit{b - Top Right}) Similar to (a) except at 8$\mu$m with a flux range
    between 39 and 400 MJy sr$^{-1}$. 
    (\textit{c -  Bottom  Left}) Similar to (a) except at 20cm with a flux range 
between 
    -5 and 20 mJy beam$^{-1}$. 
    (\textit{d - Bottom Right})
    Contours of 4.5$\mu$m emission from g12 at  9, 11, 15, 20, 30, 50, 100, 150, 300 
and 500 MJy sr$^{-1}$.
Symbols are as defined in Figure 17.
}
\label{fig:37}
\end{figure}

\begin{figure}
\ContinuedFloat
\centering
\includegraphics[origin=c,scale=0.3, angle=270]{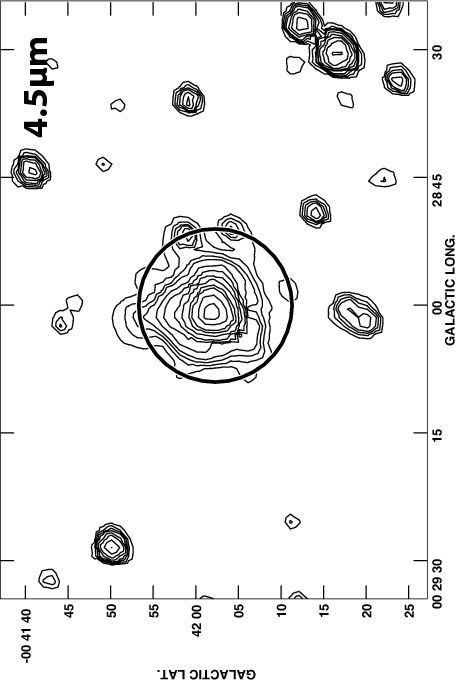}
\includegraphics[origin=c,scale=0.3, angle=270]{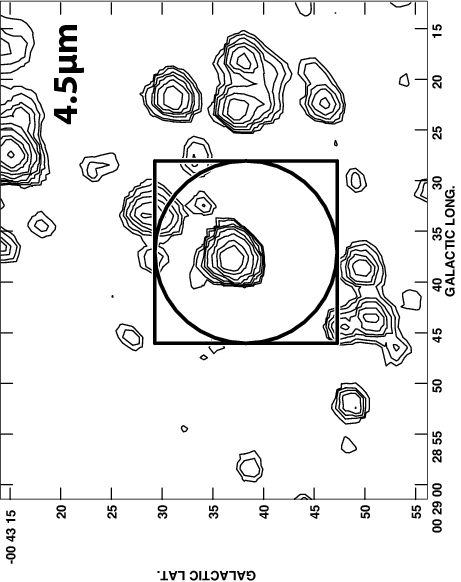}
\caption{
    (\textit{e -  Left}) 
           Contours of 4.5$\mu$m  emission from g13  with the same levels as those of 
(d).  (\textit{f - Right}) 
    Contours of 4.5$\mu$m emission from g14  with the same levels as those of (d). 
Symbols are as defined in Figure 17.
}
\label{fig:37}
\end{figure}

\clearpage
\begin{figure}
\centering
\includegraphics[origin=c,scale=0.3, angle=270]{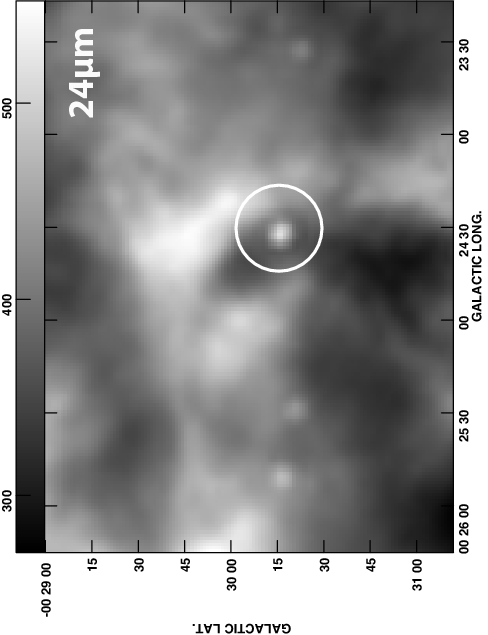}
\includegraphics[origin=c,scale=0.3, angle=270]{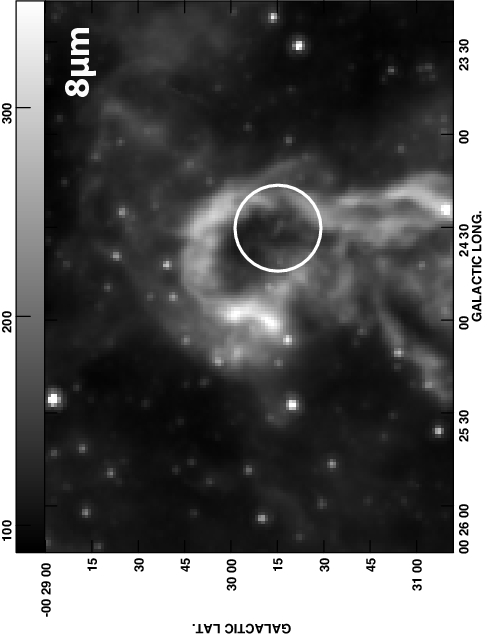}
\includegraphics[origin=c,scale=0.3, angle=270]{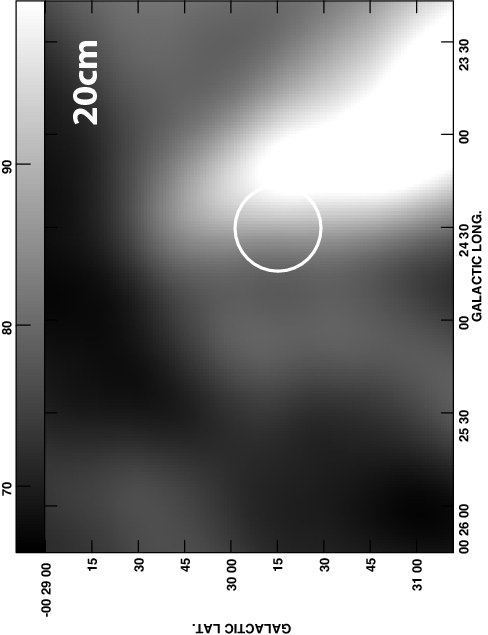}
\includegraphics[origin=c,scale=0.3, angle=270]{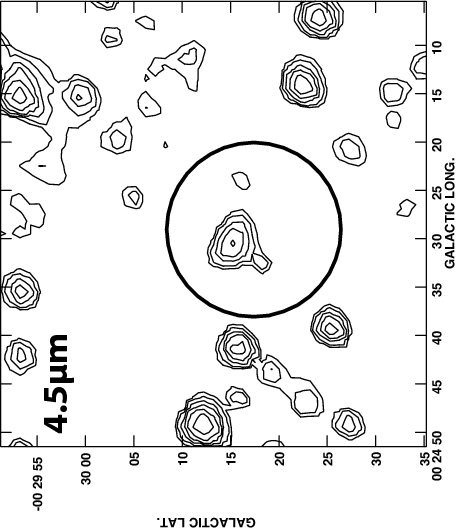}
  \caption{
    (\textit{a - Top Left}) A grayscale image of g15 at 24$\mu$m with a flux 
range 
between  272 and  551 MJy sr$^{-1}$. 
    (\textit{b - Top Right}) Similar to (a) except at 8$\mu$m with a flux range 
between 88 and 350 MJy sr$^{-1}$. 
    (\textit{c - Bottom Left}) A 20cm continuum image with a flux range between  
     66 and 100 mJy beam$^{-1}$. 
    (\textit{d - Bottom Right}) Contours of 4.5$\mu$m emission from g15 with  
    contours at levels of 9, 11, 15, 20, 30, 50, 100, 150, 300 and 500 
      MJy sr$^{-1}$.
Symbols are as defined in Figure 17.
}
\label{fig:38}
\end{figure}

\begin{figure}
\centering
\includegraphics[origin=c,scale=0.35, 
angle=0]{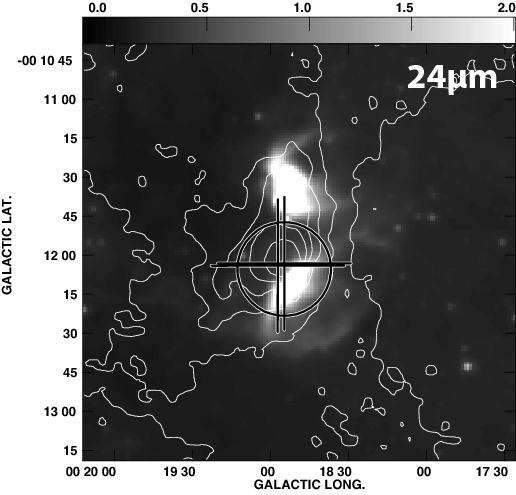}
\includegraphics[origin=c,scale=0.35, 
angle=0]{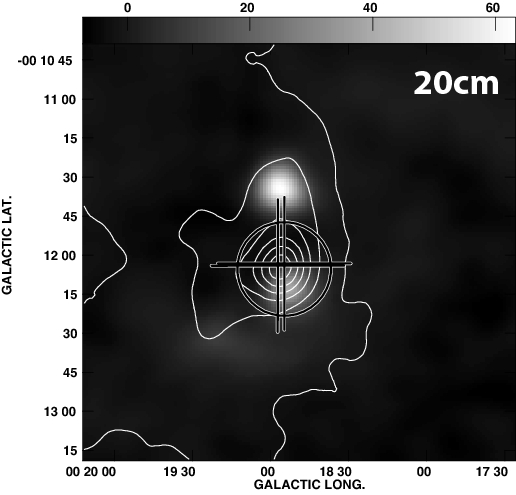}
\includegraphics[origin=c,scale=0.35, angle=270]{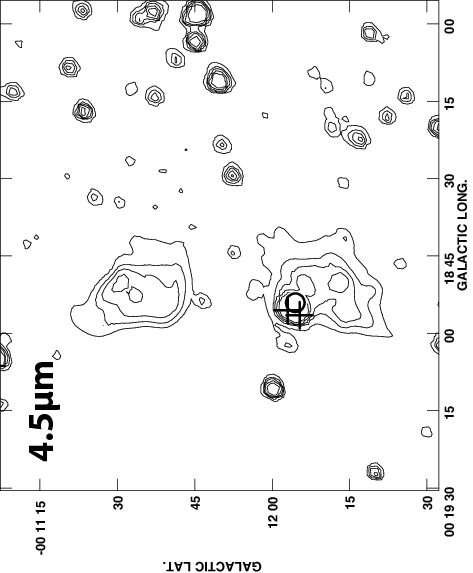}
\caption{
    (\textit{a -Top Left})
    Contours of 450$\mu$m emission from g17 at  8, 10, 13, 
16, 19, 25 Jy/beam are superimposed on a 8$\mu$m image.
    (\textit{b - Top Right}) 
    Contours of 850$\mu$m emission from g17 at  0.5, 1, 2, 3, 4, 5, 6, 
7 Jy beam$^{-1}$ are superimposed on a 20cm continuum image.
    (\textit{c - Bottom})   
    Contours of 4.5$\mu$m emission from g17 at levels of 60, 100, 140, 200 and 
300 MJy sr$^{-1}$.
Symbols are as defined in Figure 17.
}
\label{fig:39}
\end{figure}

\begin{figure}
\centering
\includegraphics[origin=c,scale=0.45, angle=0]{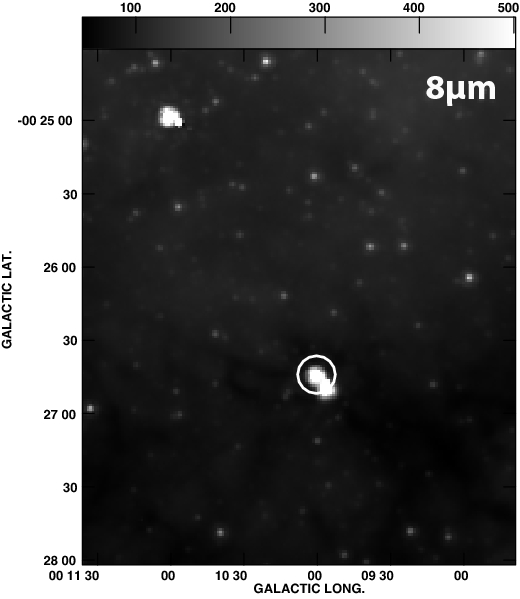}
\includegraphics[origin=c,scale=0.45, angle=0]{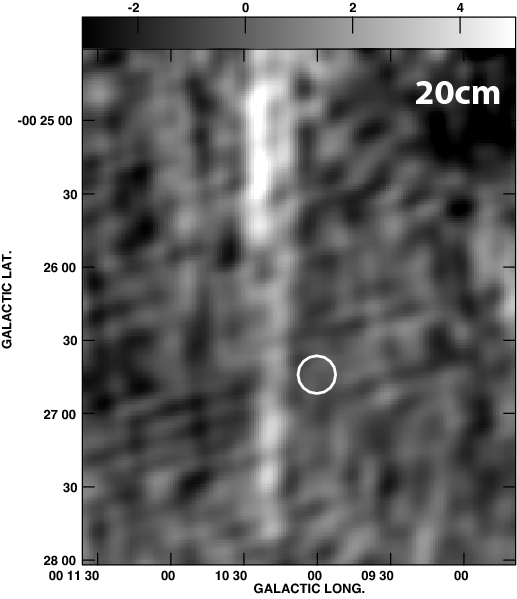}
  \caption{
    (\textit{a - Top}) A greyscale 8$\mu$m image of the region toward g18
 with flux range between  45 and  500 MJy sr$^{-1}$
    (\textit{b - Bottom}) A greyscale 20cm continuum image toward g18 with 
  flux range between  -3 and  5  mJy beam$^{-1}$. 
Symbols are as defined in Figure 17.
}
\label{fig:40}
\end{figure}

\begin{figure}
\centering
\includegraphics[origin=c,scale=0.35, angle=0]{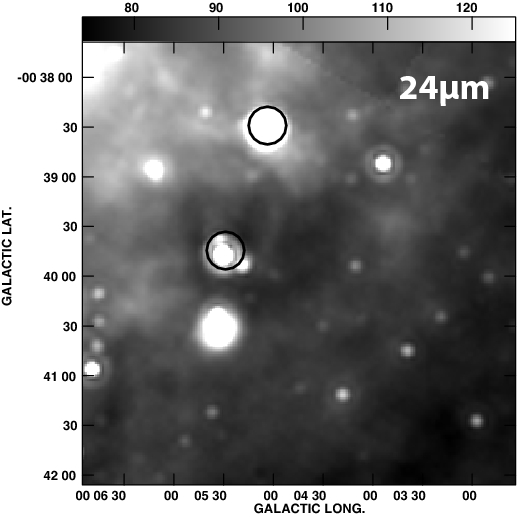}
\includegraphics[origin=c,scale=0.35, angle=0]{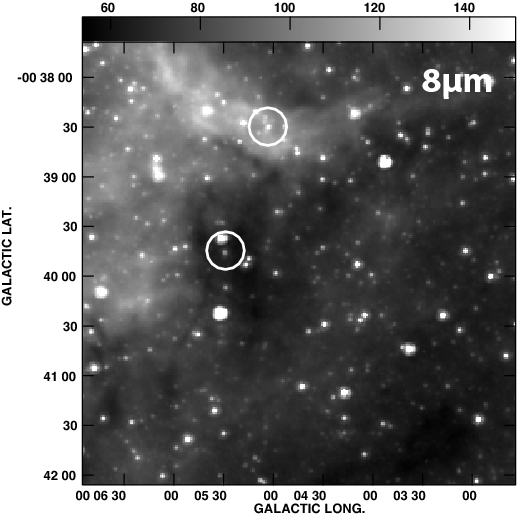}
\includegraphics[origin=c,scale=0.35, angle=0]{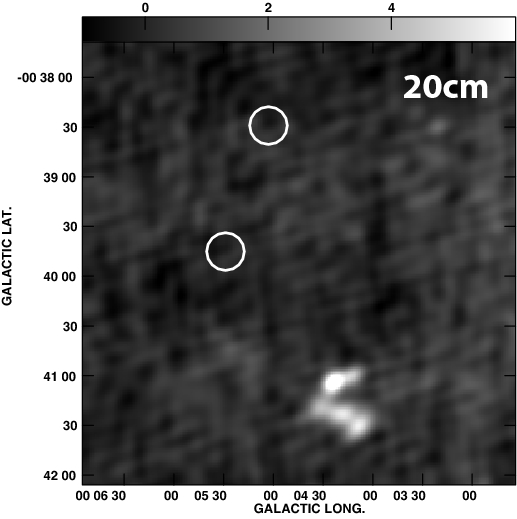}
\includegraphics[origin=c,scale=0.3, angle=270]{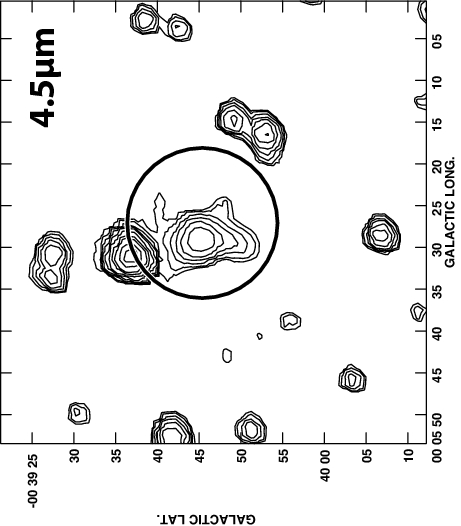}
\includegraphics[origin=c,scale=0.3, angle=0]{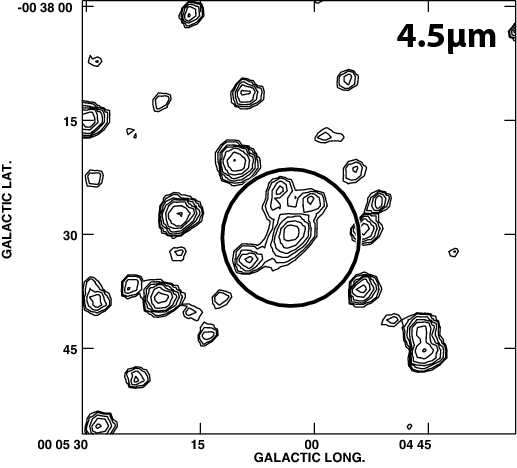}
  \caption{
    (\textit{a - Top Left}) 
A greyscale 24$\mu$m image of g19 and g20 with 
flux range between  74 and  125  MJy sr$^{-1}$. 
    (\textit{b - Top Right}) 
A greyscale 8$\mu$m image of g19 and g20 with 
flux range between  54 and  150  MJy sr$^{-1}$. 
    (\textit{c - Middle Left}) 
A greyscale 20cm continuum  image of g19 and g20 with 
flux range between  -1 and  6   mJy beam$^{-1}$. 
    (\textit{d - Middle Right}) 
    Contours of 4.5$\mu$m emission from g19 at 18, 22, 30, 40, 60, 
100, 200, 300, 600 and 1000 MJy sr$^{-1}$.
    (\textit{e - Bottom})
    Contours of 4.5$\mu$m emission from g20 at 18, 22, 30, 40, 60, 
100, 200, 300, 600 and 1000 MJy sr$^{-1}$.
Symbols are as defined in Figure 17.
}
\label{fig:41}
\end{figure}
\vfill\eject

\begin{figure}
\centering
\includegraphics[origin=c,scale=0.3, angle=270]{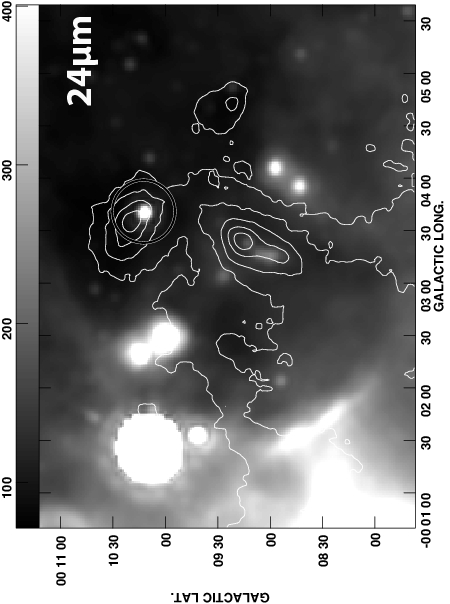}
\includegraphics[origin=c,scale=0.3, angle=270]{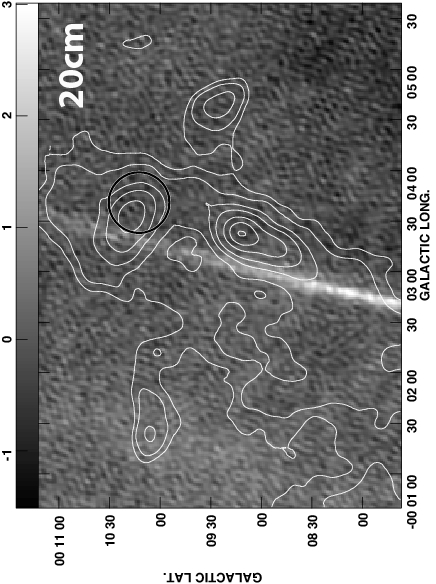}
\includegraphics[origin=c,scale=0.3, angle=0]{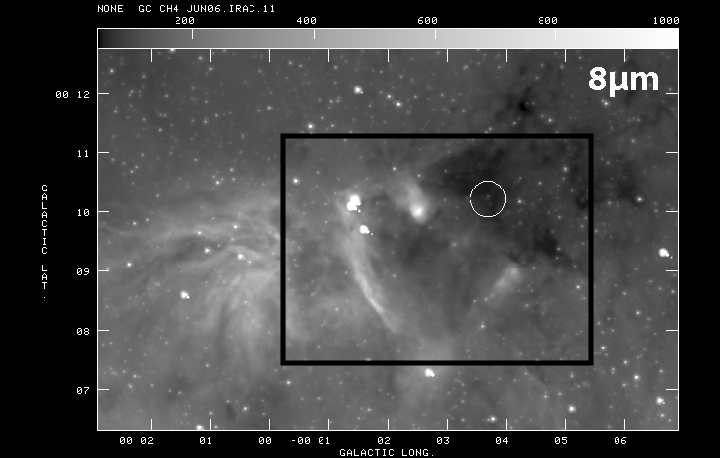}
\includegraphics[origin=c,scale=0.3, angle=0]{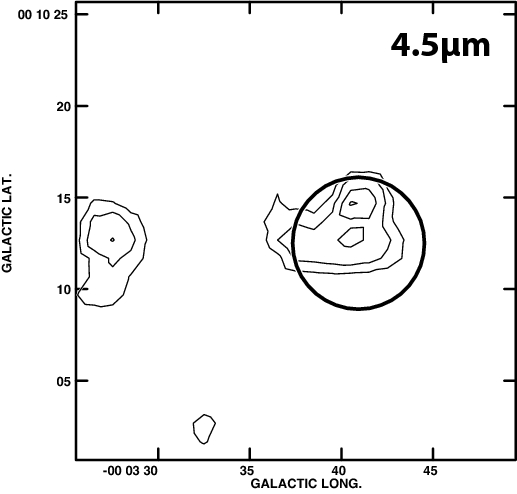}
  \caption{
    (\textit{a - Top Left}) 
    Contours of 450$\mu$m emission from g22 at  
2.5, 3, 3.5, 4, 5, 6 and 7 Jy beam$^{-1}$  are superimposed on a 24$\mu$m 
continuum image.
    (\textit{b - Top Right})
    Contours of 850$\mu$m emission from g22 at  
2.5, 3, 3.5, 4, 5, 6 and 7 Jy beam$^{-1}$  are superimposed on a 20cm  
continuum image.
    (\textit{c - Bottom Left}) 
A greyscale 8$\mu$m image of a large region 
presenting   the distribution of PAH emission in the vicinity of g22. 
The rectangular box drawn on this figure corresponds to the area shown 
in (a) and (b).  
  (\textit{d - Bottom Right}) 
    Contours of 4.5$\mu$m emission from g22 is shown at 15, 20, 30, 
40, 50, 60 MJy sr$^{-1}$. 
Symbols are as defined in Figure 17.
}
\label{fig:42}
\end{figure}

\begin{figure}
\centering
\includegraphics[scale=0.3,angle=270]{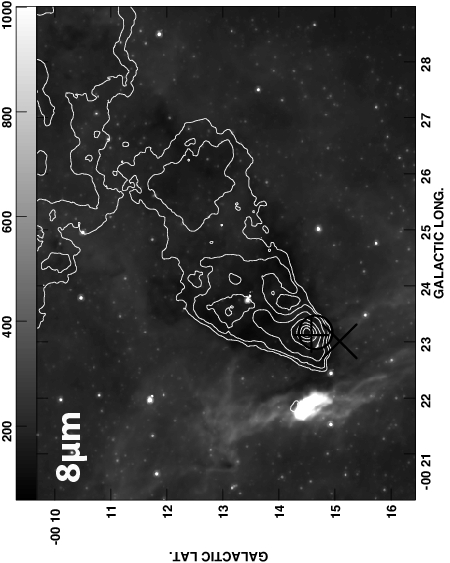}
\includegraphics[scale=0.3,angle=270]{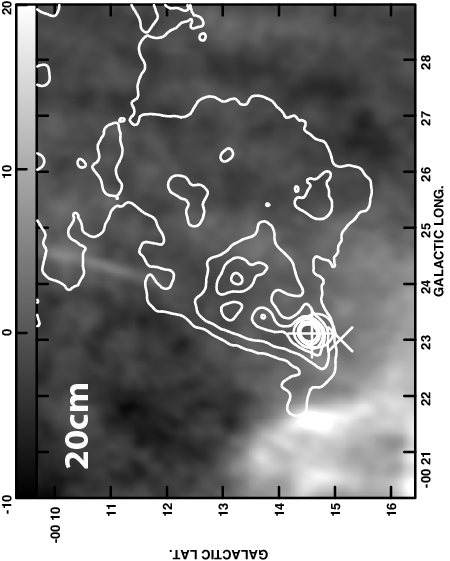}
\includegraphics[scale=0.3]{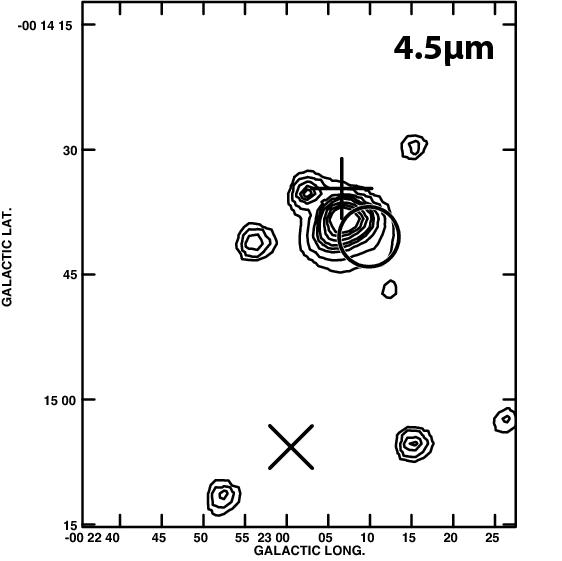}
  \caption{
    (\textit{a - Top Left}) 
    Contours of 450$\mu$m emission from g26 at  
3, 4.5, 8, 15, 25, 40 and 60 Jy beam$^{-1}$  are superimposed on a 8$\mu$m 
continuum image.
    (\textit{b - Top Right}) Contours of 850$\mu$m emission from g26 
at 2, 3, 4, 5, 7  and 9 Jy beam$^{-1}$ are 
superimposed on a grayscale 20cm continuum image.   
  (\textit{c - Bottom})  
    Contours of 4.5$\mu$m emission from g26 is shown at 50, 100, 200, 
300, 400, 600, 800 and 1000 MJy sr$^{-1}$.
Symbols are as defined in Figure 17.
}
\label{fig:43}
\end{figure}

\begin{figure}
\centering
\includegraphics[scale=0.3,angle=0]{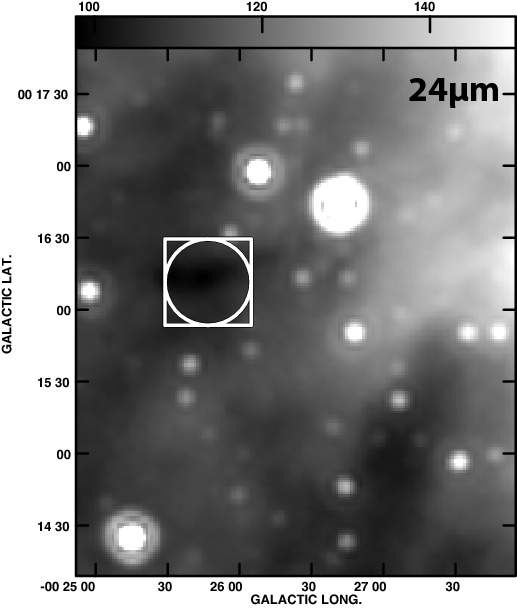}
\includegraphics[scale=0.3,angle=0]{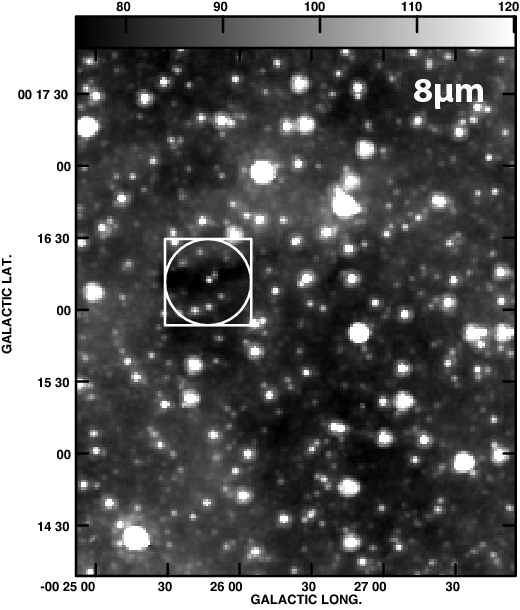}\\
\includegraphics[scale=0.3]{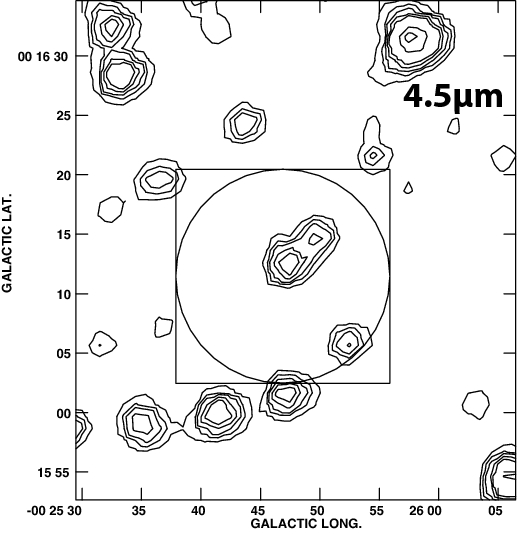}
  \caption{
    (\textit{a -  Top Left})  A 24$\mu$m greyscale image of
 the region in the vicinity 
of g28 which lies at the center of the circle and square  symbols. 
The flux range is between 98 and 150 MJy sr$^{-1}$. 
    (\textit{b - Top Right}) Similar to (a) except that the emission is seen at 
8$\mu$m. The  flux 
range is between   75 and  120 MJy sr$^{-1}$
    (\textit{c - Bottom})  Contour levels of emission at 4.5$\mu$m 
are set at 25, 35,  50, 75, 100, 300 and 500 
MJy sr$^{-1}$. 
Symbols are as defined in Figure 17.
}
\label{fig:44}
\end{figure}

\vfill\eject
\begin{deluxetable}{lrrrrrrrrrrrrrrrrrrrrrrrrr}
\tablecaption{The Glactic coordinates and flux density of sources in G359.43+0.02}
\tabletypesize{\scriptsize}
\tablewidth{0pt}
\tablehead{\colhead{\#} & \colhead{GLON} & \colhead{GLAT} & \colhead{S$_{6cm}\pm\sigma$} &
\colhead{S$_{20cm}\pm\sigma$}
& \colhead{$\theta_{maj} \arcsec$} & \colhead{$\theta_{min} \arcsec$}}
\startdata
 1  & 359.522356 & 0.012511  &   6.5$\pm$1.6 & 4.1$\pm2.2$      & 8.8 & 2.0 \\
 2  & 359.495378 & 0.038569  &   5.2$\pm2.4$ &                  & 12.9 & 10.2 \\
 3  & 359.485425 & 0.083103  &   6.5$\pm3.6$ & 3.0$\pm1.8$      & 7.9 & 3.8 \\
 4  & 359.472092 & 0.052111  &   5.7$\pm2.2$ & 5.3$\pm2.4$     & 6.9 & 0.4 \\
 5\tablenotemark{b}& 359.456575 & 0.019586  &   3.6$\pm1$ &  2.8$\pm1.7$ & 4.2& --- \\
 6\tablenotemark{a}  & 359.429236 & 0.035489  &   3.4$\pm1.2$ &  4.4$\pm0.9$ & 5.8 & --- \\
 7  & 359.434825 & 0.022075  &  12.6$\pm3.4$  &  --- & 19 & 11 \\
 8  & 359.432242 & 0.022986  &   14.4$\pm4$ &  --- & 21.3 & 11.1 \\
 9  & 359.423431 & 0.015069  &   6.9$\pm1.7$  &  4.8$\pm2.5$  & 11 & 7 \\
10  & 359.439063 & 0.005402  &   76.0 & 96$\pm9$ &                 & 900$''^2$ & \\
11\tablenotemark{c}  & 359.413464 & 0.032297  &   1.4$\pm1.7$  &  1.3$\pm2$ & 11.1 & --- \\
11e\tablenotemark{c} &            &           &  49$\pm20$  & 40.3$\pm12$  & 1472$''^2$ & \\
12  & 359.403742 & 0.000797  &   13.8$\pm2.8$ &  6.0$\pm2.3$  & 13.4 & 11.6\\
13  & 359.404019 & -0.000811 &  9.5$\pm8.6$ &  9.7$\pm5.2$  & 11.4 & 9.6 \\
14  & 359.419481 & 0.001042  &   &   &   &  \\
15  & 359.390961 & 0.035469  &   4.3$\pm1.3$  & 9.5$\pm4.2$ & 8.4 & 5.9 \\
16  & 359.373078 & 0.017686  &    &  --- & 9.0 & 8.2 \\
17  & 359.363663 & -0.000190 &  25$\pm3.2$  & 30.8$\pm5.6$ & \\
\enddata
\tablenotetext{b}{Infrared counterpart 300758} \tablenotetext{a}{Infrared counterpart 284291}
\tablenotemark{c}{source 11 has a central compact source  and an extended component}
\end{deluxetable}

\vfill\eject
\begin{deluxetable}{lrrrrrrrrr}
\tablecaption{The Parameters of the SED Fits to Candidate YSOs in G359.43+0.02}
\tabletypesize{\scriptsize}
\tablewidth{0pt}
\tablehead{
  \colhead{SST GC \#} &
  \colhead{$l$} &
  \colhead{$b$} &
  \colhead{$\chi^2_{min}$} & 
  \colhead{nfits} &   
  \colhead{$< \av >$} & 
  \colhead{$< \mstar >$} & 
  \colhead{$ < \lstar > $} & 
  \colhead{$ < \mdote >$}  & 
  \colhead{ $<$Stage$>$}}
\startdata
251914 & 359.4130 & 0.09095 &  0.03  &  310 &  32.1 &   8.0$\pm $1.3 &  1.4$\pm 
$1.2E+03 & 2.7$\pm$ 4.3E-04 &  I   \\
255492 & 359.4145 & 0.08451 &  0.00  &  736 &  31.6 &   5.0$\pm $1.6 &  6.5$\pm 
$7.8E+02 & 1.1$\pm$ 3.5E-04 &  I  \\
272626 & 359.4288 & 0.05876 &  4.57  &  240 &  18.8 &  10.1$\pm $0.7 &  6.5$\pm $1.2E+03 & 0.0$\pm $0.0E+00 &  II \\
276597 & 359.4709 & 0.07663 &  5.97  &   52 &  15.0 &   7.5$\pm $0.5 &  2.3$\pm $0.7E+03 & 0.0$\pm $0.0E+00 &  III\\
284291 \tablenotemark{a}
       & 359.4287 & 0.03534 &  0.07  &  204 &  37.6 &   9.5$\pm $1.0 &  5.2$\pm $1.9E+03 & 0.8$\pm $3.8E-04 &  I  \\
297164 & 359.4352 & 0.01400 &  0.00  &  739 &  36.2 &   8.4$\pm $0.9 &  2.4$\pm $1.2E+03 & 1.3$\pm $3.5E-04 &  I  \\
297494 & 359.4688 & 0.03400 &  0.00  &  703 &  34.9 &   7.5$\pm $0.9 &  1.7$\pm $1.0E+03 & 1.7$\pm $4.6E-04 &  I  \\
300758 \tablenotemark{b}
       & 359.4562 & 0.01993 &  1.66  &   59 &  17.6 &   8.8$\pm $1.1 &  3.3$\pm $1.6E+03 & 5.1$\pm $5.2E-05 &  I  \\
\enddata
\tablenotetext{a}{Radio counterpart 6} \tablenotetext{b}{Radio counterpart 5}
\end{deluxetable}

\begin{deluxetable}{lrrrrrrrrr}
\tablecaption{The Parameters of the SED Fits to Candidate YSOs in $l < 0\arcdeg$}
\tabletypesize{\scriptsize}
\tablewidth{0pt}
\tablehead{
  \colhead{SST GC \#} &
  \colhead{$l$} &
  \colhead{$b$} &
  \colhead{$\chi^2_{min}$} & 
  \colhead{nfits} &   
  \colhead{$< \av >$} & 
  \colhead{$< \mstar >$} & 
  \colhead{$ < \lstar > $} & 
  \colhead{$ < \mdote >$}  & 
  \colhead{ $<$Stage$>$}}
\startdata
      116907 & 358.7362 &  0.0045 &   0.80  &   13 &  16.3 &   7.7$\pm $0.5 &  1.2$\pm $0.3E+03&  7.0$\pm $9.5E-04&  I   \\ 
      119577 & 358.8940 &  0.0943 &  18.80  &    3 &  15.0 &   7.0$\pm $0.0 &  4.7$\pm $0.1E+02&  2.6$\pm $3.5E-04&  I   \\ 
      119780 & 358.8605 &  0.0729 &   0.07  &  471 &  29.7 &   4.7$\pm $0.8 &  4.2$\pm $3.3E+02&  0.2$\pm $1.8E-04&  I   \\ 
      121566 & 358.8108 &  0.0369 &   5.64  &   10 &  15.5 &   6.9$\pm $0.0 &  1.5$\pm $0.0E+03&  0.0$\pm $0.0E-00&  III \\ 
      122375 & 358.7448 & -0.0062 &   0.00  & 1072 &  28.1 &   4.1$\pm $1.6 &  2.5$\pm $2.7E+02&  0.7$\pm $1.7E-04&  I   \\ 
      126327 & 358.8288 &  0.0347 &   7.12  &    6 &  15.0 &   7.0$\pm $0.1 &  4.9$\pm $0.4E+02&  3.1$\pm $3.3E-04&  I   \\ 
      127127 & 358.7587 & -0.0108 &   0.00  &  485 &  34.7 &   5.1$\pm $1.5 &  0.8$\pm $1.0E+03&  0.3$\pm $2.8E-09&  II  \\ 
      127297 & 358.7373 & -0.0245 &   0.00  & 1537 &  38.9 &   8.9$\pm $1.0 &  3.4$\pm $1.4E+03&  1.1$\pm $3.0E-04&  I   \\
\enddata
\end{deluxetable}

\begin{deluxetable}{lrrrrrrrrr}
\tablecaption{The Parameters of the SED
Fits to Candidate YSOs in $l > 0\arcdeg$}
\tabletypesize{\scriptsize}
\tablewidth{0pt}
\tablehead{
  \colhead{SST GC \#} &
  \colhead{$l$} &
  \colhead{$b$} &
  \colhead{$\chi^2_{min}$} & 
  \colhead{nfits} &   
  \colhead{$< \av >$} & 
  \colhead{$< \mstar >$} & 
  \colhead{$ < \lstar > $} & 
  \colhead{$ < \mdote >$}  & 
  \colhead{ $<$Stage$>$}}
\startdata
      625950 &   0.4807 &  0.1243 &   2.90 &     2 &  15.0 &  8.8$\pm $0.1 &  1.4$\pm $0.2E+03 & 1.5$\pm $0.8E-03 &  I  \\ 
      630016 &   0.5217 &  0.1431 &   0.35 &   220 &  36.5 & 10.2$\pm $1.5 &  3.8$\pm $1.4E+03 & 0.4$\pm $1.1E-03 &  I  \\ 
      630178 &   0.3082 &  0.0130 &   1.89 &    43 &  15.6 &  6.7$\pm $0.4 &  1.0$\pm $0.6E+03 & 0.6$\pm $1.7E-04 &  I  \\ 
      630480 &   0.3268 &  0.0238 &   5.97 &    49 &  15.0 &  8.3$\pm $0.7 &  3.5$\pm $0.9E+03 & 0.0$\pm $0.0E-00 &  II \\ 
      630841 &   0.3311 &  0.0258 &   0.00 &  1087 &  33.4 &  7.7$\pm $1.0 &  2.4$\pm $1.2E+03 & 1.0$\pm $4.2E-04 &  I  \\ 
      630883 &   0.1551 & -0.0813 &   7.85 &    62 &  15.1 &  9.6$\pm $0.6 &  5.7$\pm $1.2E+03 & 0.0$\pm $0.0E-00 &  II \\ 
      637548 &   0.5413 &  0.1433 &   6.75 &     7 &  15.0 &  6.2$\pm $0.3 &  2.8$\pm $0.4E+02 & 2.7$\pm $3.8E-04 &  I  \\ 
      639320 &   0.3763 &  0.0402 &  10.22 &     5 &  21.6 & 10.0$\pm $0.7 &  3.2$\pm $0.6E+03 & 7.4$\pm $2.7E-04 &  I  \\
\enddata
\end{deluxetable}

\begin{deluxetable}{lllll}
\tablecaption{Table of Sources* with 4.5$\mu$m  Excess Emission}
\tabletypesize{\scriptsize} \tablewidth{0pt}
\tablehead{\colhead{Source} &  \colhead{Galactic} & 
\colhead{$\alpha_{J2000}$} & 
\colhead{$\delta_{J2000}$} & \colhead{Description} \\  \colhead{} & 
\colhead{l b} & 
\colhead{h m s} & \colhead{$\arcdeg$ $\arcmin$ $\arcsec$} & \colhead{}}
\startdata
0  &1.041-0.072 & 17 48 21.37 & -28 05 00.7  & point\\
1  &0.955-0.786 & 17 50 56.34 & -28 31 24.4  & PN, extended?\\
2  &0.868-0.697 & 17 50 23.32 & -28 33 11.6  & PN, point\\
3  &0.826-0.211 & 17 48 23.59 & -28 20 20.7  & point\\
4  &0.780-0.740 & 17 50 21.09 & -28 39 03.2  & PN, extended\\
5  &0.707+0.408 & 17 45 42.73 & -28 07 12.1  & extinction of a dark cloud?\\
6  &0.692-0.046 & 17 47 26.27 & -28 22 06.7  & extended\\
7  &0.679-0.038 & 17 47 22.65 & -28 22 31.6  & point\\
\enddata
\ * Visually identified from IRAC data. Many point sources may be multiple or extended 
sources at higher resolution.
\end{deluxetable}

\begin{deluxetable}{lllllll}
\tablecaption{Table of 70 $\mu$m sources}
\tabletypesize{\scriptsize} \tablewidth{0pt}
\tablehead{
 \colhead{source} & 
 \colhead{l} & \colhead{b} &
 \colhead{$\alpha_{J2000}$} &
 \colhead{$\delta_{J2000}$} &
 \colhead{Flux (mJy)} &
 \colhead{S/N}}
\startdata
1  &  0.95549  &  -0.78554  & 17 50 56.34 & -28 31 24.4  &   446 &  9.19 
\\ 
22*&  0.83579  &  +0.18444  & 17 46 52.86 & -28 07 34.8  & 98080 & 43.32 
\\ 
3  &  0.82608  &  -0.21067  & 17 48 23.59 & -28 20 20.7  & --    &  --   
\\ 
5  &  0.70721  &  +0.40788  & 17 45 42.73 & -28 07 12.1  & --    &  --   
\\ 
10*&  0.49639  &  +0.18764  & 17 46 04.07 & -28 24 52.8  & 76270 & 39.54 
\\ 
14 &  0.47696  &  -0.72730  & 17 49 35.52 & -28 54 15.8  & --    &  --   
\\ 
24 &359.90724  &  -0.30319  & 17 46 35.05 & -29 10 23.2  & --    &  --   
\\ 
28 &359.57031  &  +0.26985  & 17 43 32.37 & -29 09 40.8  &  2540 &  6.53 
\\ 
31 &359.19896  &  +0.04097  & 17 43 32.35 & -29 35 51.2  & --    &  --   
\\ 
\enddata 
\ * Source \# given is for table of methanol class II maser sources.
\end{deluxetable}
\begin{deluxetable}{llllll}
\tablecaption{Table of Methanol Class I Maser Sources}
\tabletypesize{\scriptsize} \tablewidth{0pt}
\tablehead{
 \colhead{Source} &
 \colhead{Galactic} &
 \colhead{$\alpha_{J2000}$} &
 \colhead{$\delta_{J2000}$} &
 \colhead{v$_{LSR}$}&
 \colhead{Reference}
 \\
 \colhead{} &
 \colhead{l b} &
 \colhead{h m s} &
 \colhead{$\arcdeg \arcmin \arcsec$} &
 \colhead{km s$^{-1}$}&
 \colhead{}}
\startdata
0  & 359.617-0.252  &  17 45 41.32  &  -29 23 40.5  &  +19.3  & S94,V00 \\
1  & 359.898-0.066  &  17 45 38.21  &  -29 03 27.7  &  +21    & H90,V00 \\
2  & 359.982-0.075  &  17 45 52.01  &  -28 59 26.7  &  +44.5  & H90,V00 \\
3  & 0.682-0.026  &  17 47 20.19 & -28 22 03.2  &+54 &    M97  \\
4  & 0.678-0.023  &  17 47 18.80 & -28 22 08.3  &+68 &    M97  \\
5  & 0.677-0.023  &  17 47 18.71 & -28 22 09.5  &+70 &    M97  \\
6  & 0.679-0.026  &  17 47 19.72 & -28 22 09.5  &+66 &    M97  \\
7  & 0.679-0.026  &  17 47 19.74 & -28 22 09.6  &+66 &    M97  \\
\enddata

References:
H90 = Haschick, Menten \& Baan 1990 (44 GHz);
S94 = Slysh et al. 1994 (44 GHz);
M97 = Mehringer et al. 1997 (44 GHz, multiple sources in Sgr B2);
V00 = Val'tts et al. 2000 (95 GHz)
\end{deluxetable}

\begin{deluxetable}{llllll}
\tablecaption{Table of Methanol Class II Maser Sources}
\tabletypesize{\scriptsize} \tablewidth{0pt}
\tablehead{
 \colhead{Source} &
 \colhead{Galactic} &
 \colhead{$\alpha_{J2000}$} &
 \colhead{$\delta_{J2000}$} &
 \colhead{v$_{LSR}$}&
 \colhead{Reference}
 \\
 \colhead{} &
 \colhead{l b} &
 \colhead{h m s} &
 \colhead{$\arcdeg \arcmin \arcsec$} &
 \colhead{km s$^{-1}$}&
 \colhead{}}
\startdata
0  &359.138+0.031  &  17:43:25.69  &  -29:39:17.4  &  -4.0 &  C96,W98 \\
1  &359.436-0.104  &  17:44:40.60  &  -29:28:16.0  & -52.0 &  C96 \\
2  &359.436-0.102  &  17:44:40.21  &  -29:28:12.5  & -53.6 &  C96 \\
3  &359.615-0.243  &  17:45:39.09  &  -29:23:30.0  & +22.5 &  C96 \\
4  &359.970-0.457  &  17:47:20.17  &  -29:11:59.4  & +23.0 &  C96 \\
5  &  0.212-0.001  &  17:46:07.63  &  -28:45:20.9  & +49.2 &  C96,W98 \\
6  &  0.315-0.201  &  17:47:09.13  &  -28:46:15.7  & +18.0 &  C96,W98 \\
7  &  0.316-0.201  &  17:47:09.33  &  -28:46:16.0  & +21.0 &  C96 \\
\enddata

References: C96 = Caswell 1996; W98 = Walsh et al. 1998
\end{deluxetable}

\begin{deluxetable}{lrrrrrrrrrr}
\tablecaption{The Parameters of SED Fits to 4.5$\mu$m Excess 
 Sources near the Galactic Center}
\tabletypesize{\scriptsize}
\tablewidth{0pt}
\tablehead{
  \colhead{Green \#} &
  \colhead{SST GC \#} &
  \colhead{$l$} &
  \colhead{$b$} &
  \colhead{$\chi^2_{min}$} & 
  \colhead{nfits} &   
  \colhead{$<$A$_v >$} & 
  \colhead{$< \msol >$} & 
  \colhead{$ < \lsol > $} & 
  \colhead{$ < \dot{M} >$}  & 
  \colhead{ $<$Stage$>$}}
\startdata
 g0 & 909173 &   1.0407 & -0.0714  &  4.36 &  31 & 42.4 & 16.1$\pm $3.5 & 2.8$\pm $1.4E+04 &0.0$\pm $0.0E+00 &III  \\
 g3 & 913321 &   0.8262 & -0.2108  & 11.45 &   4 & 43.6 & 10.6$\pm $0.0 & 7.6$\pm $0.0E+03 &0.0$\pm $0.0E+00 &II   \\
 g6 & 803187 &   0.6934 & -0.0451  &  1.71 &  46 & 46.7 & 29.9$\pm $8.2 & 1.4$\pm $1.0E+05 &1.3$\pm $0.6E-03 &I    \\
 g7 & 794899 &   0.6796 & -0.0376  &  7.33 &  20 &  7.5 &  9.4$\pm $1.2 & 3.1$\pm $0.7E+03 &5.8$\pm $5.8E-04 &I    \\
 g9 & 789483 &   0.6667 & -0.0352  &  1.37 & 169 & 27.0 & 21.7$\pm $7.0 & 6.1$\pm $6.4E+04 &1.6$\pm $1.5E-03 &I    \\
g10 & 798750 &   0.6655 & -0.0535  &  1.96 &  12 &  5.1 &  9.5$\pm $0.7 & 2.0$\pm $0.5E+03 &3.5$\pm $8.4E-04 &I    \\
g16 & 639320 &   0.3763 &  0.0402  & 10.22 &   9 & 14.0 & 10.1$\pm $0.6 & 3.2$\pm $0.6E+03 &8.1$\pm $2.7E-04 &I    \\
g23 & 530763 & 359.9316 & -0.0626  &  2.72 &  32 & 17.9 & 12.0$\pm $3.2 & 9.9$\pm $4.6E+03 &1.3$\pm $0.6E-03 &I    \\
\enddata
\end{deluxetable}

\begin{deluxetable}{lrrrrrrrrrr}
\tablecaption{The Parameters of SED Fits to 4.5$\mu$m Excess Sources foreground to the Galactic Center}
\tabletypesize{\scriptsize}
\tablewidth{0pt}
\tablehead{
  \colhead{Green \#} &
  \colhead{SST GC \#} &
  \colhead{$l$} &
  \colhead{$b$} &
  \colhead{$\chi^2_{min}$} & 
  \colhead{nfits} &   
  \colhead{$< \av >$} & 
  \colhead{$< \mstar >$} & 
  \colhead{$ < \lstar > $} & 
  \colhead{$ < \mdote >$}  & 
  \colhead{ $<$Stage$>$}}
\startdata
 g1 & 1060982 &   0.9555 & -0.7853 &  0.96 &   12 &  5.6  &  2.4$\pm $0.6 & 2.4$\pm $0.7E+01 & 6.2$\pm $2.4E-06  & I     \\
 g2 & 1043897 &   0.8679 & -0.6963 &  1.97 &  547 & 33.1  &  5.5$\pm $1.3 & 8.0$\pm $5.7E+02 & 1.0$\pm $3.0E-05  & I     \\
 g4 & 1042278 &   0.7800 & -0.7395 &  0.77 &  161 & 10.3  &  6.9$\pm $1.9 & 8.9$\pm $8.0E+02 & 9.5$\pm $6.3E-05  & I     \\
 g5 &  532335 &   0.7062 &  0.4075 &  1.11 &   89 &  4.7  &  6.0$\pm $3.0 & 1.7$\pm $4.4E+03 & 1.2$\pm $1.4E-04  & I     \\
g11 &  946558 &   0.5418 & -0.4759 &  0.25 &  152 &  5.0  &  5.2$\pm $2.5 & 0.7$\pm $2.4E+03 & 1.1$\pm $1.2E-04  & I     \\
g12 &  992035 &   0.5172 & -0.6569 &  0.00 & 2643 &  9.9  &  2.1$\pm $1.2 & 3.1$\pm $3.4E+01 & 3.4$\pm $7.3E-05  & I     \\
g13 &  997790 &   0.4835 & -0.7004 &  1.82 &    4 &  4.7  &  9.9$\pm $1.7 & 3.3$\pm $2.5E+03 & 0.6$\pm $1.0E-03  & I     \\
g14 & 1003258 &   0.4771 & -0.7268 &  1.69 &   24 &  4.9  &  7.4$\pm $1.7 & 1.0$\pm $1.3E+03 & 2.7$\pm $1.4E-04  & I     \\
\enddata
\end{deluxetable}

\end{document}